\documentclass[11pt]{article}
\usepackage{geometry}
\usepackage{amsmath, amsthm, amssymb, mathrsfs, graphicx, caption, float, multirow, upgreek}

\usepackage{graphicx, caption, float}

\allowdisplaybreaks

\usepackage{fourier, anyfontsize}

\usepackage{fancyhdr}
\usepackage{hyperref}
\usepackage[dvipsnames]{xcolor}
\hypersetup{colorlinks=true, linkcolor=Cerulean, citecolor=Cerulean}

\usepackage{chngcntr, etoolbox}
\counterwithin*{equation}{section}
\counterwithin*{equation}{subsection}
\renewcommand\theequation{\ifnumgreater{\value{subsection}}{0}{\thesubsection.}{\thesection.}\arabic{equation}}

\newcommand{\tabincell}[2]{\begin{tabular}{@{}#1@{}}#2\end{tabular}}

\newtheoremstyle{theorem}
  {10pt}
  {10pt}
  {\sl}
  {\parindent}
  {\bf}
  {. }
  { }
  {}
\theoremstyle{theorem}
\newtheorem{theorem}{Theorem}
\newtheorem{proposition}{Proposition}
\newtheorem{corollary}{Corollary}

\begin{document}
\title{\textbf{Regulating stochastic clocks}}
\author{Zhe Fei\thanks{Department of Finance, Boston University Questrom School of Business, Boston, MA, USA.}~~\thanks{Email: \underline{zhefei@bu.edu}} \and Weixuan Xia\footnotemark[1]~~\thanks{Corresponding author. Email: \underline{gabxia@bu.edu}}}
\date{Started 2021 \\ This version: \today}
\maketitle

\begin{abstract}
  Stochastic clocks represent a class of time change methods for incorporating trading activity into continuous-time financial models, with the ability to deal with typical asymmetrical and tail risks in financial returns. In this paper we propose a significant improvement of stochastic clocks for the same objective but without decreasing the number of trades or changing the trading intensity. Our methodology targets any L\'{e}vy subordinator, or more generally any process of nonnegative independent increments, and is based on various choices of regulating kernels motivated from repeated averaging. By way of a hyperparameter linked to the degree of regulation, arbitrarily large skewness and excess kurtosis of returns can be easily achieved. Generic-time Laplace transforms, characterizing triplets, and cumulants of the regulated clocks and subsequent mixed models are analyzed, serving purposes ranging from statistical estimation and option price calibration to simulation techniques. Under specified jump--diffusion processes and tempered stable processes, a robust moment-based estimation procedure with profile likelihood is developed and a comprehensive empirical study involving S\&P500 and Bitcoin daily returns is conducted to demonstrate a series of desirable effects of the proposed methods. \medskip
\end{abstract}

\newcommand{\dd}{{\rm d}}
\newcommand{\pd}{\partial}
\newcommand{\ii}{{\rm i}}
\newcommand{\PP}{\mathsf{P}}
\newcommand{\Q}{\mathsf{Q}}
\newcommand{\E}{\textsf{E}}
\newcommand{\Var}{\textsf{Var}}
\newcommand{\Skew}{\textsf{Skew}}
\newcommand{\EKurt}{\textsf{EKurt}}
\newcommand{\Gf}{\mathrm{\Gamma}}
\newcommand{\Li}{\mathrm{Li}}
\newcommand{\F}{\mathrm{F}}
\newcommand{\G}{\mathrm{G}}
\newcommand{\D}{\varDelta}
\renewcommand{\Re}{\mathrm{Re}}
\renewcommand{\Im}{\mathrm{Im}}
\renewcommand{\Xi}{\varXi}

\medskip

\section{Introduction}\label{sec:1}

Financial returns have long been documented to have excessive skewness and kurtosis relative to those of an (un)conditional Gaussian distribution, leading to asymmetrical and tail risks; see, among others, [Bollerslev, 1987] \cite{B4}, [Mittnik et al., 2000] \cite{MPR}, [Jondeau and Rockinger, 2003] \cite{JR}, and [Ornthanalai, 2014] \cite{O}. This phenomenon appears to be even more salient in the recently burgeoning cryptocurrency market which exhibits very high volatility due to its heavy reliance on supply and demand and market attention; see [Troster et al., 2019] \cite{TTSM}, [Gkillas and Katsiampa, 2018] \cite{GK}, [Chaim and Laurini, 2019] \cite{CL}, and [Liu and Tsyvinski, 2021] \cite{LT}, e.g., for the major cryptocurrencies including Bitcoin.

In continuous-time modeling, a well-known method for incorporating distributional asymmetry and heavy tails is to replace the usual, constantly forward-moving clock under which the underlying stochastic process is monitored with a generally independent nondecreasing stochastic process. The latter process is termed a ``stochastic clock'' and can be linked to trading volumes or the number of trades (see [Clark, 1973] \cite{C2}, [An\'{e} and Geman, 2000] \cite{AG}, [Geman et al., 2001] \cite{GMY}, and [Geman, 2008] \cite{G1}). Stemming from the celebrated Dambis--Dubins--Schwarz theorem of time change, [Monroe, 1978] \cite{M} first gave the possibility to represent any semimartingale as a time-changed Brownian motion. When the stochastic clock is modeled by a nonnegative process with independent and stationary increments, namely a L\'{e}vy subordinator, such a representation has had striking consequences in generating a variety of purely discontinuous L\'{e}vy processes as mixed models, which have flourished for the last two decades in time series econometrics, option pricing, and portfolio management, some recent contributions including [Madan et al., 2019] \cite{MRS}, [Aguilar et al., 2020] \cite{AKK}, and [Fallahgoul and Loeper, 2021] \cite{FL}.

The key idea behind using stochastic clocks in the absence of a drift component in producing large skewness and kurtosis is rather comprehensible: Pure-jump stochastic clocks progress in a staircase-like manner and are less active -- and less uniform -- than the usual ``sloped'' clock of calendar time, and hence possess relatively slower local speed, representing trading activity that does not take place in a perfectly continuous manner, which enables them to adequately tell apart periods of intensive trading and other relatively calmer ones. Perhaps the two most predominant clocks of this type are the gamma process and the inverse Gaussian process, which were studied in [Madan et al., 1998] \cite{MCC} and [Barndorff-Nielsen, 1997] \cite{B-N}, respectively, to time-change a drifted Brownian motion for building discontinuous models in dealing with skewed and leptokurtic financial returns.\footnote{The same processes also play an important role in many nonfinancial fields, especially in reliability engineering to model the degradation phenomena of structural components (see the overview by [Ye and Xie, 2015] \cite{YX}).}

Noteworthily, when working with ordinary stochastic clocks under L\'{e}vy subordinators, the asymmetrical and tail risks of returns are largely controlled by the number of trades occurring for a significant period of time (i.e., the aggregate level of trading activity) through an infinite-divisibility (i.e., time-multiplicative) parameter (as mentioned in the overview in [Schoutens, 2003, \text{Chap.} 4] \cite{S2}), or the long-term trading intensity (equivalently, the variation of trading volumes) via an $\alpha$-stability index (in the case of the stable family; see [Samorodnitsky and Taqqu, 1994] \cite{ST} e.g.). More specifically, the matching of a high empirical kurtosis level will require decreasing the value of the infinite-divisibility parameter which is a direct measurement of the amount of jumps,\footnote{The intuition is that as such a parameter tends to zero in value, the corresponding distribution becomes degenerate.} or altering the stability index which is heavily tied to the microstructure or the local regularity of sample paths. Such an operation seems totally innocuous when return distributions are analyzed statically, after imposing assumptions of \text{i.i.d.} samples. However, in reality, the aggregate number of trades cannot be expected to be unlimitedly small, or close to zero, leading effectively to degeneracy, or trading halts; on the other hand, the long-term trading intensity is generally specified a priori without being optimized and may be estimated in a dynamical fashion using techniques of empirical power variations (see, e.g., [A\"{\i}t-Sahalia and Jacod, 2009] \cite{AJ}, [Todorov and Tauchen, 2011] \cite{TT}, [Jing et al., 2012] \cite{JKLM}, and [Todorov, 2021] \cite{T}), which of course varies with the frequencies at which data are observed. With restrictions on these two important dimensions, nearly all the noticed L\'{e}vy models will not function well in coping with diverse asymmetrical and tail risks of returns.

In connection with this, we are motivated to consider yet another possibility for the same purpose, by changing the size of local movements of a stochastic clock instead of decreasing their amount or adjusting the corresponding activity level. From the perspective of trading activity, this would mean that the spread of trade sizes may be enlarged at will without having to alter the level of trading intensity or stick to an implausibly small number of trades. In doing so, we prefer to operate in a continuous-time environment because discrete-time analogies are easy to develop subsequently, whereas it will be understandably difficult to go in the opposite direction, e.g., to ensure infinite divisibility in finite dimensions.

Besides, consideration of such a possibility will be beneficial in unraveling the microstructure of returns from low-frequency data via a connection to phenomena that trigger abnormal trading volumes, with ample empirical evidence. For instance, the goal to increase the sparsity of trade size distributions is in agreement with the presence of large speculators in traditional and cryptocurrency markets (mentioning, e.g., [Chang et al., 1997] \cite{CPS} and [Blau, 2017] \cite{B3}, respectively) or the prevalence of trade-size clustering (referring to [Alexander and Peterson, 2008] \cite{AP} and [Chen, 2019] \cite{C1}). In this regard, the linkage between skewed and leptokurtic returns and the theoretical moments of the driving stochastic clock is simply embodied by the famous price--volume relationship (see, e.g., [Karpoff, 1987] \cite{K1} and [Richardson and Smith, 1994] \cite{RS}).

With the foregoing aspects in mind, in this paper we aim to propose a unified approach towards regulating any existing stochastic clock without impairing important fundamental properties. In this context, the act of ``regulating'' can be understood in the ordinary sense -- a stochastic clock is continually adjusted with the intent of achieving a desirable local speed, making it more reliable for capturing the differences in trading activity over time. As a starting point, however, our methodology is concentrated on the aforesaid L\'{e}vy subordinators but will be easily extendable to processes of independent but time-inhomogeneous increments, or Sato processes (see [Sato, 2006] \cite{S1} and [Eberlein and Madan, 2009] \cite{EM}), all of which can be subsequently applied to build structurally more complex stochastic clocks with mean-reverting acceleration and continuous movements that can deal with volatility clustering (see [Carr and Wu, 2004] \cite{CW}). The regulated stochastic clocks have essentially the same functionalities as the originals, and can be used to time-change diffusion processes to spawn multifarious semimartingales with discontinuities, albeit with much larger capacity in dealing with asymmetrical and tail risks.

That being said, our approach can also be exploited to generate a rich class of unexplored L\'{e}vy processes comfortable for financial modeling. As expected, this newfound dimension is encoded into a hyperparameter representing the degree of regulation, i.e., how much slower the regulated clock is compared with the unregulated one, and kept as independent as possible from the roles of other shape parameters. The resultant processes are shown to be analytically tractable whether it be statistical estimation or risk-neutral valuation of options that is of interest.

The remainder of this paper is structured as follows. In Section \ref{sec:2} we present our main methodology in three recipes, all of which serve to regulate a stochastic clock to any intended degrees. The effect of clock regulation on the jump behaviors depicting corresponding evolution of trading activity is analyzed (Theorem \ref{thm:1}, Theorem \ref{thm:2}, and Corollary \ref{cor:1}) and the impact on asymmetry and tail heaviness of the original distribution is examined in detail (Corollary \ref{cor:2}). Then, we present in Section \ref{sec:3} two simple ways for utilizing the (un)regulated stochastic clocks in order to construct real-valued mixed models. In Section \ref{sec:4}, we demonstrate clock regulation on two important special cases featuring jump--diffusion models and purely discontinuous models and discuss their potential applications in finance as well as other fields. Section \ref{sec:5} outlines two possible candidates for optimizing the degree of regulation: one employing a robust moment-based estimation procedure combined with profile log-likelihood and the other using option price calibration with numerical Fourier inversion techniques, which underlie our empirical study in Section \ref{sec:6} concerning S\&P500 and Bitcoin returns and Bitcoin options. Conclusions are drawn in Section \ref{sec:7} along with a summary of the properties of the three recipes and highlights on future research directions. All mathematical proofs are presented at the end of the paper, in Appendix \hyperref[A]{A}.

\medskip

\section{Regulated stochastic clocks}\label{sec:2}

As already noted, we shall develop our methodology using L\'{e}vy processes exclusively. By having independent and stationary increments, such processes are easy to manage and sit at the core of statistical estimation with \text{i.i.d.} temporal data. To begin with, we outline some basic intuition.

Let $X\equiv(X_{t})_{t\geq0}$ be a L\'{e}vy subordinator (nondeterministic by default), acting as a stochastic clock. The objective of regulation lies in properly compressing the amplitude of its moderate jumps (for moderate values the jump component $|\Delta X|$), and an attendant consequence is that the finite-dimensional distribution of $X$ will tend to have greater asymmetry and a heavier right tail, as measured by the skewness and kurtosis. Thus, suppose that $\phi_{X_{t}}(u):=\E e^{-uX_{t}}$ is the Laplace transform (LT) of the random variable $X_{t}$ for generic $t>0$, assumed to be definable in a neighborhood of the origin (in $\mathbb{C}$), and then with the canonical expansion
\begin{equation}\label{2.1}
  \log\phi_{X_{t}}(u)=\sum^{\infty}_{m=0}\frac{(-u)^{m}K_{t}(m)}{m!},\quad u\in\mathbb{C}
\end{equation}
in terms of corresponding cumulants $K_{t}(m)$, $m\in\mathbb{N}$ (with $K_{t}(0)=0$), the objective is converted to finding, for a fixed $u$, a linear functional $\mathcal{I}:\mathbb{C}^{\{\alpha u:\alpha\in[0,1]\}}\mapsto\mathbb{C}$ such that, from (\ref{2.1}),
\begin{equation*}
  (\mathcal{I}\log\phi_{X_{t}})(u)=\sum^{\infty}_{m=0}\frac{C_{m}(-u)^{m}K_{t}(m)}{m!},
\end{equation*}
where the time-invariant factors $C_{m}$'s satisfy $|C_{m}|\leq1$ and $C_{m+1}\leq C_{m}$, $\forall m$. These relations will immediately render cumulants of larger (up to four) orders more reduced. Although there are obviously infinitely many possible ways to construct such a functional, $\mathcal{I}$ should be conceptually matched with a recipe operating on the sample paths $[0,t]\ni s\mapsto X_{s}$ to which the collection $\big\{\phi^{\alpha}_{X_{t}}:\alpha\in[0,1]\big\}$ of LTs correspond; otherwise temporal structures represented by path properties will be difficult to justify if possible at all. In what follows we are going to present three utterly simple and closely related recipes of this nature, by exploiting continuous temporal averaging.

\subsection{Repeated averaging-induction: Three recipes}\label{sec:2.1}

We design our \textbf{first recipe} as follows. Set $\bar{X}^{(0)}:=X^{(0)}\equiv X$ and, for any $n\in\mathbb{N}_{++}\equiv\mathbb{N}\setminus\{0\}$, repeatedly define the running averages
\begin{equation}\label{2.1.1}
  \bar{X}^{(n)}_{0}=0;\quad\bar{X}^{(n)}_{t}:=\frac{1}{t}\int^{t}_{0}\bar{X}^{(n-1)}_{s}\dd s,\quad t>0.
\end{equation}
For each $n\geq1$, $\bar{X}^{(n)}$ is a stochastic process having $\PP$-\text{a.s.} continuous sample paths, where continuity at 0 follows by the dominated convergence theorem. More specifically, the sample paths $t\mapsto\bar{X}^{(n)}_{t}$ belong to the class $\mathcal{C}^{n-1}(\mathbb{R}_{+};\mathbb{R}_{+})$ of ($n-1$)th-order differentiable functions for $n\geq1$, $\PP$-a.s., as a result of repeated integration.

Since (\ref{2.1.1}) operates a linear functional, it is clear that for every $n\geq1$ and fixed $t>0$ the random variable $\bar{X}^{(n)}_{t}$ has an infinitely divisible distribution, which induces a unique L\'{e}vy process that is also nonnegative, which we denote as
\begin{equation}\label{2.1.2}
  X^{(n)}_{t}\overset{\text{d.}}{=}\bar{X}^{(n)}_{t},\quad t\geq0,\;n\in\mathbb{N}_{++}.
\end{equation}
By nonnegativity every $X^{(n)}$ can be used as a different stochastic clock, and will be referred to as the \textbf{type-I} regulated $X$-clock of degree $n$.

The intuition behind the first recipe is rather clear: Averaging in time has a natural effect of contracting moderate values of the sample paths and continuing to do so will tend to enlarge such an effect. To give some remarks, in the special case $n=1$, the process $\bar{X}^{(1)}=1/(\cdot)\int^{\cdot}_{0}X_{s}\dd s$ is precisely the running average of $X$, which is heavily tied to linear integral functionals of L\'{e}vy processes; such integrals have been widely applied to trace the cumulative patterns of various stochastic systems; some notable examples include the total cost of random epidemics (see, e.g., [Downton, 1972] \cite{D} and [Pollett, 2003] \cite{P}), cumulation of count data generalized to non-integer values ([Orsingher and Polito, 2013] \cite{OP}) or count data with overdispersion ([Xia, 2019] \cite{X1}), and the modeling of structural degradation phenomena in the presence of memory effects (see [Tseng and Peng, 2004] \cite{TP} and [Xia, 2021, \text{Sect.} 4.1] \cite{X2}).

Note that the averaging--induction procedure also works for any integer $n\geq1$, and hence it is intuitive to consider inducing a L\'{e}vy process for each $n$ using (\ref{2.1.2}), before initiating a subsequent averaging step (\ref{2.1.1}). In this way the integrand in the averaging step can be made a L\'{e}vy process even for $n\geq2$, and each resultant averaged process will correspond to the linear functional of a certain L\'{e}vy process. Such an idea leads to our \textbf{second recipe}. Likewise, set $\tilde{X}^{(0)}:=Y^{(0)}\equiv X$, and then repeatedly utilize the first recipe for every $n\in\mathbb{N}_{++}$ in the following fashion:
\begin{equation}\label{2.1.3}
  \tilde{X}^{(n)}_{0}=0;\quad\tilde{X}^{(n)}_{t}:=\frac{1}{t}\int^{t}_{0}Y^{(n-1)}_{s}\dd s,\quad t>0,
\end{equation}
where $Y^{(n)}$'s are induced L\'{e}vy processes such that
\begin{equation*}
  Y^{(n)}_{t}\overset{\text{d.}}{=}\tilde{X}^{(n)}_{t},\quad t\geq0,\;n\in\mathbb{N}_{++}.
\end{equation*}
It is clear that the averaged processes $\bar{X}^{(1)}$ and $\tilde{X}^{(1)}$ are indistinguishable, and so are the induced processes $X^{(1)}$ and $Y^{(1)}$, but for $n\geq2$ they are different with dissimilar path properties -- the sample paths $t\mapsto\tilde{X}^{(n)}_{t}$ will remain in the class $(\mathcal{C}\setminus\mathcal{C}^{1})(\mathbb{R}_{+};\mathbb{R}_{+})$ ($\PP$-a.s.) for $n\geq2$.

It can be shown (see the proof of Theorem \ref{thm:1} in Appendix \hyperref[A]{A}) that the second recipe can be essentially thought to average the log-LT, while in comparison the first recipe, by construction, has the averaging effect on the sample paths. We will refer to the process $Y^{(n)}$, $n\geq1$, as the \textbf{type-II} regulated $X$-clock of degree $n$.

In the averaging step (\ref{2.1.1}), if we stack up the time scale ($1/t$) outside of the integral operators for every $n\geq2$, the averaging procedure becomes a folded cumulation with a one-time power scaling, and we shall have our \textbf{third recipe}. Set $\breve{X}^{(0)}:=Z^{(0)}\equiv X$ and define the quasi-averages
\begin{equation}\label{2.1.4}
  \breve{X}^{(n)}_{0}=0;\quad\breve{X}^{(n)}_{t}:=\frac{1}{t^{n}}\underbrace{\int\cdots\int^{t}_{0}}_{n}X_{s}\underbrace{\dd s\cdots\dd s}_{n},\quad t>0,\;n\in\mathbb{N}_{++},
\end{equation}
which has an infinitely divisible distribution for the same reason as in (\ref{2.1.1}), and induce a unique nonnegative L\'{e}vy process,
\begin{equation*}
  Z^{(n)}_{t}\overset{\text{d.}}{=}\breve{X}^{(n)}_{t},\quad t\geq0,\;n\in\mathbb{N}_{++},
\end{equation*}
referred to as the \textbf{type-III} regulated $X$-clock of degree $n$.

It is obvious that the processes $\breve{X}^{(n)}$ and $\bar{X}^{(n)}$ have the same degree of path smoothness but are guaranteed to coincide (up to indistinguishability) only if $n=1$. Besides, the distribution of $\breve{X}^{(n)}_{t}$ is also infinitely divisible for every $n\geq1$ and fixed $t$.

We present some general formulae for the LTs of the three types of regulated stochastic clocks in Theorem \ref{thm:1}, which are naturally extended to non-integer-valued $n>0$ as well. This aspect hints at the extension of the three recipes to the sense of averaging of fractional degrees and will be considered in the optimization of the regulation degree $n$.

\begin{theorem}\label{thm:1}
For generic $t>0$ and any $n>0$, the LTs of $\bar{X}^{(n)}_{t}$, $\tilde{X}^{(n)}_{t}$, and $\breve{X}^{(n)}_{t}$, are given by
\begin{align}\label{2.1.5}
  \phi_{\bar{X}^{(n)}_{t}}(u)&:=\E e^{-u\bar{X}^{(n)}_{t}}=\exp\bigg(t\int^{1}_{0}\log\phi_{X_{1}}\bigg(\bigg(1-\frac{\Gf(n,-\log s)}{\Gf(n)}\bigg)u\bigg)\dd s\bigg), \nonumber\\
  \phi_{\tilde{X}^{(n)}_{t}}(u)&:=\E e^{-u\tilde{X}^{(n)}_{t}}=\exp\bigg(\frac{t}{\Gf(n)}\int^{1}_{0}(-\log v)^{n-1}\log\phi_{X_{1}}(uv)\dd v\bigg), \nonumber\\
  \phi_{\breve{X}^{(n)}_{t}}(u)&:=\E e^{-u\breve{X}^{(n)}_{t}}=\exp\bigg(t\int^{1}_{0}\log\phi_{X_{1}}\bigg(\frac{s^{n}u}{\Gf(n+1)}\bigg)\dd s\bigg),\quad\Re u\geq0,
\end{align}
respectively, where $\Gf(\cdot)$ and $\Gf(\cdot,\cdot)$ symbolize, respectively, the usual gamma function and the (upper) incomplete gamma function.
\end{theorem}

In any case, the recipes have given rise to three families of stochastic clocks, controlled by the degree $n$ of regulation. The effect of operating the three recipes on the unregulated clock $X$ is twofold -- on path properties and distributional properties, as we are bound to investigate next.

\subsection{Characterizing triplets}\label{sec:2.2}

From their primary motivation, the three recipes are all supposed to compress the jumps of $X$ into extreme values. To confirm such effects, it is necessary to derive the characterizing triplets of the regulated clocks $X^{(n)}$'s, $Y^{(n)}$'s, and $Z^{(n)}$'s governing their sample path properties. Doing so will also be helpful for developing simulation methods, the main reason being that the LTs that we have obtained in (\ref{2.1.5}), having exponentiated integrals, are generally difficult to invert analytically, thus circumventing the implementation of inverse sampling methods.

First, there is no Brownian component in $X^{(n)}$'s, $Y^{(n)}$'s or $Z^{(n)}$'s and we assume without loss of generality that there is no drift as well, so that we can concentrate on the corresponding L\'{e}vy measure. More specifically, suppose that the LT $\phi_{X_{1}}$ takes the following general form due to the L\'{e}vy--Khintchine representation:
\begin{equation*}
  \phi_{X_{1}}(u)=\exp\int^{\infty}_{0}(e^{-uz}-1)\nu(\dd z),\quad\Re u\geq0,
\end{equation*}
with the characterizing triplet $(0,0,\nu)$, for some L\'{e}vy measure\footnote{The notation $\nu_{X}$ will occasionally be used for the L\'{e}vy measure of $X$ in the mathematical proofs (Appendix \hyperref[A]{A}) to avoid ambiguity.} $\nu$ imposed on $\mathcal{B}(\mathbb{R}_{+})$ such that $\nu(\{0\})=0$ and $\int^{\infty}_{0}(z\wedge1)\nu(\dd z)<\infty$. Then, the characterizing triplets of $X^{(n)}$, $Y^{(n)}$, and $Z^{(n)}$, given $n>0$, are denoted as $(0,0,\bar{\nu}^{(n)})$, $(0,0,\tilde{\nu}^{(n)})$, and $(0,0,\breve{\nu}^{(n)})$ respectively, which make the representations
\begin{align}\label{2.2.1}
  \phi_{\bar{X}^{(n)}_{1}}(u)&=\exp\int^{\infty}_{0}(e^{-uz}-1)\bar{\nu}^{(n)}(\dd z), \nonumber\\
  \phi_{\tilde{X}^{(n)}_{1}}(u)&=\exp\int^{\infty}_{0}(e^{-uz}-1)\tilde{\nu}^{(n)}(\dd z), \nonumber\\
  \phi_{\breve{X}^{(n)}_{1}}(u)&=\exp\int^{\infty}_{0}(e^{-uz}-1)\breve{\nu}^{(n)}(\dd z).
\end{align}

\begin{theorem}\label{thm:2}
For any $n>0$, we have
\begin{align}\label{2.2.2}
  \bar{\nu}^{(n)}(\dd z)&=\Gf(n)\int^{\infty}_{1}\frac{\nu(\dd(yz))}{y^{2}\mathrm{Q}^{n-1}(n,1-1/y)}\dd y, \nonumber\\
  \tilde{\nu}^{(n)}(\dd z)&=\frac{1}{\Gf(n)}\int^{\infty}_{1}\frac{(\log y)^{n-1}\nu(\dd(yz))}{y^{2}}\dd y, \nonumber\\
  \breve{\nu}^{(n)}(\dd z)&=\int^{\infty}_{\Gf(n+1)}\bigg(\frac{\Gf(n+1)}{y}\bigg)^{1/n}\frac{\nu(\dd(yz))}{ny}\dd y,\quad z\geq0,
\end{align}
where $\mathrm{Q}(n,s)$ is the inverse regularized (upper) incomplete gamma function.\footnote{It is understood as the unique solution $x\geq0$ to the transcendental equation $\Gf(n,x)/\Gf(n)=s$ for $s\in[0,1)$; see [DiDonato and Morris, 1986] \cite{DM}.}

Furthermore, if $\nu$ is absolutely continuous with respect to Lebesgue measure, then
\begin{align}\label{2.2.3}
  \bar{\ell}^{(n)}(z)&:=\frac{\bar{\nu}^{(n)}(\dd z)}{\dd z}=\Gf(n)\int^{\infty}_{1}\frac{\ell(yz)}{y\mathrm{Q}^{n-1}(n,1-1/y)}\dd y, \nonumber\\
  \tilde{\ell}^{(n)}(z)&:=\frac{\tilde{\nu}^{(n)}(\dd z)}{\dd z}=\frac{1}{\Gf(n)}\int^{\infty}_{1}\frac{(\log y)^{n-1}\ell(yz)}{y}\dd y, \nonumber\\
  \breve{\ell}^{(n)}(z)&:=\frac{\breve{\nu}^{(n)}(\dd z)}{\dd z}=\int^{\infty}_{\Gf(n+1)}\bigg(\frac{\Gf(n+1)}{y}\bigg)^{1/n}\frac{\ell(yz)}{n}\dd y,\quad z\geq0,
\end{align}
with $\nu(\dd z)=\ell(z)\dd z$.
\end{theorem}

We remark that the L\'{e}vy measures $\bar{\nu}^{(n)}$'s, $\tilde{\nu}^{(n)}$'s, and $\breve{\nu}^{(n)}$'s from (\ref{2.2.2}), for $n\geq1$, are all absolutely continuous because of integration, and thus their associated L\'{e}vy densities on the left-hand side of (\ref{2.2.3}) exist unconditionally. More importantly, we have the following equalities among their Blumenthal--Getoor indices ([Blumenthal and Getoor, 1961] \cite{BG}).

\begin{corollary}\label{cor:1}
For any $n>0$, it holds that
\begin{equation}\label{2.2.4}
  \mathfrak{B}(X^{(n)}):=\inf\bigg\{p>0:\int^{1}_{0}|z|^{p}\bar{\nu}^{(n)}(\dd z)<\infty\bigg\}=\mathfrak{B}(X),
\end{equation}
and similarly $\mathfrak{B}(Y^{(n)})=\mathfrak{B}(Z^{(n)})=\mathfrak{B}(X)$.
\end{corollary}

Corollary \ref{cor:1} means that none of the three recipes alters the path regularity of $X$, which is beyond doubt a very desirable property as there is never an intention of changing the jump activity of the underlying models connected with trading intensities (recalling Section \ref{sec:1}). On the other hand, all of the induced L\'{e}vy measures govern jumps of arbitrarily large amplitudes, whereas their actual weights (namely likelihood) will depend on the functional form of $\nu$ (or that of the LT), subject to further adjustment of its parameters, and through the foregoing relations. This aspect will be clearer as we move to specific distribution families of interest in Section \ref{sec:4}.

\begin{figure}[H]
  \centering
  \includegraphics[scale=0.4]{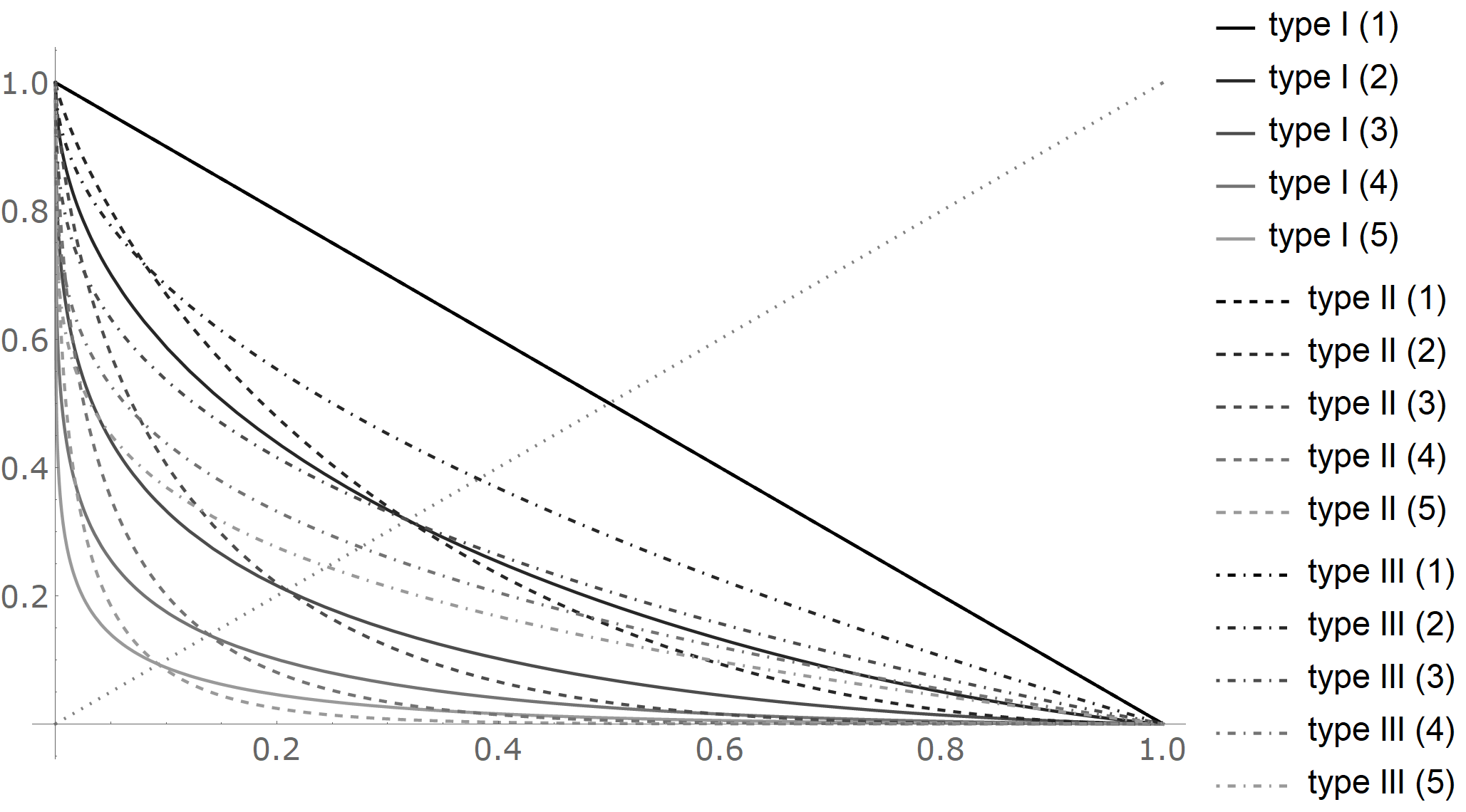}
  \caption{Regulation impact on induced L\'{e}vy measures}
  \label{fig:1}
\end{figure}

To gain additional insight into how the three recipes distort the L\'{e}vy measures $\bar{\nu}^{(n)}$, $\tilde{\nu}^{(n)}$, and $\breve{\nu}^{(n)}$ from a comparative angle, it is first observed from (\ref{2.2.2}) in Theorem \ref{thm:2} that these L\'{e}vy measures do coincide if and only if $n=1$. Upon the substitution $yz\mapsto y$, they can be viewed as fractional integrals of the L\'{e}vy measure $\nu$ of $X$ under different kernels, which with $y=1$ fixed boil down to the functions $\Gf(n)/\mathrm{Q}^{n-1}(n,1-z)$, $(-\log z)^{n-1}/\Gf(n)$, and $z^{1/n-1}/n$, respectively, defined for $z\in(0,1)$. Interestingly, we then observe that the first two functions are precisely the derivatives in magnitude of $e^{-\mathrm{Q}(n,1-z)}$ and $1-\Gf(n,-\log z)/\Gf(n)$, respectively, which happen to be inverses of each other. Put differently, the effects on the (average-induced) L\'{e}vy measures from the first two recipes are perfectly complementary, in the sense that neither dominates the other. Differently, the third function integrates in magnitude to $1-z^{1/n}$, which is not in direct comparison with the others, for the third recipe, strictly speaking, does not embody a running average for $n\neq1$. This discussion is further illustrated in Figure \ref{fig:1}, with five integer degrees $n\in\mathbb{N}\cap[1,5]$ of regulation. We see that the first recipe, which averages the sample paths, has a larger impact (distortion) when $\nu$ is concentrated on small values of $\mathbb{R}_{+}$, relative to the second, which at bottom averages the log-LT. In addition, the larger the regulation degree, the faster the impact differences are reflected.

\subsection{Asymmetry and tail heaviness}\label{sec:2.3}

The static effect from operating the three recipes is to enlarge the asymmetry and (right) tail heaviness of the distribution of the unregulated clock $X_{t}$ for any fixed $t>0$. With the LTs in (\ref{2.1.5}) at hand, computation of the statistics of the corresponding regulated clocks $X^{(n)}_{t}$'s and $Y^{(n)}_{t}$'s is straightforward. More particularly, we shall focus on the impact of regulation on skewness and excess kurtosis. Let us restate from (\ref{2.1}) that the corresponding cumulants are defined to be the series coefficients $\{\bar{K}_{t}(m)\}^{\infty}_{m=0}$ and $\{\tilde{K}_{t}(m)\}^{\infty}_{m=0}$ in
\begin{equation*}
  \log\phi_{\bar{X}_{t}}(u)=\sum^{\infty}_{m=0}\frac{(-u)^{m}\bar{K}_{t}(m)}{m!},\quad \log\phi_{\tilde{X}_{t}}(u)=\sum^{\infty}_{m=0}\frac{(-u)^{m}\tilde{K}_{t}(m)}{m!},\quad \log\phi_{\breve{X}_{t}}(u)=\sum^{\infty}_{m=0}\frac{(-u)^{m}\breve{K}_{t}(m)}{m!},
\end{equation*}
with $\bar{K}_{t}(0)=\tilde{K}_{t}(0)=\breve{K}_{t}(0)=0$. The next corollary explains how the cumulants vary with the regulation degree $n$.

\begin{corollary}\label{cor:2}
For any $n>0$, we have the cumulant reduction relations
\begin{equation}\label{2.3.1}
  \bar{K}^{(n)}_{t}(m)=C_{m,n}K_{t}(m),\quad\tilde{K}^{(n)}_{t}(m)=\frac{K_{t}(m)}{(m+1)^{n}},\quad \breve{K}^{(n)}_{t}(m)=\frac{K_{t}(m)}{(mn+1)\Gf^{m}(n+1)},
\end{equation}
with the sequence
\begin{equation}\label{2.3.2}
  C_{m,n}=\int^{1}_{0}\bigg(1-\frac{\Gf(n,-\log s)}{\Gf(n)}\bigg)^{m}\dd s,\quad m\in\mathbb{N}_{++},
\end{equation}
which is rational for $n\in\mathbb{N}_{++}$.
\end{corollary}

Using Corollary \ref{cor:2}, the relations for the mean, variance, skewness and excess kurtosis of the type-I $n$-degree regulated $X$-clocks are given by
\begin{align}\label{2.3.3}
  &\E X^{(n)}_{t}=C_{1,n}\E X_{t},\quad\Var X^{(n)}_{t}=C_{2,n}\Var X_{t}, \nonumber\\
  &\Skew X^{(n)}_{t}=\frac{C_{3,n}}{C^{3/2}_{2,n}}\Skew X_{t},\quad\EKurt X^{(n)}_{t}=\frac{C_{4,n}}{C^{2}_{2,n}}\EKurt X_{t}.
\end{align}
Since the integrand in (\ref{2.3.2}) admits the series representation ([Abramowitz and Stegun, 1972, \text{Eq.} 6.5.29] \cite{AS})
\begin{equation*}
  1-\frac{\Gf(n,-\log s)}{\Gf(n)}=\frac{s(-\log s)^{n}}{\Gf(n)}\sum^{\infty}_{k=0}\frac{(-\log s)^{k}}{(n)_{k+1}}=\frac{s(-\log s)^{n}}{\Gf(n+1)}(1+O(n^{-1})),\quad\text{as }n\rightarrow\infty,
\end{equation*}
carrying out the integral over $s\in(0,1)$ gives
\begin{equation}\label{2.3.4}
  C_{m,n}=O\bigg(n^{(1-m)/2}\bigg(\frac{m}{m+1}\bigg)^{mn}\bigg),\quad\text{as }n\rightarrow\infty,
\end{equation}
and hence the sequence decays exponentially as $n$ increases for $m=1$ and (strictly) faster for $m\geq2$. This indicates that the first recipe shrinks the cumulants of $X$ with acceleration -- more aggressively so as regulation continues with $n$. Note that for $m=1$ the exponential decay is in fact an equality $C_{1,n}=2^{-n}$. Looking at (\ref{2.3.3}) and (\ref{2.3.4}) we can also claim that the type-I regulated clocks have at least exponentially decaying mean and variance but faster-than-exponentially increasing skewness and excess kurtosis, with respect to the degree $n$ of regulation, other things (e.g., parameter values) held equal.

In the case of the type-II regulated clocks, the counterpart of (\ref{2.3.3}) for $Y^{(n)}_{t}$ is
\begin{equation}\label{2.3.5}
  \E Y^{(n)}_{t}=\frac{1}{2^{n}}\E X_{t},\quad\Var Y^{(n)}_{t}=\frac{1}{3^{n}}\Var X_{t},\quad\Skew Y^{(n)}_{t}=\bigg(\frac{3\sqrt{3}}{4}\bigg)^{n}\Skew X_{t},\quad\EKurt Y^{(n)}_{t}=\bigg(\frac{9}{5}\bigg)^{n}\EKurt X_{t}.
\end{equation}
Intuitively, this signifies that the cumulant-reduction effect of the second recipe is uniform, in the sense that for each additional round of regulation, the mean and variance are reduced by fixed proportions ($1/2$ and $1/3$, respectively) while the skewness and excess kurtosis are simultaneously enlarged by fixed proportions ($3\sqrt{3}/4$ and $9/5$, respectively). In consequence, for the type-II regulated $X$-clocks the mean and variance decay exactly exponentially and so do the skewness and excess kurtosis increase, other things equal.

As for the type III, we have similarly from Corollary \ref{cor:2} the quotient $\breve{K}^{(n)}_{t}(m)/K_{t}(m)=1/((mn+1)\Gf^{m}(n+1))$, $m\in\mathbb{N}_{++}$, and subsequently the following moment relations for $Z^{(n)}_{t}$:
\begin{align}\label{2.3.6}
  &\E Z^{(n)}_{t}=\frac{1}{\Gf(n+2)}\E X_{t},\quad\Var Z^{(n)}_{t}=\frac{1}{(2n+1)\Gf^{2}(n+1)}\Var X_{t}, \nonumber\\
  &\Skew Z^{(n)}_{t}=\frac{(2n+1)^{3/2}}{3n+1}\Skew X_{t},\quad\EKurt Z^{(n)}_{t}=\frac{(2n+1)^{2}}{4n+1}\EKurt X_{t}.
\end{align}
Upon comparing the above quotient with (\ref{2.3.4}), with $\Gf(n+1)=O(\sqrt{n}(n/e)^{n})$ as $n\rightarrow\infty$ it is seen that the third recipe generates an even severer cumulant-reduction effect than the first recipe does. However, the skewness and excess kurtosis are enlarged with deceleration, noted that $(2n+1)^{3/2}/(3n+1)=O(n^{1/2})$ and $(2n+1)^{2}/(4n+1)=O(n)$, as $n\rightarrow\infty$. Put together, the type-III regulated $X$-clocks have mean and variance of faster-than-exponential decay whereas the skewness and excess kurtosis exhibit slower-than-linear growth, which properties are in stark contrast with those of the first two types.

On paper, if we do not alter the parameter (if any) values of $X$, then all three recipes are able to effectively enlarge the asymmetry and tail heaviness of its distribution, with the first having the most significant effect, whereas the downside is that the mean and variance are reduced at the same time. The latter effect is, of course, undesirable for applications, and eluding it will require at least two independent parameters so that the mean and variance can be held constant across the regulation degree $n$. More specifically, when a real-valued model is to be constructed from $X$ (see Section \ref{sec:3}), with the aid of an additional location parameter the mean can be readily fixed by way of centralization, and so the other parameter will most likely (and ideally) be a scale parameter, to which the standard deviation is proportional and the skewness and excess kurtosis are immune. Such a requirement becomes rather standard in statistical inference, and in consequence the skewness and excess kurtosis relations stated in (\ref{2.3.3}), (\ref{2.3.5}), and (\ref{2.3.6}) provide useful implications. We remark, however, that if the mean is to be kept invariant to regulation by way of a shape parameter of $X$ (oftentimes existent for one-sided distributions) then changing its value is likely to enlarge skewness and excess kurtosis at rates different from those in (\ref{2.3.3}) and (\ref{2.3.5}), despite exponential growth rates. Such a situation cannot be analyzed from a universal viewpoint and will again depend on the functional form of the LT $\phi_{X_{1}}$, with respect to the shape parameter. More details are provided in Section \ref{sec:4}.

\begin{figure}[H]
  \centering
  \includegraphics[scale=0.4]{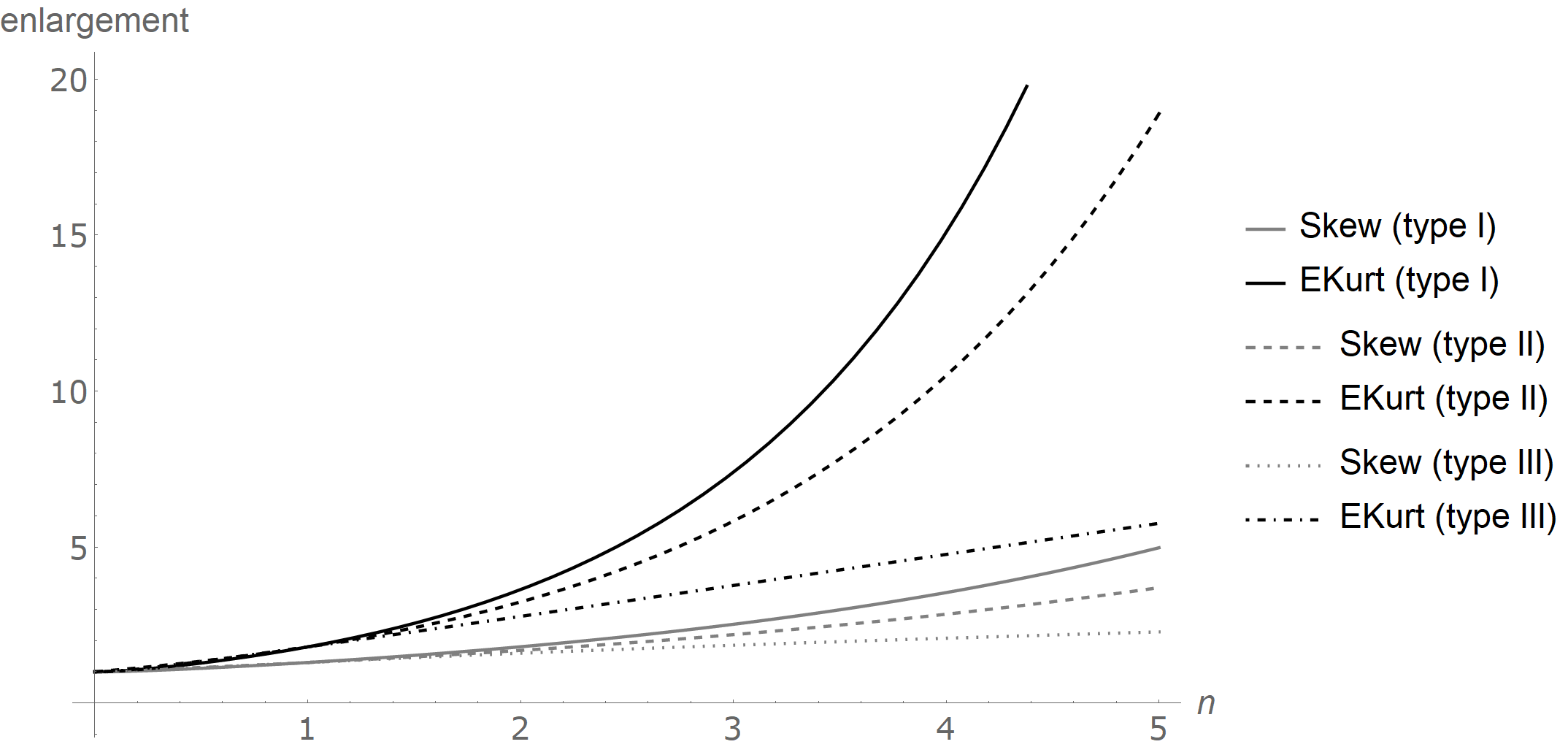}
  \caption{Enlargement effect on asymmetry and tail heaviness with respect to regulation degree}
  \label{fig:2}
\end{figure}

Similarly, the results for the statistics stated in this section can be naturally extended for fractional degrees $n>0$ of regulation. In Figure \ref{fig:2} we give a joint plot of the enlargement effect on asymmetry and tail heaviness for the two recipes with real regulation degrees $n\in(0,5]$, by using (\ref{2.3.3}), (\ref{2.3.5}), and (\ref{2.3.6}) in sequence.

\medskip

\section{Time-changed processes}\label{sec:3}

As aforementioned, the functionality of the regulated clocks $X^{(n)}$'s, $Y^{(n)}$'s, and $Z^{(n)}$'s is meant to be the same as that of $X$ -- mathematically, to adjust the time underlying a (continuous) base process and generate a wide class of models for real-valued temporal data such as financial returns. In the following we shall consider two well-known base processes for the mixture, which are also shown to retain, leastways in part, the enlargement effect on asymmetry and tail heaviness through the clock regulation.

\subsection{Constant mixtures}\label{sec:3.1}

The simplest base process is beyond doubt a (deterministic) drift $(t)$ up to positive scaling, and time-changing it by the clock $X$ gives none but $X$ itself. Then, in order to create a real-valued random variable from $X$ it suffices to take the difference of two of its independent copies. Specifically, let $X'$ be an independent copy of $X$, and then define the difference process
\begin{equation}\label{3.1.1}
  \xi^{X}_{t}=\mu t+\kappa_{1}X_{t}-\kappa_{2}X'_{t},\quad t\geq0,
\end{equation}
where $\mu\in\mathbb{R}$ is a location parameter and $\kappa_{1},\kappa_{2}\geq0$ are additional scale parameters. Then $\xi^{X}$ is a purely discontinuous real-valued L\'{e}vy process admitting the LT (fixing $t>0$)
\begin{equation}\label{3.1.2}
  \phi_{\xi^{X}_{t}}(u):=\E e^{-u\xi^{X}_{t}}=e^{-\mu tu}\phi_{X_{t}}(\kappa_{1}u)\phi_{X_{t}}(-\kappa_{2}u),\quad u\in\ii\mathbb{R},
\end{equation}
where the domain of $\Re u$ can be extended to a neighborhood of $0$ provided that (\ref{2.1}) holds.

Note that since the integral operator $1/t\int^{t}_{0}\dd s$ is for every $t>0$ a linear isometry over the space $\mathbb{L}^{\infty}(\mathbb{R}_{+};\mathbb{R}_{+})$ of bounded functions, considering such independent differences under the $n$-degree regulated clocks is no different from operating the three recipes on the initial difference process $\xi^{X}$. As a result, we can define the corresponding difference processes under each $X^{(n)}$, $Y^{(n)}$, and $Z^{(n)}$, denoted $\xi^{(n),X}$, $\xi^{(n),Y}$, and $\xi^{(n),Z}$, respectively, in the same way as in (\ref{3.1.1}).

\begin{corollary}\label{cor:3}
For generic $t>0$ and any $n>0$, the LTs of the constant mixtures subject to clock regulation are given by
\begin{align}\label{3.1.3}
  \phi_{\xi^{(n),X}_{t}}(u)&=e^{-\mu tu}\phi_{\bar{X}^{(n)}_{t}}(\kappa_{1}u)\phi_{\bar{X}^{(n)}_{t}}(-\kappa_{2}u), \nonumber\\
  \phi_{\xi^{(n),Y}_{t}}(u)&=e^{-\mu tu}\phi_{\tilde{X}^{(n)}_{t}}(\kappa_{1}u)\phi_{\tilde{X}^{(n)}_{t}}(-\kappa_{2}u), \nonumber\\
  \phi_{\xi^{(n),Z}_{t}}(u)&=e^{-\mu tu}\phi_{\breve{X}^{(n)}_{t}}(\kappa_{1}u)\phi_{\breve{X}^{(n)}_{t}}(-\kappa_{2}u),\quad u\in\ii\mathbb{R}.
\end{align}
\end{corollary}

Besides, by independence the $m$th cumulant of $\xi^{(n),X}_{t}$ for fixed $t>0$ is nothing but $\mu t\mathbb{1}_{\{1\}}(m)+(\kappa^{m}_{1}+(-\kappa_{2})^{m})\bar{K}^{(n)}_{t}(m)$ and similarly for $\xi^{(n),Y}_{t}$ and $\xi^{(n),Z}_{t}$. If the mean and variance of $\xi^{(n),X}_{t}$ and $\xi^{(n),Y}_{t}$ are kept constant by adjusting, respectively, the location parameter $\mu$ and a scale parameter of $X$, then the skewness and excess kurtosis increase at precisely the same rates as those of $X$ stated in (\ref{2.3.3}), (\ref{2.3.5}), and (\ref{2.3.6}), respectively. Therefore, the constant mixture approach also preserves enlargement effects on asymmetry and tail heaviness of the constituent subordinator $X$, regardless of the additional parameters $\mu$, $\kappa_{1}$, and $\kappa_{2}$.

Simplicity notwithstanding, a major drawback of this approach, due to linearity, is that it inevitably inherits the path regularity of $X$ into the processes $\xi^{(n),X}$'s, $\xi^{(n),Y}$'s, and $\xi^{(n),Z}$'s, as the next corollary shows.

\begin{corollary}\label{cor:4}
For any $n>0$, we have $\mathfrak{B}(\xi^{(n),X})=\mathfrak{B}(\xi^{(n),Y})=\mathfrak{B}(\xi^{(n),Z})=\mathfrak{B}(\xi^{X})=\mathfrak{B}(X)$.
\end{corollary}

However, in some situations, e.g., as mentioned in [Todorov and Tauchen, 2011] \cite{TT}, a suitable model needs to have sample paths of infinite variation. This is clearly not possible via (\ref{3.1.1}), because by nonnegativity $\mathfrak{B}(X)\in[0,1)$, thus leading us to consider a second approach.

\subsection{Gaussian mixtures}\label{sec:3.2}

Alternatively, one can construct a real-valued random variable by using it as a scale parameter for an independent Gaussian random variable -- a technique called ``Gaussian mixture,'' which is implemented here by way of using $X$ to time-change a Brownian motion with drift (see, e.g., [Barndorff-Nielsen, 1997, \text{Sect.} 3.1] \cite{B-N} and [Madan et al., 1998, \text{Sect.} 2] \cite{MCC}). To be precise, let $W$ be a standard Brownian motion independent of $X$, and define the mixed process
\begin{equation}\label{3.2.1}
  \Xi^{X}_{t}=\mu t+\theta X_{t}+W_{X_{t}},\quad t\geq0,
\end{equation}
where $\mu\in\mathbb{R}$ is a location parameter and $\theta\in\mathbb{R}$ is a Brownian drift parameter representing signed random momenta. The dispersion of the Brownian motion can be fixed at $1$ for simplicity as long as $X$ has a scale parameter, which is already able to capture jump volatility. Then, $\Xi^{X}$ is also a purely discontinuous real-valued L\'{e}vy process. Upon applying the tower property of conditional expectations, the LT of $\Xi^{X}_{t}$ for fixed $t>0$ is found to be
\begin{equation}\label{3.2.2}
  \phi_{\Xi^{X}_{t}}(u):=\E e^{-u\Xi^{X}_{t}}=e^{-\mu tu}\phi_{X_{t}}\bigg(\theta u-\frac{u^{2}}{2}\bigg),\quad u\in\ii\mathbb{R}.
\end{equation}
Again, a partial domain extension into $\mathbb{C}\setminus(\ii\mathbb{R})$ is possible with (\ref{2.1}).

In a similar fashion, we can time-change the Brownian motion with drift by the regulated clocks, though this would not be equivalent to operating the two recipes on $\Xi^{X}$. The resultant processes are denoted as, respectively, $\Xi^{(n),X}$, $\Xi^{(n),Y}$, and $\Xi^{(n),Z}$, with the LTs ($t>0$) states as follows.

\begin{corollary}\label{cor:5}
For generic $t>0$ and any $n>0$, the LTs of the Gaussian mixtures subject to clock regulation are given by
\begin{align}\label{3.2.3}
  \phi_{\Xi^{(n),X}_{t}}(u)&=e^{-\mu tu}\phi_{\bar{X}^{(n)}_{t}}\bigg(\theta u-\frac{u^{2}}{2}\bigg), \nonumber\\
  \phi_{\Xi^{(n),Y}_{t}}(u)&=e^{-\mu tu}\phi_{\tilde{X}^{(n)}_{t}}\bigg(\theta u-\frac{u^{2}}{2}\bigg), \nonumber\\
  \phi_{\Xi^{(n),Z}_{t}}(u)&=e^{-\mu tu}\phi_{\breve{X}^{(n)}_{t}}\bigg(\theta u-\frac{u^{2}}{2}\bigg),\quad u\in\ii\mathbb{R}.
\end{align}
\end{corollary}

From Corollary \ref{cor:5}, an application of Fa\`{a} di Bruno's formula yields the cumulants of $\Xi^{(n),X}_{t}$ in terms the finite sums
\begin{equation*}
  (-1)^{m}\frac{\dd^{m}}{\dd u^{m}}\log\phi_{\Xi^{(n),X}_{t}}(u)\bigg|_{u=0}=\mu t\mathbb{1}_{\{1\}}(m)+\sum^{m}_{k=1}\mathrm{B}_{m,k}(\theta,1)\bar{K}^{(n)}_{t}(k),\quad m\in\mathbb{N}_{++},
\end{equation*}
where $\mathrm{B}_{m,k}(\theta,1)$'s are partial Bell polynomials. In particular, by the order of $m\in\{1,2,3,4\}$ the first four cumulants are: $\mu t+\theta\bar{K}^{(n)}_{t}(1)$, $\bar{K}^{(n)}_{t}(1)+\theta^{2}\bar{K}^{(n)}_{t}(2)$, $3\theta\bar{K}^{(n)}_{t}(2)+\theta^{3}\bar{K}^{(n)}_{t}(3)$, and $3\bar{K}^{(n)}_{t}(2)+6\theta^{2}\bar{K}^{(n)}_{t}(3)+\theta^{4}\bar{K}^{(n)}_{t}(4)$, in order. Similar results hold for $\Xi^{(n),Y}$ and $\Xi^{(n),Z}$. Notably, the elegance of the Gaussian mixture (\ref{3.2.1}) lies in that the additional parameter $\theta$ in its own right acts as a scaling factor on $X$, so that scaling $\Xi^{X}$ is fundamentally linked to doing $X$, making it a feasible task to control the variance of $\Xi^{(n),X}$ (and that of $\Xi^{(n),Y}$ and $\Xi^{(n),Z}$) to be invariant to clock regulation, while the invariance of the mean is taken care of by the location parameter $\mu$. The consequence is that the skewness and kurtosis relations in (\ref{2.3.3}), (\ref{2.3.5}), and (\ref{2.3.6}) set the upper bounds for those (lower-bounded by 1) of the Gaussian-mixed processes, whose exact rates of increase are subject to the value of $\theta$.

The main reason to consider Gaussian mixtures is that they allow for processes with infinite-variation sample paths. Indeed, it can be shown (see Appendix \hyperref[A]{A}) that the Blumenthal-Getoor indices of the Gaussian mixtures are precisely double that of $X$.

\begin{corollary}\label{cor:6}
For any $n>0$ we have $\mathfrak{B}\big(\Xi^{(n),X}\big)=\mathfrak{B}\big(\Xi^{(n),Y}\big)=\mathfrak{B}\big(\Xi^{(n),Z}\big) =\mathfrak{B}\big(\Xi^{X}\big)=2\mathfrak{B}(X)$.
\end{corollary}

Therefore, as long as $\mathfrak{B}(X)\geq1/2$, $\Xi^{X}$, and hence $\Xi^{(n),X}$, $\Xi^{(n),Y}$, and $\Xi^{(n),Z}$, $n>0$, on the regulated clocks, will all have jumps of infinite variation. The same idea goes for the Gaussian mixtures on the regulated clocks. This property overcomes the limitation of the constant mixtures and is particularly attractive when flexibility to include highly active jumps in asset returns is required for model formulation.

\medskip

\section{Some important special cases}\label{sec:4}

We now consider two specific distributions of the unregulated clock $X$ for practical interests. By infinite divisibility we fix time at $t=1$ unless otherwise specified. Explicit formulae will be presented for the distributions of all three types of regulated clocks, as well as their time-changed processes.

\subsection{Poisson process}\label{sec:4.1}

Let the LT of $X_{1}$ be given by $\phi_{X_{1}}(u)=\exp(\lambda(e^{-u}-1))$ with shape parameter $\lambda>0$ (Poisson intensity). Then we have the following specialized formulae.

\begin{proposition}\label{pro:1}
If $X$ is a Poisson process with intensity $\lambda$, then for any $n>0$,
\begin{align}\label{4.1.1}
  \phi_{\bar{X}^{(n)}_{1}}(u)&=\exp\bigg(\lambda\bigg(\Gf(n)\int^{1}_{0}\frac{e^{-uz}}{\mathrm{Q}^{n-1}(n,1-z)}\dd z-1\bigg)\bigg), \nonumber\\
  \phi_{\tilde{X}^{(n)}_{1}}(u)&=\exp\bigg(\frac{1}{\Gf(n)}\int^{1}_{0}(-\log v)^{n-1}(e^{-uv}-1)\dd v\bigg)\overset{n\in\mathbb{N}_{++}}{=}\exp(\lambda(\;_{n}\F_{n}(1,\dots,1;2,\dots,2;-u)-1)), \nonumber\\
  \phi_{\breve{X}^{(n)}_{1}}(u)&=\exp\bigg(\lambda\bigg(\bigg(\frac{\Gf(n+1)}{u}\bigg)^{1/n}\frac{\Gf(1/n)-\Gf(1/n,u/\Gf(n+1))}{n}-1\bigg)\bigg),\quad u\in\mathbb{C},
\end{align}
where $\;_{\cdot}\F_{\cdot}(\cdots;\cdots;\cdot)$ denotes the generalized hypergeometric function (see [Slater, 1966] \cite{S3}); also,
\begin{align}\label{4.1.2}
  \bar{\ell}^{(n)}(z)&=\frac{\lambda\Gf(n)\mathbb{1}_{(0,1]}(z)}{\mathrm{Q}^{n-1}(n,1-z)}, \nonumber\\
  \tilde{\ell}^{(n)}(z)&=\frac{\lambda(-\log z)^{n-1}\mathbb{1}_{(0,1]}(z)}{\Gf(n)}, \nonumber\\
  \breve{\ell}^{(n)}(z)&=\frac{\lambda(\Gf(n+1)z)^{1/n}\mathbb{1}_{(0,1/\Gf(n+1)]}(z)}{nz},\quad z>0.
\end{align}
\end{proposition}

It is an easy implication from Proposition \ref{pro:1} that the type-I regulated clock $X^{(n)}$ is of compound Poisson type: In light of the discussion in Subsection \ref{sec:2.2} and with inverse sampling in mind,
\begin{equation}\label{4.1.3}
  X^{(n)}_{t}\overset{\text{d.}}{=}\sum^{X_{t}}_{k=1}\bigg(1-\frac{\Gf(n,-\log U_{k})}{\Gf(n)}\bigg),\quad t\geq0,
\end{equation}
where $\{U_{k}\}^{\infty}_{k=1}$ is a sequence of \text{i.i.d.} standard uniform random variables. The transformed variable representing the jump amplitudes reduces to the standard uniform for $n=1$ and takes values in $[0,1]$ for all values of $n$. In particular, $X^{(1)}$ is a compound Poisson-uniform process, which result has been discovered as a special case of fractionally integrated Poisson processes in [Orsingher and Polito, 2013] \cite{OP}.

The first LT (\ref{4.1.1}) cannot be evaluated explicitly\footnote{Apparently, this is caused by the lack of an explicit antiderivative of the function $e^{-s}s^{s}$, $s\in[0,1)$. Even so, one can always derive alternative series representations for the log-LT.} except for $n=1$, in which case it is $\phi_{\bar{X}^{(1)}_{1}}(u)=\exp(\lambda((1-e^{-u})/u-1))$. Nonetheless, the first integral in (\ref{4.1.1}) is very amenable to numerical computations using a standard Gauss--Kronrod quadrature rule because the integrand is an increasing function (of $s\in[0,1)$) with values in $[e^{-u},1]$ for any $u\geq0$.

Likewise, under the second recipe, the following representation exists for $Y^{(n)}$:
\begin{equation}\label{4.1.4}
  Y^{(n)}_{t}\overset{\text{d.}}{=}\sum^{X_{t}}_{k=1}e^{-\mathrm{Q}(n,1-U_{k})},\quad t\geq0,
\end{equation}
which is also a compound Poisson process. Similar to the type I, the jump amplitudes are all valued in $[0,1]$, and are standard uniformly distributed for $n=1$.

For the same reason as the first, the second LT in (\ref{4.1.1}) can be efficiently carried out as a numerical integral thanks to the monotonicity and boundedness of the integrand, while an explicit expression has been given for integer regulation degrees. Only for two degree values can it be expressed in terms of more elementary functions. Of course, if $n=1$, we have the compound Poisson-uniform LT $\phi_{\tilde{X}^{(1)}_{1}}(u)=\exp(\lambda((1-e^{-u})/u-1))=\phi_{\bar{X}^{(1)}_{1}}(u)$. If $n=2$, one can consult [Bateman, 1954, \text{Eq.} 4.6.1 and \text{Eq.} 4.6.2] \cite{B1} to obtain
\begin{equation*}
  \phi_{\tilde{X}^{(2)}_{1}}(u)=\exp\bigg(\lambda\bigg(\frac{\log u+\Gf(0,u)-\Gf'(1)}{u}-1\bigg)\bigg),
\end{equation*}
where $-\Gf'(1)\approx0.577216$ is known as the Euler--Mascheroni constant.

Due to simplicity, the third LT in (\ref{4.1.1}) (under the third recipe) permits explicit evaluation for any real regulation degree, from which we have the representation
\begin{equation}\label{4.1.5}
  Z^{(n)}_{t}\overset{\text{d.}}{=}\frac{1}{\Gf(n+1)}\sum^{X_{t}}_{k=1}(1-U_{k})^{n}\overset{\text{d.}}{=} \frac{1}{\Gf(n+1)}\sum^{X_{t}}_{k=1}B_{k,(1/n,1)},\quad t\geq0,
\end{equation}
showing that $Z^{(n)}$ is a compound Poisson process with $1/\Gf(n+1)$-scaled beta($1/n,1$)-distributed jumps. Unlike the other two types, due to such scaling the support of the jump amplitudes of $Z^{(n)}$ shrinks with $n$, and converges to the origin in its lower limit.

According to (\ref{4.1.3}), (\ref{4.1.4}), and (\ref{4.1.5}), we conclude that when $X$ is a Poisson process, all three types of regulated clocks follow compound Poisson processes with bounded jumps. These representations provide easy ways for conducting simulations. Also, the two corresponding transformations under the first two recipes form inverse functions of each other, in line with the demonstration by Figure \ref{fig:1}.

In financial applications, compound Poisson processes with finite jump amplitudes, coupled with an independent Brownian motion, form adequate jump--diffusion models for log-price dynamics when sudden movements in returns are believed not to be arbitrarily large in reality; see [Yan and Hanson, 2006] \cite{YH} and also [Baustian et al., 2017] \cite{BMPS}. These models are built from uniform distributions and are deemed to enlarge the tail heaviness of Gaussian distributions to an extent comparable to normally or double-exponentially distributed jumps (e.g., [Kou, 2002] \cite{K3}). Noted that $X^{(1)}$ is a compound Poisson-uniform process, they can be simply expressed in terms of
\begin{equation}\label{4.1.6}
  \xi^{(1),X/b}_{t}+\sigma W_{t}=\mu t+\sigma W_{t}+\sum^{X_{t}}_{k=1}\frac{U_{k}}{b},\quad t\geq0
\end{equation}
for a rate (reciprocal scale) parameter $b>0$ and a Brownian dispersion parameter $\sigma>0$. Since the continuous part $(\mu t+\sigma W_{t})$ is immune to regulation by the stability of the Gaussian distribution, clock regulation can still be understood on the entire jump--diffusion process $\xi^{(1),X/b}_{t}+\sigma W_{t}$. Viewing from the representations (\ref{4.1.3}), (\ref{4.1.4}), and (\ref{4.1.5}), the effect of regulation is then to assign greater (probabilistic) weights to smaller price jumps, and the consequent jump amplitude distribution, which possesses a strictly decreasing density over a finite domain, is a ``give-and-take'' between the uniform and the exponential family.

\begin{figure}[H]
  \centering
  \includegraphics[scale=0.21]{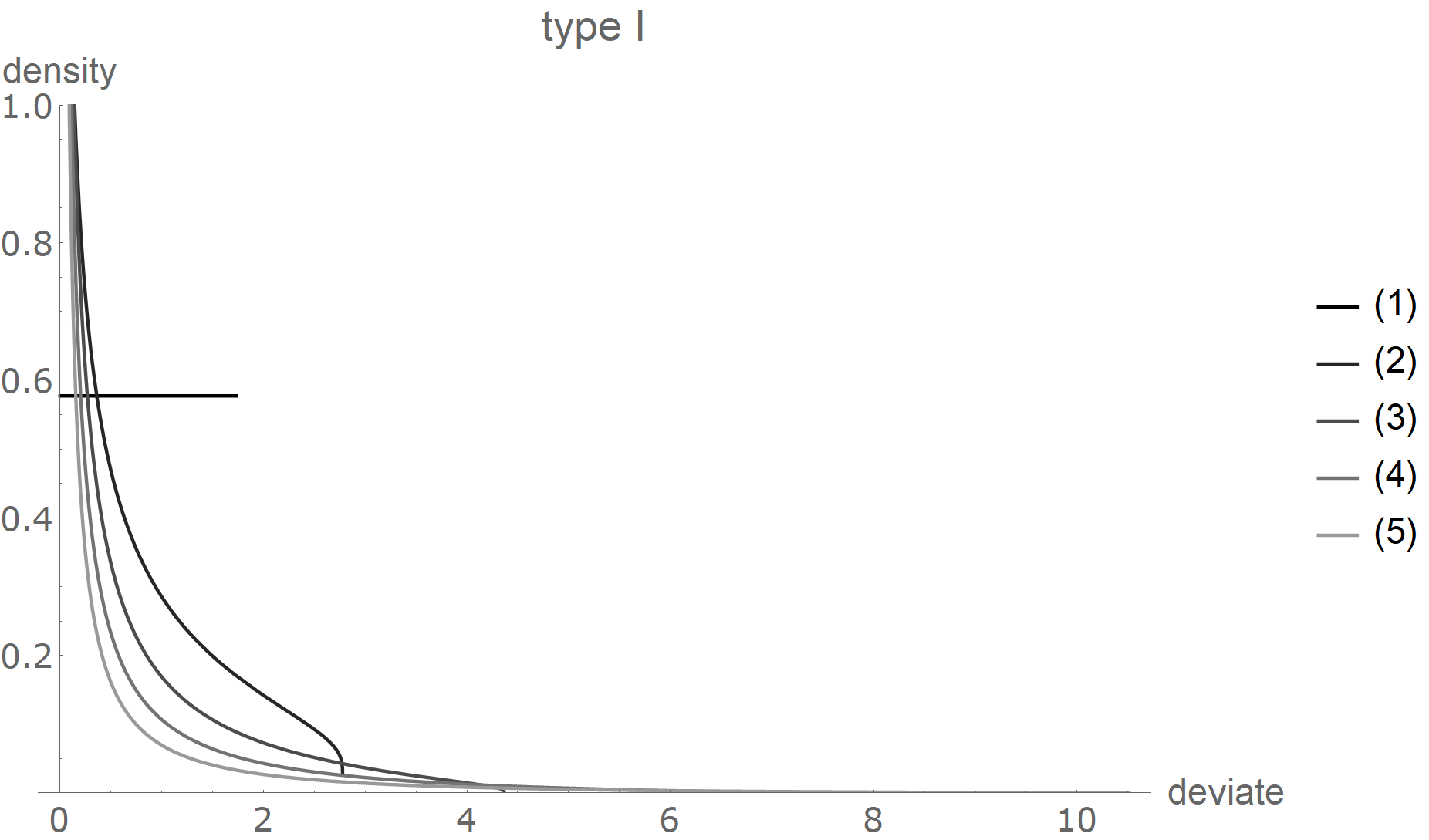}
  \includegraphics[scale=0.21]{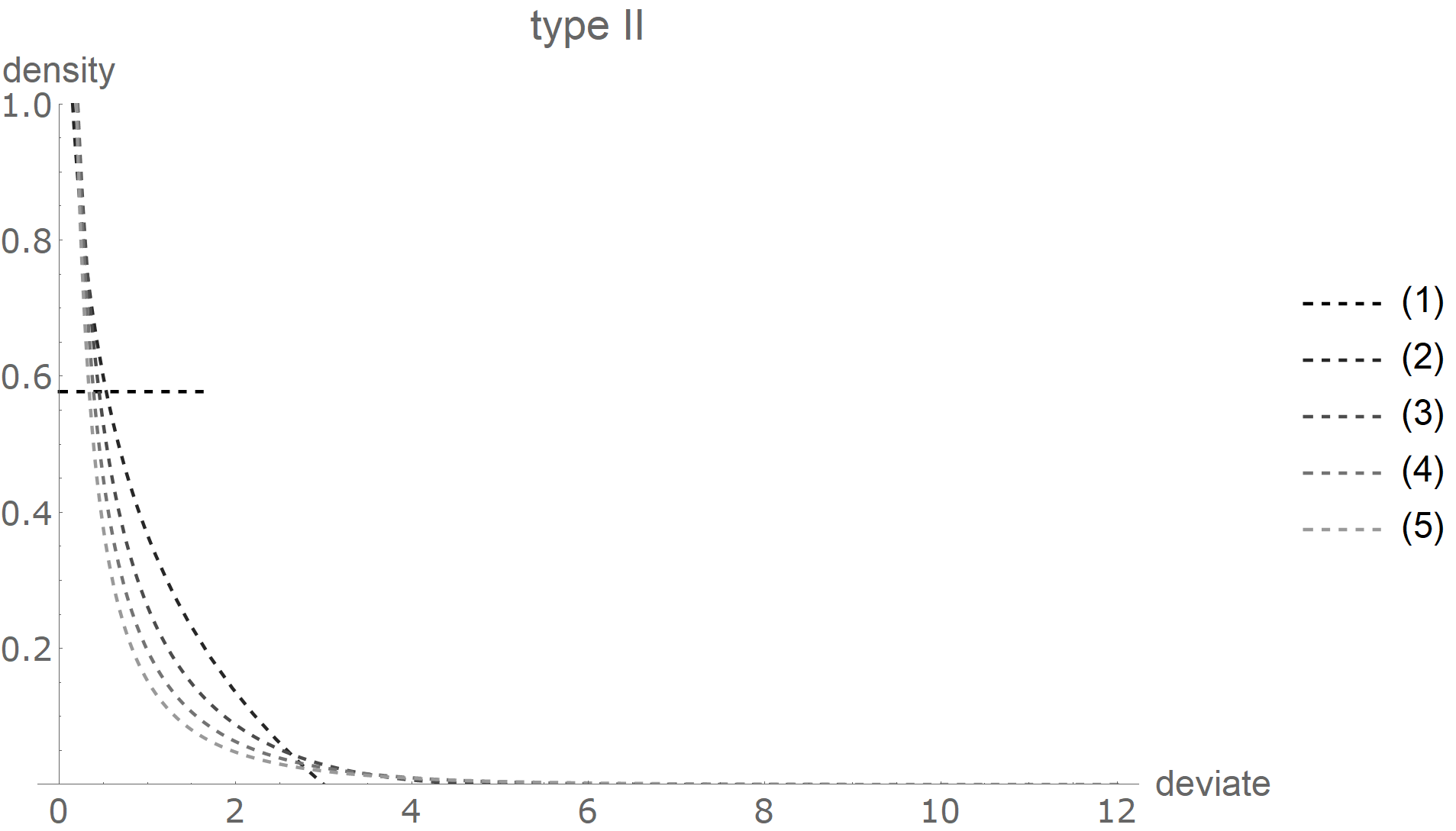}
  \includegraphics[scale=0.21]{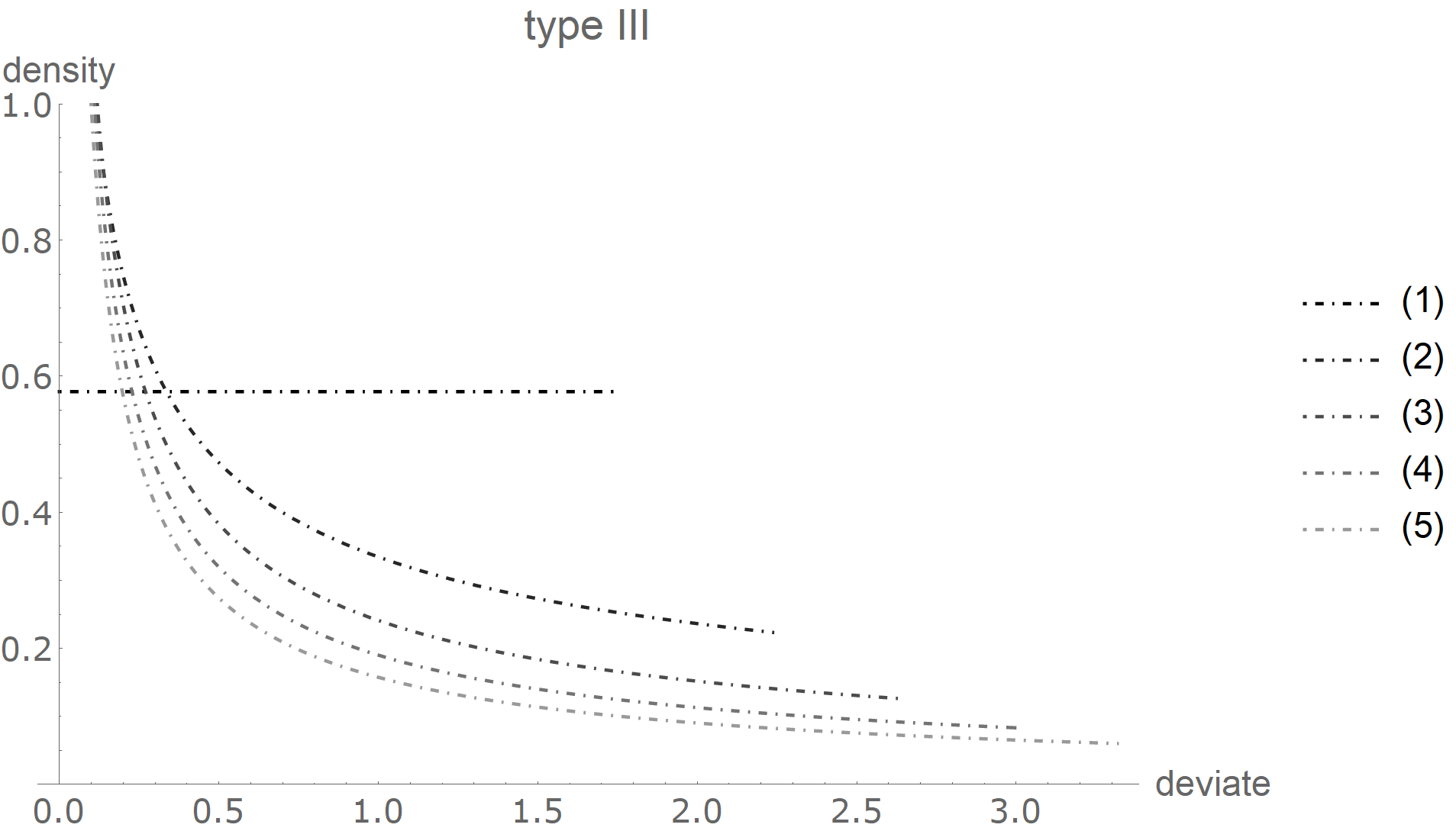}
  \caption{Distorted uniform distributions of (positive) jumps for fixed variance}\label{fig:3}
\end{figure}

To give an example, let us consider the scaled clock $X/b_{n}$, where $b_{n}$ is for every $n>0$ a rate parameter (assuming $b_{0}=1$ for simplicity), and derive from it the regulated clocks $X^{(n)}$, $Y^{(n)}$, and $Z^{(n)}$ accordingly, whose constant mixtures are constructed following Subsection \ref{sec:3.1}, with $\mu=0$ and $\kappa_{1}=\kappa_{2}=1$. Then the processes $\xi^{(n),X}$, $\xi^{(n),Y}$, and $\xi^{(n),Z}$ are centered symmetric compound Poisson processes with intensity $\lambda$ and positive and negative jumps equally distributed in magnitude. To fix the variance of $\xi^{(n),X}$ (at generic time) over $n$, it suffices to set $b_{n}=\sqrt{C_{2,n}}$, for $\xi^{(n),Y}$, $b_{n}=3^{-n/2}$, and for $\xi^{(n),Z}$, $b_{n}=1/(\sqrt{2n+1}\Gf(n+1))$. Then, as $n$ increases, the domain of jump amplitudes spreads out, albeit bounded, and the probability of smaller jumps also rises, as Figure \ref{fig:3} shows. Therefore, $\xi^{(n),X}$, $\xi^{(n),Y}$, and $\xi^{(n),Z}$ all constitute a rich class of desirable models, when return jumps are known to be bounded in size but unevenly distributed.

Moreover, if $X$ and $X'$ are two independent Poisson processes with possibly different intensities, say $\lambda,\lambda'>0$, the modified constant mixture\footnote{The superscript is expanded to include the process $X'$ because here it is not an exact copy of $X$.} $\xi^{X,X'}/b$ with parameters $b>0$, $\mu=0$, and $\kappa_{1}=\kappa_{2}=1$ is seen to be a scaled Skellam process which has jumps of fixed amplitude (equal to $1/b$). This type of discrete-valued processes have gained popularity in modeling tick-by-tick price changes with high-frequency data, as pioneered in [Barndorff-Nielsen et al., 2011] \cite{B-NPS}. Recent developments have dealt with stochastic-volatility extensions for memory effects; see, e.g., [Kerss et al., 2014] \cite{KLS}, and [Gupta et al., 2020] \cite{GKL}, the latter also having studied associated running average processes beside discussing natural applications to scoring in ball sports. In connection with this, the process $\bar{X}^{(n)}-\bar{X}'^{(n)}$ ($n\in\mathbb{N}_{++}$), up to scaling by a positive number, defines exactly the $n$-degree running average of $\xi^{X}$ (by linearity) and enables analysis of arbitrarily long-term behaviors of such phenomena (tick price changes versus scoring gaps). On the other hand, the scaled L\'{e}vy processes $\xi^{(n),X,X'}/b$, $\xi^{(n),Y,Y'}/b$, and $\xi^{(n),Z,Z'}/b$ of possibly fractional degrees $n>0$, when applied for the same modeling purposes, allow for imprecision (e.g., ambiguity between change versus no change) in possible outcomes, by assigning positive probability to jumps of amplitudes strictly less than $1/b$. In consequence, these possible outcomes are extended to real values to balance out false judgments, but still cannot fall out of their constructional range (tick sizes or score scales). In particular, the extent of imprecision is directly related to $n$ -- if $n\in(0,1)$, ambiguity in favor of no change stays minor. Obviously, in the limit as $n\searrow0$ the original discrete-valued processes are restored.

\subsection{Tempered stable subordinators}\label{sec:4.2}

Tempered stable distributions are an infinitely divisible family that was initially introduced in [Koponen, 1995] \cite{K2} from a procedure incorporating the finite moments property of Gaussian processes into L\'{e}vy stable processes, where it bore the name ``truncated L\'{e}vy flights.'' It is most commonly known in its one-sided form, characterized by up to three parameters $a>0$ (shape), $b>0$ (rate) and $c\in[0,1)$ (family), with which the corresponding L\'{e}vy processes are often called tempered stable subordinators and so are readily usable as stochastic clocks. We refer to [Schoutens, 2003, \text{Sect.} 5.3.6] \cite{S2}, [Rosi\'{n}ski, 2007] \cite{R}, and [K\"{u}chler and Tappe, 2013] \cite{KT} for a comprehensive treatment of these processes; see also [Grabchak, 2021] \cite{G2} for their discrete-valued analogy. The tempered stable family includes several commonly used subordinators as special cases, such as the gamma process ($c=0$) and the inverse Gaussian process ($c=1/2$). If $X$ is a tempered stable subordinator with parametrization $\{a,b,c\}$, its LT (at $t=1$) is given by
\begin{equation*}
  \phi_{X_{1}}(u)=
  \begin{cases}
    \exp(a\Gf(-c)((u+b)^{c}-b^{c}))\quad&\text{if }c\in(0,1)\\
    \displaystyle \bigg(\frac{u}{b}+1\bigg)^{-a}\quad&\text{if }c=0,
  \end{cases}
  \quad u\in\mathbb{C}\setminus(-\infty,-b],
\end{equation*}
which is right-continuous at $c=0$. The L\'{e}vy density $\ell$ of $X$ exists and equals $\ell(z)=ae^{-bz}/z^{c+1}$, $z>0$.

For $n=1$, the indistinguishable order-1 regulated clocks $X^{(1)}$, $Y^{(1)}$, and $Z^{(1)}$ have been studied in depth in [Xia, 2021] \cite{X2}, by the name of ``average-tempered stable subordinators.'' For all other values of $n>0$, in the presence of the gamma-related functions, for the type I it does not seem possible to further write the corresponding formulae in (\ref{2.1.5}) and (\ref{2.2.3}) explicitly, whereas both in the present form can be computed numerically with high efficiency thanks to the positivity and boundedness of the integrands.

In contrast, the formulae in (\ref{2.1.5}) and (\ref{2.2.3}) for the type-II clocks are reducible to some closed-form expressions provided $n\in\mathbb{N}_{++}$, by exploiting series representations of the integrands; further, due to the constructional simplicity of the type III, its corresponding formulae readily permit explicit evaluation, even for fractional degrees, which results are given in Proposition \ref{pro:2}.

\begin{proposition}\label{pro:2}
If $X$ is a tempered stable subordinator with with parametrization $\{a,b,c\}$, then for any $n>0$,
\begin{align}\label{4.2.1}
  \phi_{\tilde{X}^{(n)}_{1}}(u)&=
  \begin{cases}
    \displaystyle \exp\bigg(ab^{c}\Gf(-c)\bigg(\;_{n+1}\F_{n}\bigg(1,\dots,1,-c;2,\dots,2;-\frac{u}{b}\bigg)-1\bigg)\bigg)\quad&\text{if }c>0 \\
    \displaystyle e^{an}\bigg(\frac{u}{b}+1\bigg)^{-a(b/u+1)}\exp\Bigg(\frac{ab}{u}\sum^{n}_{j=2}\Li_{j}\bigg(-\frac{u}{b}\bigg)\Bigg)\quad&\text{if }c=0,
  \end{cases}
  \nonumber\\
  &\quad u\in\mathbb{C}\setminus(-\infty,-b],\;n\in\mathbb{N}_{++}, \nonumber\\
  \phi_{\breve{X}^{(n)}_{t}|c>0}(u)&=
  \begin{cases}
    \displaystyle \exp\bigg(atb^{c}\Gf(-c)\bigg(\;_{2}\F_{1}\bigg(-c,\frac{1}{n};\frac{1}{n}+1;-\frac{u}{b\Gf(n+1)}\bigg)-1\bigg)\bigg)\quad&\text{if }c>0 \\
    \displaystyle \bigg(1+\frac{u}{b\Gf(n+1)}\bigg)^{-at}\exp\bigg(\frac{atnu}{b\Gf(n+2)} \;_{2}\F_{1}\bigg(1,\frac{1}{n}+1;\frac{1}{n}+2;-\frac{u}{b\Gf(n+1)}\bigg)\bigg)\quad&\text{if }c=0, \\
  \end{cases}
  \nonumber\\
  &\quad u\in\mathbb{C}\setminus(-\infty,-b\Gf(n+1)],\;n>0,
\end{align}
where $\Li_{j}(\cdot)=\sum^{\infty}_{k=1}(\cdot)^{k}k^{-j}$ the poly-logarithm, and where analytic continuation is understood in each slit region $(D_{0}(b))^{\complement}\setminus(-\infty,-b]$ and $(D_{0}(b\Gf(n+1)))^{\complement}\setminus(-\infty,-b\Gf(n+1)]$, $n>0$, respectively, and
\begin{align}\label{4.2.2}
  \tilde{\ell}^{(n)}(z)&=ab^{c+1}\G^{n+1,0}_{n,n+1}\bigg(
  \begin{array}{cc}
    1,\dots,1 \\
    0,\dots,0,-c-1
  \end{array}
  \bigg|bz\bigg), \quad z>0,\;n\in\mathbb{N}_{++},\nonumber\\
  \breve{\ell}^{(n)}(z)&=\frac{ab^{c}(b\Gf(n+1)z)^{1/n}\Gf(-c-1/n,b\Gf(n+1)z)}{nz},\quad z>0,\;n>0,
\end{align}
where $\G^{\cdot,\cdot}_{\cdot,\cdot}(\cdots|\cdot)$ denotes the Meijer G-function (see the proof in Appendix \hyperref[A]{A}).
\end{proposition}

Note that if $n=1$, then with $\;_{2}\mathrm{F}_{1}(1,-c;2;-u/b)=(b^{-c}(u+b)^{c+1}-b)/((c+1)u)$ and $\sum^{1}_{j=2}\equiv0$ the elementary formulae follow:
\begin{equation*}
  \phi_{\tilde{X}^{(1)}_{1}|c>0}(u)=\exp\frac{a\Gf(-c)((u+b)^{c+1}-b^{c}((c+1)u+b))}{(c+1)u},\quad \phi_{\tilde{X}^{(1)}_{1}|c=0}(u)=e^{a}\bigg(\frac{u}{b}+1\bigg)^{-a(b/u+1)},
\end{equation*}
as were initially obtained in [Xia, 2021, \text{Thm.} 1 and \text{Corol.} 2] \cite{X2}. For rational values of the family parameter $c\in(0,1)$, it is possible to rewrite (\ref{4.2.1}) in terms of multiply nested sums of poly-logarithms; see, e.g., [Kalmykov and Kniehl, 2009] \cite{KK}. However, it is our preference not to straggle into those representations which are extremely lengthy in general, as interest is more for computational efficiency. For example, the following elementary formula applies to $n=2$ in the special case $c=1/2$, corresponding to the inverse Gaussian process:
\begin{equation*}
  \phi_{\tilde{X}^{(2)}_{1}|c=\frac{1}{2}}(u)=\bigg(\frac{1}{2}\bigg(\sqrt{\frac{u}{b}+1}+1\bigg)\bigg)^{8a\sqrt{b^{3}\pi}/(3u)} \exp\bigg(2a\sqrt{b\pi}\bigg(\frac{4b}{9u}\bigg(4-\bigg(\frac{u}{b}+4\bigg)\sqrt{\frac{u}{b}+1}\bigg)+1\bigg)\bigg).
\end{equation*}
It can be easily verified using the antiderivative of $(\log v)\sqrt{uv+b}$ in $v\in(0,1)$; see [Gradshteyn and Ryzhik, 2007, \text{Eq.} 2.725] \cite{GR}.

Also, we note that none of the L\'{e}vy densities given in (\ref{4.2.2}) is integrable on $\mathbb{R}_{++}$, meaning that the processes $Y^{(n)}$'s all have essentially infinite jump activity, as discussed earlier; in particular, for the tempered stable subordinator $\mathfrak{B}(X^{(n)})=\mathfrak{B}(Y^{(n)})=\mathfrak{B}(Z^{(n)})=\mathfrak{B}(X)=c\in[0,1)$. In the special case $n=1$, the first formula in (\ref{4.2.2}) reduces to
\begin{equation*}
  \tilde{\ell}^{(1)}(z)=ab^{c+1}\Gf(-c-1,bz),\quad z>0,
\end{equation*}
corresponding to the L\'{e}vy density derived in [Xia, 2021, \text{Corol.} 1] \cite{X2}, while for $n\geq2$, the formula (\ref{4.2.2}) stands and cannot be simplified to more elementary functions.\footnote{The first formula (\ref{4.2.2}) is not really useful beyond deciphering the functional form of the induced L\'{e}vy densities; more specifically, because the G-function is defined through complex line integration by most computer algebra systems, its implementation is understandably much less computationally efficient relative to its numerical counterpart (via (\ref{2.2.3}) by the quadrature rule), which also accommodates fractional values of $n$.}

Unlike the other two, the third recipe benefits by having no effect of increasing the computational complexity of the LT when operating on a tempered stable subordinator,\footnote{Numerical computations through the formulae (\ref{4.2.1}) and (\ref{4.2.2}) still intensify with the regulation degree $n\in\mathbb{N}_{++}$ which controls the dimensionality of the hypergeometric functions.} and by this advantage the formulae (\ref{4.2.1}) can be confidently applied to characteristic function-based optimization problems such as calibrating the parameters of a market model on strings of derivatives prices using numerical Fourier transforms.

Because the LTs $\phi_{\bar{X}^{(n)}_{t}}$, $\phi_{\tilde{X}^{(n)}_{t}}$, and $\phi_{\breve{X}^{(n)}_{t}}$ (fixing $t>0$) are all analytic in the complex plane except for a branch point on the negative real axis, the corresponding inverses (i.e., (a posteriori existent) density functions) can be found via a standard deformation argument employing a counterclockwise keyhole contour avoiding the cut extending to $-\infty$. Consequently, we obtain the following formulae:
\begin{align}\label{4.2.3}
  f_{\bar{X}^{(n)}_{t}}(x)&:=\frac{\PP\big\{\bar{X}^{(n)}_{t}\in\dd x\big\}}{\dd x}=\frac{1}{\pi}\int^{\infty}_{b}\dd u\;\exp\big(\Re\log\phi_{\bar{X}^{(n)}_{t}}(e^{\ii\pi}u)-ux\big)\sin\big(-\Im\log\phi_{\bar{X}^{(n)}_{t}}(e^{\ii\pi}u)\big), \nonumber\\
  f_{\tilde{X}^{(n)}_{t}}(x)&:=\frac{\PP\big\{\tilde{X}^{(n)}_{t}\in\dd x\big\}}{\dd x}=\frac{1}{\pi}\int^{\infty}_{b}\dd u\;\exp\big(\Re\log\phi_{\tilde{X}^{(n)}_{t}}(e^{\ii\pi}u)-ux\big)\sin\big(-\Im\log\phi_{\tilde{X}^{(n)}_{t}}(e^{\ii\pi}u)\big), \nonumber\\
  f_{\breve{X}^{(n)}_{t}}(x)&:=\frac{\PP\big\{\breve{X}^{(n)}_{t}\in\dd x\big\}}{\dd x}=\frac{1}{\pi}\int^{\infty}_{b\Gf(n+1)}\dd u\;\exp\big(\Re\log\phi_{\breve{X}^{(n)}_{t}}(e^{\ii\pi}u)-ux\big)\sin\big(-\Im\log\phi_{\breve{X}^{(n)}_{t}}(e^{\ii\pi}u)\big)
\end{align}
for $x,t>0$, where the real and imaginary parts are efficiently computable according to Theorem \ref{thm:1}.\footnote{From a computational viewpoint, the formulae in (\ref{4.2.3}) are stabler than performing a numerical inverse Fourier transform and allow the density functions to evaluated (or plotted) on a fine grid.} In particular, for $n\in\mathbb{N}_{++}$ we have, after the substitution $u/b\mapsto1/y$
\begin{align}\label{4.2.4}
  f_{\tilde{X}^{(n)}_{t}|c>0}(x)&=\frac{be^{-atb^{c}\Gf(-c)}}{\pi}\int^{1}_{0}\frac{\dd y}{y^{2}}\exp\bigg(atb^{c}\Gf(-c)\Re\;_{n+1}\F_{n}\bigg(1,\dots,1,-c;2,\dots,2;\frac{1}{y}\bigg)-\frac{bx}{y}\bigg) \nonumber\\
  &\qquad\times\sin\bigg(-atb^{c}\Gf(-c)\Im\;_{n+1}\F_{n}\bigg(1,\dots,1,-c;2,\dots,2;\frac{1}{y}\bigg)\bigg),\quad x>0
\end{align}
and for any $n>0$, with $u/(b\Gf(n+1))\mapsto1/y$,
\begin{align}\label{4.2.5}
  f_{\breve{X}^{(n)}_{t}|c>0}(x)&=\frac{b\Gf(n+1)e^{-atb^{c}\Gf(-c)}}{\pi}\int^{1}_{0}\frac{\dd y}{y^{2}}\exp\bigg(atb^{c}\Gf(-c)\Re\;_{2}\F_{1}\bigg(-c,\frac{1}{n};\frac{1}{n}+1;\frac{1}{y}\bigg)-\frac{b\Gf(n+1)x}{y}\bigg) \nonumber\\
  &\qquad\times\sin\bigg(-atb^{c}\Gf(-c)\Im\;_{2}\F_{1}\bigg(-c,\frac{1}{n};\frac{1}{n}+1;\frac{1}{y}\bigg)\bigg),\quad x>0.
\end{align}
Despite stability, implementation of (\ref{4.2.4}) will require some effort for $n\geq2$ since the hypergeometric functions need to be computed by analytic continuation of the definitional series which are only convergent within $D_{0}(1)$, while there is no such issue for (\ref{4.2.5}) because of fixed dimensionality of the hypergeometric functions.

With its overall flexibility and explicitness, processes with marginal tempered stable distributions are of great applicability in operations research, especially in the fields of reliability engineering and mathematical finance; we mention [van Noortwijk, 2009] \cite{vN}, [Wang and Xu, 2015] \cite{WX}, and [Ye and Xie, 2015] \cite{YX} for r\'{e}sum\'{e}s on related structural degradation models and [Boyarchenko and Levendorskii, 2000] \cite{BL} and [Carr et al., 2002] \cite{CGMY} for derivatives pricing theory. For Gaussian mixtures, the path regularity is directly captured by the family parameter $c\in[0,1)$. Recalling that $\mathfrak{B}(\Xi^{X})=\mathfrak{B}(\Xi^{(n),X})=\mathfrak{B}(\Xi^{(n),Y})=\mathfrak{B}(\Xi^{(n),Z})=2c$, for any $n>0$ (by Corollary \ref{cor:6}), if $c\in[0,1/2)$, then the mixed processes $\Xi^{(n),X}$, $\Xi^{(n),Y}$, and $\Xi^{(n),Z}$ all have sample paths of finite variation, while the paths are of infinite variation for $c\in[1/2,1)$. In the special cases $c=0$ and $c=1/2$, $\Xi^{X}$ becomes the variance gamma process and the normal inverse Gaussian process, respectively.\footnote{In [Carr et al., 2002] \cite{CGMY}, two-sided tempered stable processes were studied under the name ``CGMY,'' which include the variance gamma process as a special case. However, for $c\neq0$ such processes cannot be interpreted as Gaussian mixtures of tempered stable subordinators.} With $\mathfrak{B}(X)=0$ the sample paths of the variance gamma process are of finite variation while with $\mathfrak{B}(X)=1$ those of the normal inverse Gaussian process have infinite variation. In such cases, the processes $\Xi^{(n),X}$, $\Xi^{(n),Y}$, and $\Xi^{(n),Z}$ derived from stochastic clock regulation are natural extensions of the foregoing models with arbitrarily large skewness and kurtosis, the mean and variance fixed and path regularity unchanged.

Subject to the addition parameters $\mu,\theta\in\mathbb{R}$, the density function of $\Xi^{(n),X}_{t}$ can be written either with (\ref{3.2.2}) as the Fourier inverse
\begin{align}\label{4.2.6}
  f_{\Xi^{(n),X}_{t}}(x):=\frac{\PP\big\{\Xi^{(n),X}_{t}\in\dd x\big\}}{\dd x}&=\frac{1}{\pi}\int^{\infty}_{0}\Re\big(e^{-\ii ux}\phi_{\Xi^{(n),X}_{t}}(-\ii u)\big)\dd u \nonumber\\
  &=\frac{1}{\pi}\int^{\infty}_{0}\Re\bigg(e^{\ii(\mu t-x)u}\phi_{\bar{X}^{(n)}_{t}}\bigg(-\ii\theta u+\frac{u^{2}}{2}\bigg)\bigg)\dd u,\quad x\in\mathbb{R},\;t>0
\end{align}
or, after applying the Bayes theorem to (\ref{4.2.3}), as the marginalization
\begin{align}\label{4.2.7}
  f_{\Xi^{(n),X}_{t}}(x)&=\int^{\infty}_{0}\frac{\dd w}{\sqrt{2\pi w}}e^{-(x-\theta w)^{2}/(2w)}f_{\bar{X}^{(n)}_{t}}(w) \nonumber\\
  &=\frac{1}{\pi}\int^{\infty}_{b}\frac{\dd u}{\sqrt{2u+\theta^{2}}}\exp\big(\Re\log\phi_{\bar{X}^{(n)}_{t}}(e^{\ii\pi}u)+\theta(x-\mu t)-|x-\mu t|\sqrt{2u+\theta^{2}}\big) \nonumber\\
  &\qquad\times\sin\big(-\Im\log\phi_{\bar{X}^{(n)}_{t}}(e^{\ii\pi}u)\big),
\end{align}
where the use of Fubini's theorem in the second equality is under the condition that $c<1/2$;\footnote{It can be shown that the polarity of $\Re\log\phi_{X_{t}}(e^{\ii\pi}u)$ with $X$ being a tempered stable subordinator exactly equals that of $2c-1$, which feature is also unaffected by regulation ($n$).} similar results hold for the densities $f_{\Xi^{(n),Y}_{t}}$ and $f_{\Xi^{(n),Z}_{t}}$ in terms of $\phi_{\tilde{X}^{(n)}_{t}}$ and $\phi_{\breve{X}^{(n)}_{t}}$, respectively.

It is straightforward to keep the mean and variance of the Gaussian mixture $\Xi^{(n),Y}$ invariant to $n$ with the parametrization $\{a,b,\mu,\theta\}$. One can for example adjust the location and Brownian drift parameters $\mu$ and $\theta$ according to the rules $\mu_{n}+\theta_{n}\E X_{1}/2^{n}=[\text{target mean}]$ and $\E X_{1}/2^{n}+\theta^{2}_{n}(\Var X_{1})/3^{n}=[\text{target variance}]$;\footnote{Since we know from Subsection \ref{sec:3.2} that the value of $\theta$ directly controls the skewness, this approach can very quickly result in abnormally large asymmetry as $n$ increases.} to maintain the moderate enlargement effects in (\ref{2.3.5}), one can as well adjust the location and (tempered stable) rate parameters $\mu$ and $b$ by the rules $\mu_{n}+\theta\E X_{1}/(2^{n}b_{n})=[\text{target mean}]$ and $\E X_{1}/(2^{n}b_{n})+\theta^{2}(\Var X_{1})/(3^{n}b^{2}_{n})=[\text{target variance}]$, where $X_{1}$ has unit scale; alternatively, disregarding the additional parameters $\mu$ and $\theta$ one can change the shape and rate parameters $a$ and $b$ of the unregulated clock $X$, as discussed in Subsection \ref{sec:3.2}.\footnote{In contrast to the first, this one tends to give milder-than-moderate enlargement effects and so can fit in with a wide range of values of $n$, hence also more manageable; see Appendix \hyperref[B]{B} for details.} Along these lines one can come up with many hybrid methods for fixing the mean and variance, e.g., by changing $\mu$, $a$ and powers of $b$ simultaneously. Similar arguments go for $\Xi^{(n),X}$ and $\Xi^{(n),Z}$.

\medskip

\section{Optimal degrees of regulation}\label{sec:5}

It is time to consider the problem of finding an optimal degree $n$ of regulation with available data. Optimization may be accomplished by many different rules, depending on various aspects of the resultant distributions and data types. Instead of a comprehensive analysis, the aim of this section is to discuss two popular and comfortable methods that can justify values significantly deviating from zero (unregulated case) with due contrast control. The methods by themselves accommodate both the constant mixture $\xi^{X}$ and the Gaussian mixture $\Xi^{X}$.

\subsection{Moment-based estimation}\label{sec:5.1}

In financial modeling, it is a standard assumption that a probability distribution of asset returns can be defined by the knowledge of their first several (usually four) moments; see, e.g., [An\'{e} and Geman, 2000] \cite{AG}. This suggests that one can use the first sample moments -- in particular the sample mean, variance, skewness, and excess kurtosis -- to retrieve certain model parameters, and underlies our moment-based estimation procedure, which goes as follows.

We start by fixing a truncated and discretized domain $\mathfrak{N}\subsetneq\mathbb{R}_{+}$ of degrees of regulation, and conduct moment-based estimation of the model parameters for each $n\in\mathfrak{N}$, with $n=0$ corresponding to the unregulated case. Recall that regulation leads to simple cumulant reduction relations (Corollary \ref{cor:2}), with regulation type-specific coefficients
\begin{equation}\label{5.1.1}
  \rho(m,n):=
  \begin{cases}
    C_{m,n}\quad&\text{type I}\\
    \displaystyle \frac{1}{(m+1)^{n}}\quad&\text{type II}\\
    \displaystyle \frac{1}{(mn+1)\Gf^{m}(n+1)}\quad&\text{type III},
  \end{cases}\quad m\in\mathbb{N}_{++},\;n\geq0,
\end{equation}
so it is most convenient to construct corresponding moment conditions from the model cumulants; for similar estimation procedures in the literature we refer to [Beckers, 1981] \cite{B2} and [Bandi and Nguyen, 2003] \cite{BN}. After experimenting with every member in $\mathfrak{N}$, the optimal value of $n$ is identified (a posteriori) based on the maximal profile likelihood as a criterion for model selection. The reason behind choosing moment-based estimation over maximum likelihood estimation is twofold: It is much more computationally efficient because the density function of L\'{e}vy models are generally difficult to obtain in closed form (e.g., see [Schoutens, 2003, \text{Chap.} 4] \cite{S2}), with or without clock regulation; it is also more robust by making fewer assumptions for the underlying data-generating mechanism as only a finite number of moments are utilized.

Suppose that we have obtained an \text{i.i.d.} sample of some log-returns $\{\check{x}_{k}\}^{N}_{k=1}$, with $N\gg1$ being the sample size, at a uniform frequency $\D>0$ in fractions of a year; from now on we shall write as $[M]$, $[V]$, $[SK]$, and $[EK]$ the sample mean, variance, skewness, and excess kurtosis, respectively, and $K^{(n)}_{\D}(m)$, for $m\in\{1,2,3,4\}$, the mixed model cumulants with regulation degrees $n\geq0$. Then, the four moment conditions can be written
\begin{equation}\label{5.1.2}
  K^{(n)}_{\D}(1)=[M],\quad K^{(n)}_{\D}(2)=[V],\quad K^{(n)}_{\D}(3)=[SK][V]^{3/2},\quad\frac{K^{(n)}_{\D}(3)}{K^{(n)}_{\D}(4)}=\frac{[SK]}{[EK][V]^{1/2}}.
\end{equation}

First, we take $X$ to be a Poisson process and its constant mixture $\xi^{X}$ plus an independent Brownian motion, as discussed in Subsection \ref{sec:4.1}. From (\ref{4.1.3}), (\ref{4.1.4}), and (\ref{4.1.5}) we know that in this case the model is of jump--diffusion type with bounded (distorted uniform) jumps. For convenience, we shall focus on downside jumps, in view of excessive tail risk, by imposing the auxiliary parameter values $\kappa_{1}=0$ and $\kappa_{2}=1$ in (\ref{3.1.1}). Note that the mixed model with clock regulation has a total of four parameters to be estimated, including an infinite-divisibility parameter $\lambda_{n}>0$, a rate parameter $b_{n}>0$, a location parameter $\mu\in\mathbb{R}$, and a Brownian dispersion parameter $\sigma_{n}>0$, all of which are subscripted by the corresponding degree of regulation, $n\geq0$.

\begin{proposition}\label{pro:3}
Given any regulation degree $n\geq0$, under the Poisson-based jump--diffusion model (see (\ref{4.1.6})) with $\kappa_{1}=0$ and $\kappa_{2}=1$ and parametrization $\{\lambda_{n},b_{n},\mu_{n},\sigma_{n}\}$, the moment-based estimators are uniquely determined as
\begin{equation}\label{5.1.3}
  \hat{b}_{n}=\frac{|[SK]|}{[EK][V]^{1/2}}\frac{\rho(4,n)}{\rho(3,n)},\quad \hat{\lambda}_{n}=\frac{|[SK]|[V]^{3/2}\hat{b}^{3}_{n}}{\D\rho(3,n)},\quad \hat{\sigma}_{n}=\frac{[V]}{\D}-\rho(2,n)\frac{\hat{\lambda}_{n}}{\hat{b}^{2}_{n}},\quad \hat{\mu}_{n}=\frac{[M]}{\D}+\rho(1,n)\frac{\hat{\lambda}_{n}}{\hat{b}_{n}},
\end{equation}
where $\rho(m,n)$'s are specified in (\ref{5.1.1}).
\end{proposition}

In terms of the above estimates, the profile likelihood function under the mixed model (with an $n$-degree regulated clock) can be directly computed using the Bernoulli approximation technique in [Kou, 2002, \text{Eq.} (5)] \cite{K3} as
\begin{align*}
  \hat{g}_{\D}(n)&=\prod^{N}_{k=1}\bigg(\D\int_{\mathbb{R}}\frac{\hat{b}_{n}}{\sqrt{2\pi\hat{\sigma}^{2}_{n}}} \exp\bigg(-\frac{(\check{x}_{k}-z-\hat{\mu}_{n}\D)^{2}}{2\hat{\sigma}^{2}_{n}\D}\bigg)\ell_{n}(-\hat{b}_{n}z)\dd z\\
  &\quad+(1-\hat{\lambda}_{n}\D)\frac{1}{\sqrt{2\pi\hat{\sigma}^{2}_{n}}} \exp\bigg(-\frac{(\check{x}_{k}-\hat{\mu}_{n}\D)^{2}}{2\hat{\sigma}^{2}_{n}\D}\bigg)\bigg),
\end{align*}
provided that $\D$ is small -- e.g., as in the case of daily data, where the L\'{e}vy density $\ell_{n}$ represents any one of $\bar{\ell}_{n}$, $\tilde{\ell}_{n}$, and $\breve{\ell}_{n}$ (with scaling factor $\hat{\lambda}_{n}$) in Proposition \ref{pro:1} according to the type of regulation. Whichever model yields the largest value of $\hat{g}_{\D}(n)$ is taken as the best.

Second, we let $X$ be a tempered stable subordinator and construct its Gaussian mixture $\Xi^{X}$. The family parameter $c\in[0,1)$ is fixed a priori for its special connection to local path regularity, which leaves us with also four parameters to be estimated -- (with regulation degree subscripts) the infinite-divisibility parameter $a_{n}>0$, rate parameter $b_{n}>0$, location parameter $\mu\in\mathbb{R}$, and the Brownian drift parameter $\theta_{n}\in\mathbb{R}$. In this case, the moment-based estimators can be shown to be generally unique under the conditions in (\ref{5.1.2}).

\begin{proposition}\label{pro:4}
Given any regulation degree $n\geq0$, under the Gaussian-mixed tempered stable model with $c\in[0,1)$ and parametrization $\{a_{n},b_{n},\mu_{n},\theta_{n}\}$, the moment-based estimators are given by the conditions
\begin{align}\label{5.1.4}
  &\hat{b}_{n}=\frac{2(2-c)\rho(3,n)\hat{\theta}^{2}_{n}}{\rho(2,n)P(|\hat{\theta}_{n}|)},\quad \hat{a}_{n}=\frac{[V]\hat{b}^{1-c}_{n}}{\D\Gf(1-c)\big(\rho(1,n)+(1-c)\rho(2,n)\hat{\theta}^{2}_{n}/\hat{b}_{n}\big)}, \nonumber\\
  &\hat{\mu}_{n}=\frac{[M]}{\D}-\frac{\rho(1,n)\hat{\theta}_{n}[V]}{\D\big(\rho(1,n)+(1-c)\rho(2,n)\hat{\theta}^{2}_{n}/\hat{b}_{n}\big)},
\end{align}
where $\hat{\theta}_{n}=\mathrm{sgn}([SK])|\hat{\theta}_{n}|$ and $|\hat{\theta}_{n}|$ solves the rational equation
\begin{equation}\label{5.1.5}
  \frac{(3-c)\rho(2,n)\rho(4,n)}{4(2-c)\rho(3,n)^{2}}P^2(|\theta_{n}|)+\bigg(3-\frac{[EK]|\theta_{n}|}{2|[SK]|}\bigg)P(|\theta_{n}|) +3\bigg(1-\frac{[EK]|\theta_{n}|}{|[SK]|}\bigg)=0,
\end{equation}
with
\begin{align*}
  P(|\theta_n|)&=(|[SK]||\theta_{n}|-3)+D^{1/2}(|\theta_{n}|), \\
  D(|\theta_n|)&=(|[SK]||\theta_{n}|-3)^{2}+\frac{4(2-c)\rho(1, n)\rho(3, n)}{(1-c)\rho(2,n)^{2}}|[SK]||\theta_{n}|.
\end{align*}
\end{proposition}

With these estimates, the corresponding profile likelihood function is then simply $\hat{g}_{\D}(n)=\prod^{N}_{k=1}f_{\Xi^{(n),\#}_{\D}}(\check{x}_{k})$ (see (\ref{4.2.6}) and (\ref{4.2.7})), where $\#$ is a placeholder for $X$, $Y$, or $Z$ depending on the regulation type. Again, the best model is taken to be the one with the largest value of $\hat{g}_{\D}(n)$.

\subsection{Model calibration on option prices}\label{sec:5.2}

Alternatively, instead of estimating parameters from sample moments, one can recover them by model calibration on option price data. With the assumption of no-arbitrage, such an exercise will require that the LT be evaluated under some risk-neutral probability measure $\Q$ (other than $\PP$) under which the asset price is discounted at some (supposedly) constant risk-free rate $r\geq0$. Allowing for calibration efficiency, here we concentrate on the type-II and type-III regulated clocks which permit some explicit formulae when acting on Gaussian mixtures of tempered stable subordinators (recall Subsection \ref{sec:4.2}). By way of the ``mean-correcting measure'' outlined in [Schoutens, 2003, \text{Sect.} 6.2.2] \cite{S2}, we can then directly assume that the LT of the log asset price at time $t>0$ under the measure $\Q$ is given by
\begin{equation*}
  \varphi_{t}(u;n)=\bigg(\frac{S_{0}e^{rt}}{\phi_{\Xi^{(n),Y}_{t}}(-1)}\bigg)^{-u}\phi_{\Xi^{(n),Y}_{t}}(u)\quad\text{or}\quad \varphi_{t}(u;n)=\bigg(\frac{S_{0}e^{rt}}{\phi_{\Xi^{(n),Z}_{t}}(-1)}\bigg)^{-u}\phi_{\Xi^{(n),Z}_{t}}(u),\quad n\in\mathfrak{N},
\end{equation*}
provided that the denominator in the power is finite, where $r\geq0$ is the constant risk-free rate in the associated money market and $\mathfrak{N}$ is the same select domain of the regulation degree $n$ for experiments.

Then, the no-arbitrage price of a European-style call option written on $S$ at time $t=0$ with strike price $K$ and maturity $T>0$ fixed is given by
\begin{equation}\label{5.2.1}
  \varPi^{\text{call}}_{0}(n)=S_{0}e^{-qT}Q_{1}(n)-Ke^{-rT}Q_{2}(n),
\end{equation}
where $q>0$ is the (constant) dividend yield and
\begin{equation*}
  Q_{1}(n)=\frac{1}{2}+\frac{1}{\pi}\int^{\infty}_{0}\Re\frac{K^{-\ii u}\varphi_{T}(-\ii u-1;n)}{\ii u\varphi_{T}(-1;n)}\dd u,\quad Q_{2}(n)=\frac{1}{2}+\frac{1}{\pi}\int^{\infty}_{0}\Re\frac{K^{-\ii u}\varphi_{T}(-\ii u;n)}{\ii u}\dd u
\end{equation*}
are associated risk-neutralized in-the-money probabilities.

Suppose that we have obtained contemporaneous market quotes of $M$ call options on $S$ for various strike prices and maturities, denoted as $\check{\varPi}^{\text{call}}$. Focusing on the Gaussian mixed model $\Xi^{X}$ where $X$ is the tempered stable subordinator with a fixed family parameter $c\in[0,1)$, then after mean correction there will be only three parameters, $a_{n}$, $b_{n}$, and $\theta_{n}$ (with degree subscripts as before), whose (locally) optimal values can be found by solving the following minimization problem:
\begin{equation}\label{5.2.2}
  \{\hat{a}_{n},\hat{b}_{n},\hat{\theta}_{n}\}=\underset{a>0,b>0,\theta\in\mathbb{R}}{\arg\min} \frac{1}{M}\sum_{K,T}\frac{\big|\check{\varPi}^{\text{call}}_{0}-\varPi^{\text{call}}_{0}(n)\big|}{\check{\varPi}^{\text{call}}_{0}},
\end{equation}
where the objective function is the mean absolute percentage error (MAPE) measuring average pricing errors and the summation runs over all available strikes and maturities. The best model is then characterized by imposing the criterion $\min_{n\in\mathfrak{N}}$ on the right side of (\ref{5.2.2}). A similar exercise using put options is immediate by applying a standard parity argument to (\ref{5.2.1}). Nevertheless, compared with moment-based estimation, calibration is expected to be less robust because it generally constitutes a highly non-convex optimization problem which must be solved numerically.

\medskip

\section{Empirical modeling}\label{sec:6}

In this section we demonstrate the effect of clock regulation on real financial returns. We focus on the equity market and the cryptocurrency market; in particular, we have collected S\&P500 index and Bitcoin price data on the daily basis over the 2020 calendar year (data source: {\sl Yahoo Finance}) and have computed corresponding log-returns. The two data sets hence have sample sizes ($N$) of 252 and 366, respectively, with $\D=1/252,1/366$.

In accordance with the moment-based estimation procedures discussed in Subsection \ref{sec:5.1}, our empirical study will feature two exercises, one using the constant mixture of a Poisson process augmented by an independent Brownian motion and the other using the Gaussian mixture of a tempered stable subordinator, respectively. We highlight that the first exercise is interesting for modeling (presumably) bounded jumps in asset returns ([Yan and Hanson, 2006] \cite{YH} and [Baustian et al., 2017] \cite{BMPS}) and aims to compare the distorted jump distributions with commonly used uniform distributions; the second, on the other hand, is especially significant for possible improvement of tempered stable processes ([Rosi\'{n}ski, 2007] \cite{R} and [K\"{u}chler and Tappe, 2013] \cite{KT}) in terms of flexibility, by investigating how the degree $n$ of regulation affects the model fit subject to moment matching for different choices of the family parameter $c$.

First, we report the four sample statistics of the collected return data in Table \ref{tab:1}. Relative to the S\&P500, Bitcoin returns have clearly exhibited a much larger (absolute) skewness and excess kurtosis, speaking to high asymmetrical and tail risk.

\begin{table}[H]\small
  \centering
  \caption{Descriptive statistics of daily returns ($\sim$six significant digits)}
  \label{tab:1}
  \begin{tabular}{c|c|c|c|c}
    \hline
    Data set & mean & variance & skewness & excess kurtosis \\ \hline
    S\&P500 & 0.000564705 & 0.00047912 & $-0.861012$ & 8.46843 \\ \hline
    Bitcoin & 0.00384156 & 0.00160601 & $-4.07498$ & 50.8679 \\ \hline
  \end{tabular}
\end{table}

\begin{figure}[H]
  \centering
  \includegraphics[scale=0.3]{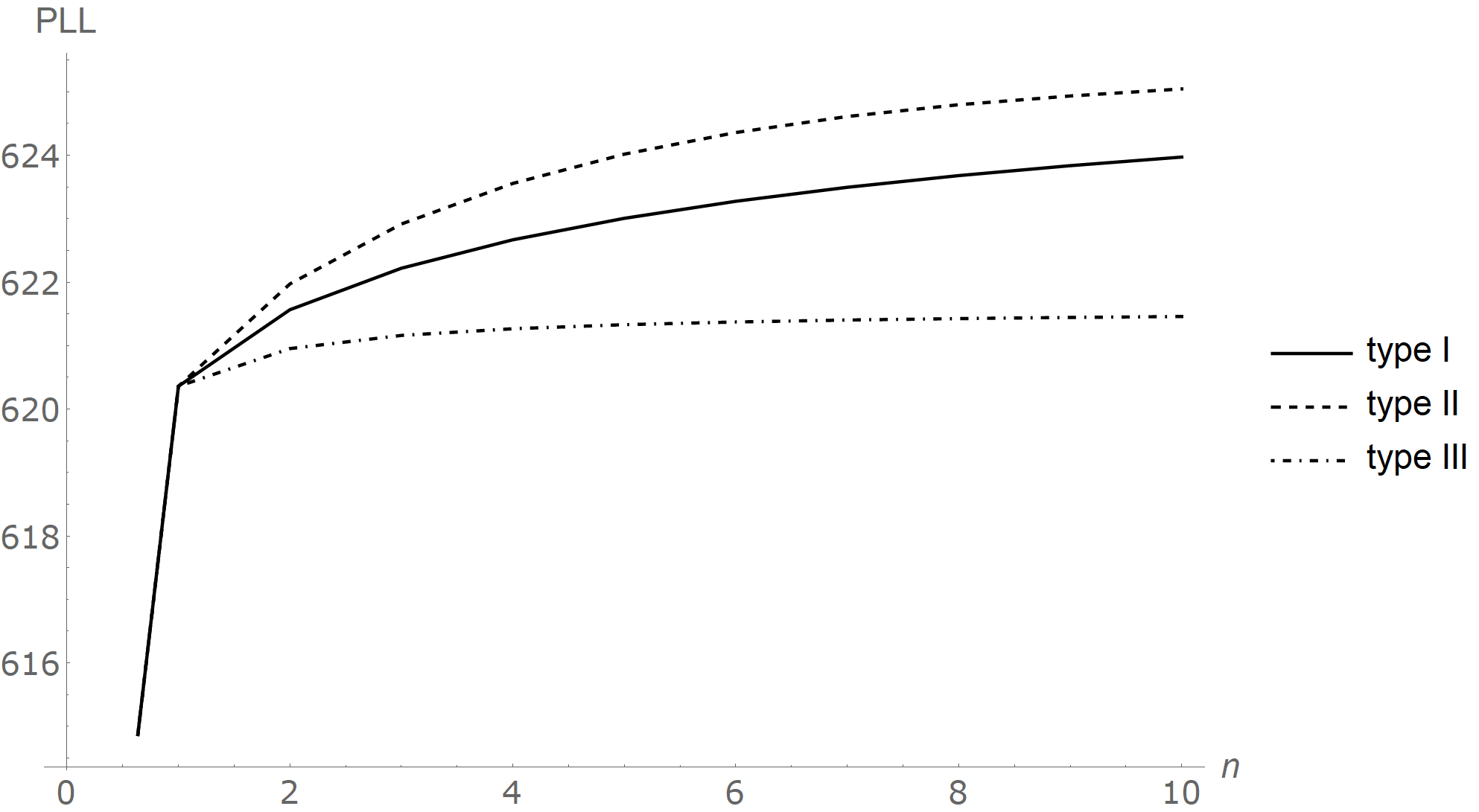}
  \includegraphics[scale=0.3]{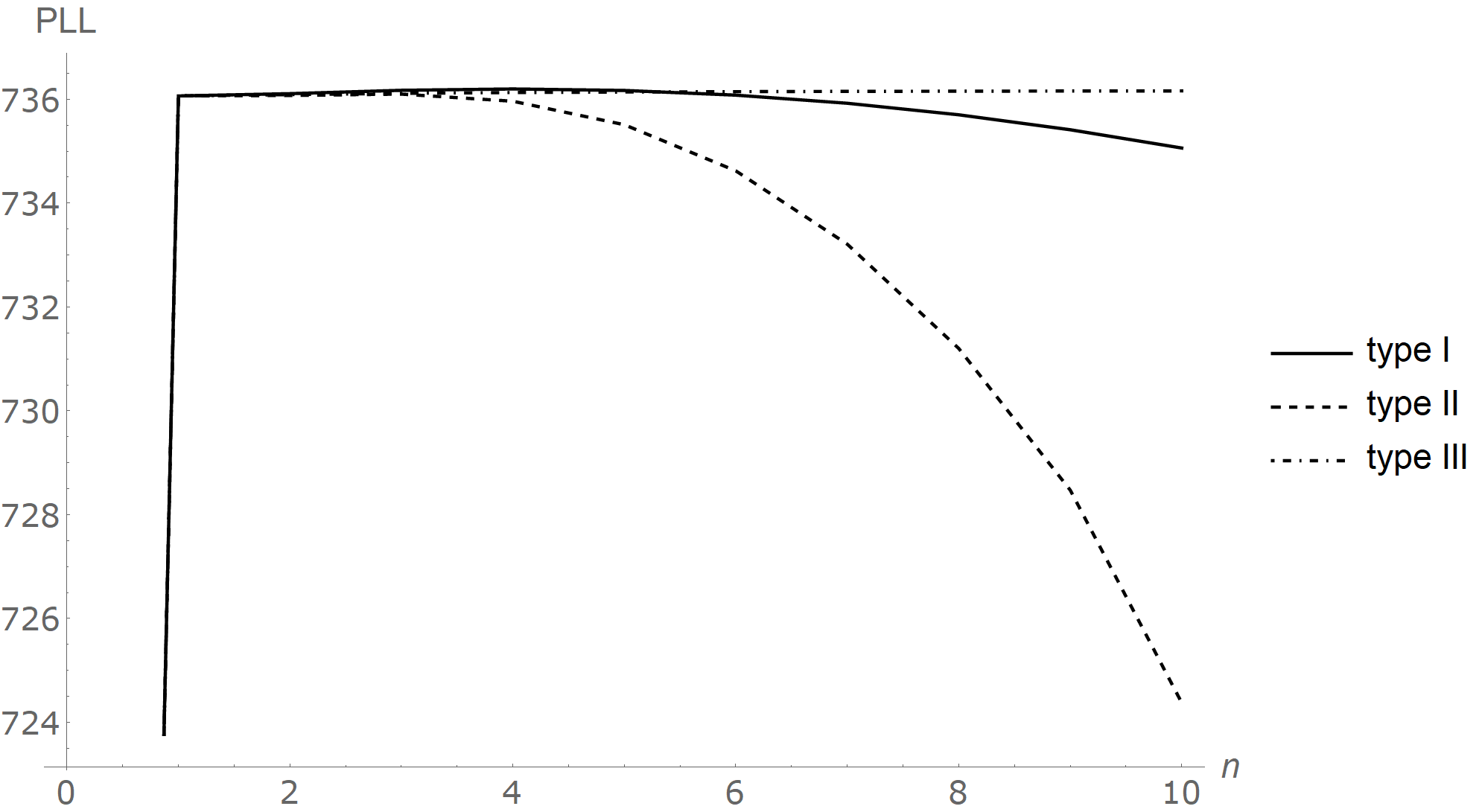}
  \caption{\label{fig:4} Jump diffusion log-likelihood with regulation degrees for S\&P500 (left) and Bitcoin (right) daily returns}
\end{figure}

In both exercises, the choice of the degree domain is $\mathfrak{N}\subsetneq[0,10]$ with fixed modulus $|\mathfrak{N}|=11$ to ease comparison. In the first exercise, the exponential jump--diffusion model ([Kou, 2002] \cite{K3}) is also included as a contrast model because the distorted uniform distributions due to regulation have exponential shapes. In the second exercise, we experiment with four values of $c$: 0 (variance gamma), 0.25, 0.5 (normal inverse Gaussian), and 0.75. Estimation results are presented in ten (sub)tables -- Table \ref{tab:2a}, Table \ref{tab:2b}, Table \ref{tab:3a}, Table \ref{tab:3b}, Table \ref{tab:3c}, Table \ref{tab:3d}, Table \ref{tab:4a}, Table \ref{tab:4b}, Table \ref{tab:4c}, and Table \ref{tab:4d}, and are comprised of parameter estimates and the profile log-likelihood (PLL), $\log\hat{g}_{\D}(n)$, for every $n\in\mathfrak{N}$. In each model setting (including each chosen value of $c$), the model with the highest PLL value for a certain regulation type is marked ``$\ast$'' and that with the highest PLL value taking into account all three types is marked ``$\star$.'' To enhance visual impact, changes in the PLL are plotted separately in Figure \ref{fig:4}, Figure \ref{fig:6a}, and Figure \ref{fig:6b}.

Furthermore, we provide Figure \ref{fig:5}, Figure \ref{fig:7a}, and Figure \ref{fig:7b} to compare the estimated density functions of the best fit model (namely the one with the highest PLL value across all three types) and the original model ($n=0$) against the empirical kernel density function which is estimated using the Gaussian kernel, i.e.,
\begin{equation*}
  \check{f}_{\D}(x)=\frac{1}{N\mathfrak{b}}\sum^{N}_{k=1}\frac{1}{\sqrt{2\pi}}e^{-(x-\check{x}_{k})^{2}/(2\mathfrak{b}^{2})},\quad x\in\mathbb{R};
\end{equation*}
the optimal bandwidth $\mathfrak{b}$ is determined by the Silverman rule of thumb:
\begin{equation*}
  \mathfrak{b}=\frac{0.9}{N^{1/5}}\min\bigg([V]^{1/2},\frac{\mathrm{IQR}(\check{x})}{1.349}\bigg),
\end{equation*}
where $\mathrm{IQR}$ is the interquartile range (the difference between the third and the first quartiles).

\clearpage

\vspace*{0.7in}
\begin{table}[H]\small
  \centering
  \ContinuedFloat*
  \caption{\label{tab:2a} Jump diffusion estimation results on S\&P500 daily returns ($\sim$six significant digits)}
  \begin{tabular}{c|c|cccc|c}
    \hline
    \multirow{11}[11]{*}{\textbf{type I}} & $n$ & $\hat{\lambda}_{n}$ & $\hat{b}_{n}$ & $\hat{\mu}_{n}$ & $\hat{\sigma}_{n}$ & PLL \\ \hline
    & 0 & 0.228049 & 4.64498 & 0.191402 & 0.331917 & 605.28 \\
    & 1 & 0.467045 & 3.71599 & 0.205148 & 0.330853 & 620.36 \\
    & 2 & 1.03524 & 3.37102 & 0.219081 & 0.330044 & 621.566 \\
    & 3 & 2.27627 & 3.11901 & 0.233531 & 0.329334 & 622.22 \\
    & 4 & 4.98085 & 2.91283 & 0.249179 & 0.328666 & 622.669 \\
    & 5 & 10.8658 & 2.73578 & 0.266423 & 0.328018 & 623.007 \\
    & 6 & 23.6553 & 2.5796 & 0.285589 & 0.327377 & 623.276 \\
    & 7 & 51.4231 & 2.43945 & 0.306992 & 0.326739 & 623.496 \\
    & 8 & 111.664 & 2.31218 & 0.330954 & 0.326099 & 623.68 \\
    & 9 & 242.275 & 2.19558 & 0.357827 & 0.325455 & 623.837 \\
    & 10 & 525.314 & 2.08804 & 0.387992 & 0.324803 & 623.972$\ast$ \\ \hline
    \multirow{11}[11]{*}{\textbf{type II}} & $n$ & $\hat{\lambda}_{n}$ & $\hat{b}_{n}$ & $\hat{\mu}_{n}$ & $\hat{\sigma}_{n}$ & PLL \\ \hline
    & 0 & 0.228049 & 4.64498 & 0.191402 & 0.331917 & 605.28 \\
    & 1 & 0.467045 & 3.71599 & 0.205148 & 0.330853 & 620.36 \\
    & 2 & 0.956508 & 2.97279 & 0.222744 & 0.329716 & 621.973 \\
    & 3 & 1.95893 & 2.37823 & 0.245267 & 0.328497 & 622.917 \\
    & 4 & 4.01189 & 1.90259 & 0.274096 & 0.327193 & 623.557 \\
    & 5 & 8.21634 & 1.52207 & 0.310998 & 0.325796 & 624.018 \\
    & 6 & 16.8271 & 1.21765 & 0.358232 & 0.3243 & 624.359 \\
    & 7 & 34.4618 & 0.974124 & 0.418691 & 0.322695 & 624.612 \\
    & 8 & 70.5779 & 0.779299 & 0.496079 & 0.320976 & 624.798 \\
    & 9 & 144.543 & 0.623439 & 0.595135 & 0.319131 & 624.935 \\
    & 10 & 296.025 & 0.498751 & 0.721927 & 0.317151 & 625.046$\star$ \\ \hline
    \multirow{11}[11]{*}{\textbf{type III}} & $n$ & $\hat{\lambda}_{n}$ & $\hat{b}_{n}$ & $\hat{\mu}_{n}$ & $\hat{\sigma}_{n}$ & PLL \\ \hline
    & 0 & 0.228049 & 4.64498 & 0.191402 & 0.331917 & 605.28 \\
    & 1 & 0.467045 & 3.71599 & 0.205148 & 0.330853 & 620.36 \\
    & 2 & 0.751093 & 1.80638 & 0.211606 & 0.330498 & 620.954 \\
    & 3 & 1.038 & 0.595511 & 0.214933 & 0.330338 & 621.161 \\
    & 4 & 1.32573 & 0.148002 & 0.216952 & 0.330247 & 621.267 \\
    & 5 & 1.6138 & 0.029492 & 0.218306 & 0.330189 & 621.33 \\
    & 6 & 1.90206 & 0.00490304 & 0.219277 & 0.330148 & 621.373 \\
    & 7 & 2.19041 & 0.000699163 & 0.220007 & 0.330118 & 621.404 \\
    & 8 & 2.47883 & 0.000087275 & 0.220576 & 0.330095 & 621.427 \\
    & 9 & 2.7673 & $9.68674\times10^{-6}$ & 0.221031 & 0.330077 & 621.445 \\
    & 10 & 3.05579 & $9.6783\times10^{-7}$ & 0.221404 & 0.330062 & 621.459 $\ast$ \\ \hline
    \textbf{exponential} & - & 2.43253 & 18.5799 & 0.273228 & 0.326566 & 624.048 \\ \hline
  \end{tabular}\\
  PLL: profile log-likelihood
\end{table}

\vspace*{0.7in}
\begin{table}[H]\small
  \centering
  \ContinuedFloat
  \caption{\label{tab:2b} Jump diffusion estimation results on Bitcoin daily returns ($\sim$six significant digits)}
  \begin{tabular}{c|c|cccc|c}
    \hline
    \multirow{11}[11]{*}{\textbf{type I}} & $n$ & $\hat{\lambda}_{n}$ & $\hat{b}_{n}$ & $\hat{\mu}_{n}$ & $\hat{\sigma}_{n}$ & PLL \\ \hline
    & 0 & 0.766746 & 1.99897 & 1.78958 & 0.62922 & 641.506 \\
    & 1 & 1.5703 & 1.59918 & 1.89698 & 0.618971 & 736.065 \\
    & 2 & 3.48068 & 1.45072 & 2.00583 & 0.611076 & 736.11 \\
    & 3 & 7.65327 & 1.34227 & 2.11873 & 0.604087 & 736.174 \\
    & 4 & 16.7466 & 1.25354 & 2.24098 & 0.597444 & 736.2$\star$ \\
    & 5 & 36.533 & 1.17735 & 2.3757 & 0.590935 & 736.171 \\
    & 6 & 79.5338 & 1.11014 & 2.52544 & 0.584454 & 736.08 \\
    & 7 & 172.894 & 1.04982 & 2.69265 & 0.577936 & 735.924 \\
    & 8 & 375.437 & 0.995048 & 2.87986 & 0.571337 & 735.702 \\
    & 9 & 814.575 & 0.944869 & 3.08981 & 0.564624 & 735.414 \\
    & 10 & 1766.21 & 0.89859 & 3.32547 & 0.557771 & 735.063 \\ \hline
    \multirow{11}[11]{*}{\textbf{type II}} & $n$ & $\hat{\lambda}_{n}$ & $\hat{b}_{n}$ & $\hat{\mu}_{n}$ & $\hat{\sigma}_{n}$ & PLL \\ \hline
    & 0 & 0.766746 & 1.99897 & 1.78958 & 0.62922 & 641.506 \\
    & 1 & 1.5703 & 1.59918 & 1.89698 & 0.618971 & 736.065 \\
    & 2 & 3.21597 & 1.27934 & 2.03445 & 0.607849 & 736.076 \\
    & 3 & 6.5863 & 1.02347 & 2.21041 & 0.595756 & 736.101$\ast$ \\
    & 4 & 13.4887 & 0.81878 & 2.43565 & 0.582581 & 735.963 \\
    & 5 & 27.6249 & 0.655024 & 2.72395 & 0.568191 & 735.511 \\
    & 6 & 56.5759 & 0.524019 & 3.09297 & 0.552428 & 734.623 \\
    & 7 & 115.867 & 0.419215 & 3.56532 & 0.535103 & 733.21 \\
    & 8 & 237.297 & 0.335372 & 4.16992 & 0.515982 & 731.201 \\
    & 9 & 485.983 & 0.268298 & 4.94382 & 0.494772 & 728.473 \\
    & 10 & 995.294 & 0.214638 & 5.9344 & 0.471097 & 724.378 \\ \hline
    \multirow{11}[11]{*}{\textbf{type III}} & $n$ & $\hat{\lambda}_{n}$ & $\hat{b}_{n}$ & $\hat{\mu}_{n}$ & $\hat{\sigma}_{n}$ & PLL \\ \hline
    & 0 & 0.766746 & 1.99897 & 1.78958 & 0.62922 & 641.506 \\
   & 1 & 1.5703 & 1.59918 & 1.89698 & 0.618971 & 736.065 \\
   & 2 & 2.52532 & 0.777379 & 1.94743 & 0.615517 & 736.092 \\
   & 3 & 3.48997 & 0.256279 & 1.97342 & 0.613954 & 736.116 \\
   & 4 & 4.45737 & 0.0636928 & 1.9892 & 0.613067 & 736.131 \\
   & 5 & 5.42592 & 0.0126919 & 1.99977 & 0.612496 & 736.141 \\
   & 6 & 6.39508 & 0.00211003 & 2.00736 & 0.612097 & 736.148 \\
   & 7 & 7.36459 & 0.000300886 & 2.01306 & 0.611803 & 736.153 \\
   & 8 & 8.33432 & 0.0000375589 & 2.01751 & 0.611578 & 736.157 \\
   & 9 & 9.30419 & $4.1687\times10^{-6}$ & 2.02107 & 0.611399 & 736.16 \\
   & 10 & 10.2742 & $4.16507\times10^{-7}$ & 2.02398 & 0.611254 & 736.162$\ast$ \\ \hline
   \textbf{exponential} & - & 8.17863 & 7.9959 & 2.42886 & 0.576157 & 735.574 \\ \hline
  \end{tabular}\\
  PLL: profile log-likelihood
\end{table}

\vspace*{0.5in}

\begin{figure}[H]
  \centering
  \includegraphics[scale=0.39]{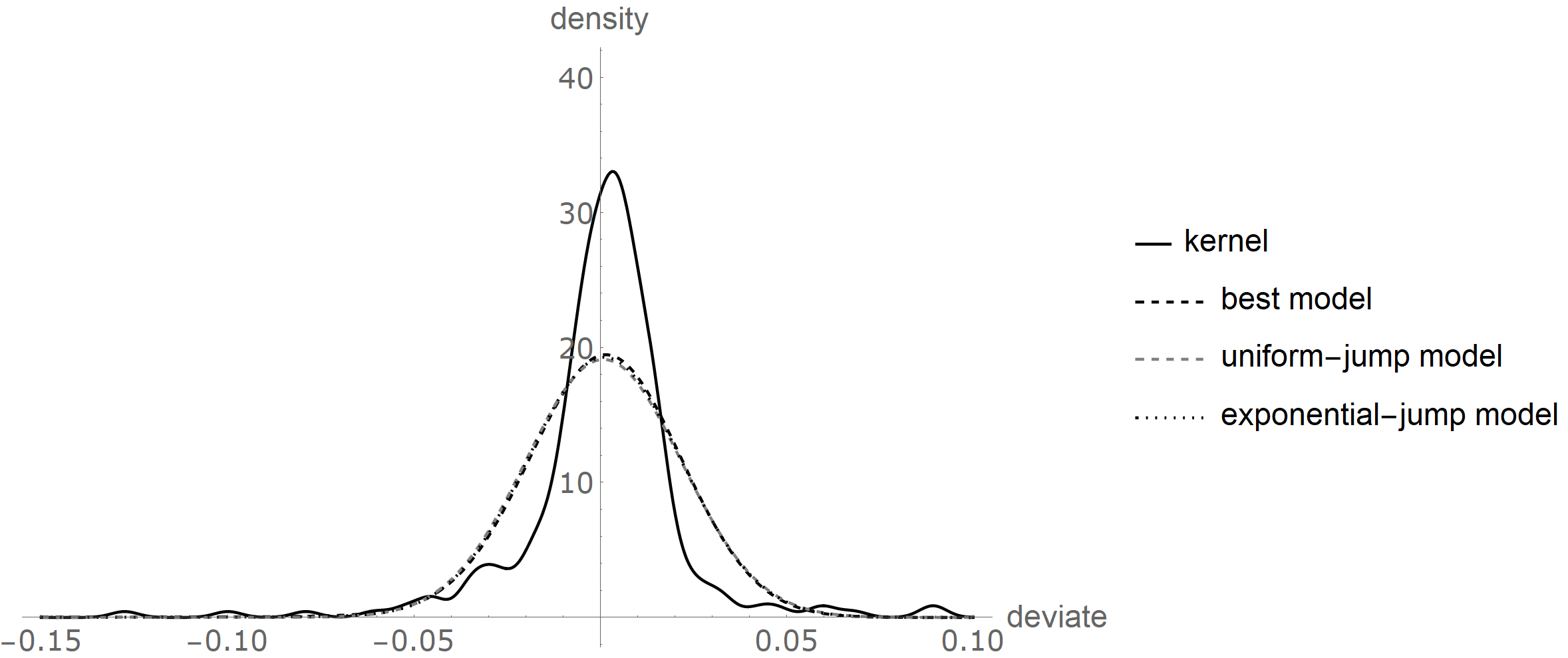}
  \includegraphics[scale=0.39]{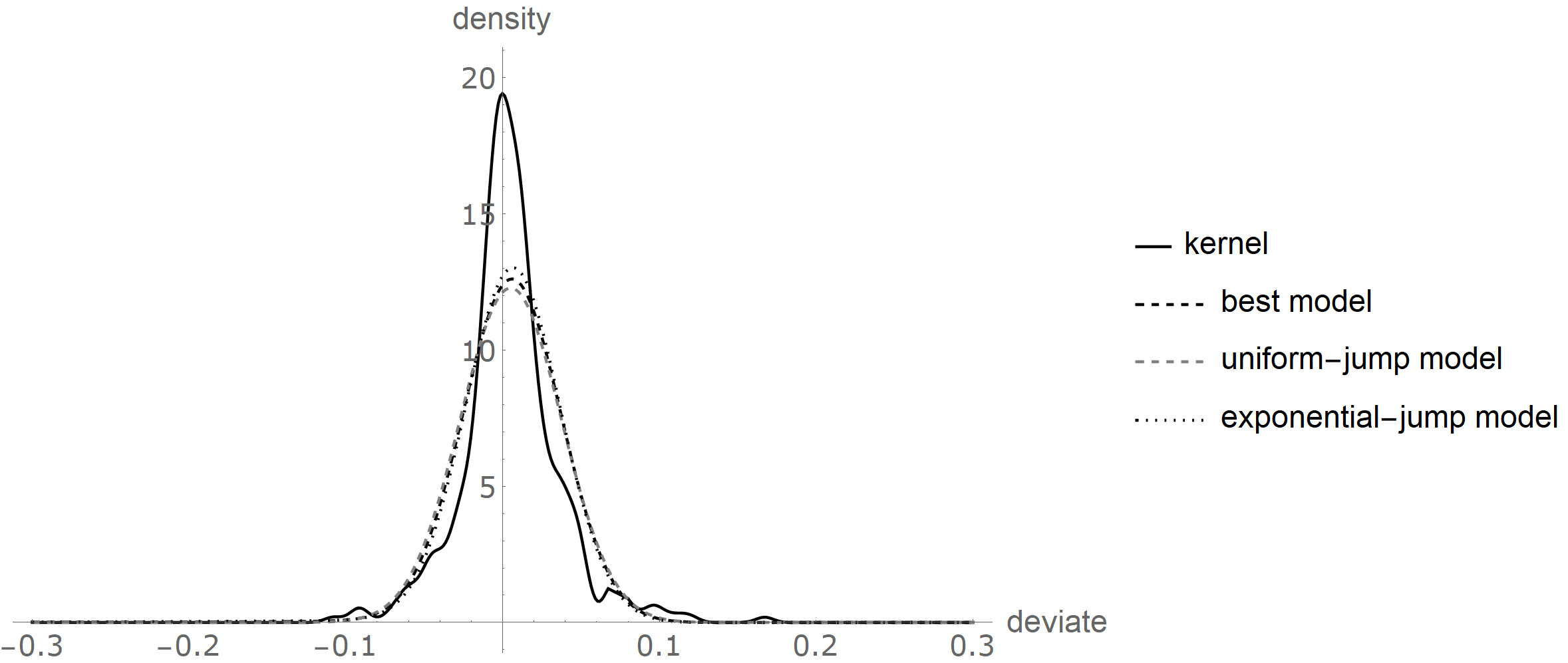}
  \caption{\label{fig:5} Jump diffusion estimated densities with regulation degrees for S\&P500 (top) and Bitcoin (bottom) daily returns}
\end{figure}

\vspace*{0.7in}
\begin{table}[H]\small
  \centering
  \ContinuedFloat*
  \caption{\label{tab:3a} Tempered stable estimation results on S\&P500 daily returns ($\sim$six significant digits)}
  \begin{tabular}{c|c|cccc|c}
    \hline
    \multicolumn{7}{c}{$c=0$ (VG)} \\ \hline
    \multirow{11}[11]{*}{\textbf{type I}} & $n$ & $\hat{a}_{n}$ & $\hat{b}_{n}$ & $\hat{\mu}_{n}$ & $\hat{\theta}_{n}$ & PLL \\ \hline
    & 0 & 94.8367 & 811.158 & 0.744442 & $-5.1502$ & 654.012 \\
    & 1 & 128.456 & 549.545 & 0.752422 & $-5.22027$ & 662.425 \\
    & 2 & 203.096 & 434.589 & 0.760763 & $-5.29356$ & 666.79 \\
    & 3 & 334.145 & 357.634 & 0.7691 & $-5.36685$ & 668.065$\ast$ \\
    & 4 & 561.015 & 300.338 & 0.777794 & $-5.44332$ & 667.986 \\
    & 5 & 954.266 & 255.534 & 0.787051 & $-5.52481$ & 667.252 \\
    & 6 & 1638.64 & 219.491 & 0.797029 & $-5.6127$ & 666.333 \\
    & 7 & 2835.01 & 189.959 & 0.807875 & $-5.70831$ & 665.136 \\
    & 8 & 4935.86 & 165.446 & 0.81974 & $-5.81298$ & 662.129 \\
    & 9 & 8641.45 & 144.906 & 0.832792 & $-5.92824$ & 662.009 \\
    & 10 & 15206.5 & 127.574 & 0.847223 & $-6.05579$ & 660.34 \\ \hline
    \multirow{11}[11]{*}{\textbf{type II}} & $n$ & $\hat{a}_{n}$ & $\hat{b}_{n}$ & $\hat{\mu}_{n}$ & $\hat{\theta}_{n}$ & PLL \\ \hline
    & 0 & 94.8367 & 811.158 & 0.744442 & $-5.1502$ & 654.012 \\
    & 1 & 128.456 & 549.545 & 0.752422 & $-5.22027$ & 662.425 \\
    & 2 & 174.38 & 373.157 & 0.761651 & $-5.30136$ & 666.218 \\
    & 3 & 237.339 & 254.059 & 0.772375 & $-5.39565$ & 667.676$\ast$ \\
    & 4 & 324.016 & 173.515 & 0.784904 & $-5.50592$ & 667.625 \\
    & 5 & 443.949 & 118.945 & 0.799644 & $-5.63577$ & 666.73 \\
    & 6 & 610.907 & 81.8998 & 0.817127 & $-5.78997$ & 665.336 \\
    & 7 & 845.057 & 56.6956 & 0.838074 & $-5.97499$ & 663.626 \\
    & 8 & 1176.48 & 39.5076 & 0.863484 & $-6.19982$ & 661.708 \\
    & 9 & 1651.07 & 27.7586 & 0.894794 & $-6.47743$ & 659.645 \\
    & 10 & 2340.97 & 19.7109 & 0.93414 & $-6.82722$ & 657.469 \\ \hline
    \multirow{11}[11]{*}{\textbf{type III}} & $n$ & $\hat{a}_{n}$ & $\hat{b}_{n}$ & $\hat{\mu}_{n}$ & $\hat{\theta}_{n}$ & PLL \\ \hline
    & 0 & 94.8367 & 811.158 & 0.744442 & $-5.1502$ & 654.012 \\
    & 1 & 128.456 & 549.545 & 0.752422 & $-5.22027$ & 662.425 \\
    & 2 & 174.86 & 249.402 & 0.756665 & $-5.25754$ & 665.585 \\
    & 3 & 223.022 & 79.5317 & 0.758921 & $-5.27736$ & 666.776 \\
    & 4 & 271.766 & 19.384 & 0.760309 & $-5.28957$ & 667.32 \\
    & 5 & 320.774 & 3.81341 & 0.761249 & $-5.29782$ & 667.615 \\
    & 6 & 369.923 & 0.628261 & 0.761926 & $-5.30377$ & 667.795 \\
    & 7 & 419.157 & 0.0889868 & 0.762437 & $-5.30826$ & 667.915 \\
    & 8 & 468.446 & 0.0110503 & 0.762836 & $-5.31177$ & 667.999 \\
    & 9 & 517.773 & 0.0012214 & 0.763157 & $-5.31459$ & 668.061 \\
    & 10 & 567.126 & 0.000121622 & 0.76342 & $-5.31691$ & 668.108$\star$ \\ \hline
  \end{tabular}\\
  VG: variance gamma | PLL: profile log-likelihood
\end{table}

\vspace*{0.7in}
\begin{table}[H]\small
  \centering
  \ContinuedFloat
  \caption{\label{tab:3b} Tempered stable estimation results on S\&P500 daily returns ($\sim$six significant digits)}
  \begin{tabular}{c|c|cccc|c}
    \hline
    \multicolumn{7}{c}{$c=0.25$} \\ \hline
    \multirow{11}[11]{*}{\textbf{type I}} & $n$ & $\hat{a}_{n}$ & $\hat{b}_{n}$ & $\hat{\mu}_{n}$ & $\hat{\theta}_{n}$ & PLL \\ \hline
    & 0 & 11.8708 & 621.521 & 0.755058 & $-5.24342$ & 668.251 \\
    & 1 & 17.7616 & 422.34 & 0.764691 & $-5.32808$ & 668.672$\star$ \\
    & 2 & 29.8475 & 335.05 & 0.774806 & $-5.41703$ & 668.277 \\
    & 3 & 51.6775 & 276.597 & 0.784966 & $-5.50645$ & 667.305 \\
    & 4 & 90.8543 & 233.057 & 0.795618 & $-5.60026$ & 666.07 \\
    & 5 & 161.326 & 198.997 & 0.807023 & $-5.70079$ & 664.739 \\
    & 6 & 288.564 & 171.591 & 0.819393 & $-5.80991$ & 661.032 \\
    & 7 & 519.198 & 149.132 & 0.832928 & $-5.92943$ & 652.583 \\
    & 8 & 938.875 & 130.495 & 0.847844 & $-6.06127$ & 660.391 \\
    & 9 & 1705.49 & 114.887 & 0.864387 & $-6.20765$ & 659.088 \\
    & 10 & 3111.25 & 101.731 & 0.882844 & $-6.37118$ & 657.695 \\ \hline
    \multirow{11}[11]{*}{\textbf{type II}} & $n$ & $\hat{a}_{n}$ & $\hat{b}_{n}$ & $\hat{\mu}_{n}$ & $\hat{\theta}_{n}$ & PLL \\ \hline
    & 0 & 11.8708 & 621.521 & 0.755058 & $-5.24342$ & 668.251 \\
    & 1 & 17.7616 & 422.34 & 0.764691 & $-5.32808$ & 668.672$\star$ \\
    & 2 & 26.6293 & 287.789 & 0.775895 & $-5.42662$ & 668.241 \\
    & 3 & 40.0192 & 196.745 & 0.789001 & $-5.54199$ & 667.224 \\
    & 4 & 60.311 & 135.026 & 0.804439 & $-5.67803$ & 665.812 \\
    & 5 & 91.1977 & 93.0987 & 0.82278 & $-5.83985$ & 664.09 \\
    & 6 & 138.465 & 64.553 & 0.844796 & $-6.03439$ & 662.253 \\
    & 7 & 211.286 & 45.0713 & 0.871567 & $-6.27138$ & 660.317 \\
    & 8 & 324.434 & 31.744 & 0.904647 & $-6.56486$ & 657.45 \\
    & 9 & 502.195 & 22.6087 & 0.946369 & $-6.93602$ & 655.531 \\
    & 10 & 785.628 & 16.3417 & 1.0004 & $-7.41838$ & 655.78 \\ \hline
    \multirow{11}[11]{*}{\textbf{type III}} & $n$ & $\hat{a}_{n}$ & $\hat{b}_{n}$ & $\hat{\mu}_{n}$ & $\hat{\theta}_{n}$ & PLL \\ \hline
    & 0 & 11.8708 & 621.521 & 0.755058 & $-5.24342$ & 668.251 \\
    & 1 & 17.7616 & 422.34 & 0.764691 & $-5.32808$ & 668.672$\star$ \\
    & 2 & 29.4917 & 191.979 & 0.769827 & $-5.37324$ & 668.582 \\
    & 3 & 50.086 & 61.2722 & 0.772562 & $-5.39729$ & 668.423 \\
    & 4 & 86.8967 & 14.9415 & 0.774248 & $-5.41212$ & 668.29 \\
    & 5 & 154.046 & 2.94048 & 0.775389 & $-5.42216$ & 668.186 \\
    & 6 & 278.893 & 0.484571 & 0.776211 & $-5.42939$ & 668.103 \\
    & 7 & 515.191 & 0.0686477 & 0.776833 & $-5.43486$ & 668.038 \\
    & 8 & 970.033 & 0.00852589 & 0.777318 & $-5.43913$ & 667.985 \\
    & 9 & 1859.66 & 0.000942492 & 0.777708 & $-5.44256$ & 667.941 \\
    & 10 & 3626.32 & 0.0000938585 & 0.778028 & $-5.44538$ & 667.904 \\ \hline
  \end{tabular}\\
  PLL: profile log-likelihood
\end{table}

\vspace*{0.7in}
\begin{table}[H]\small
  \centering
  \ContinuedFloat
  \caption{\label{tab:3c} Tempered stable estimation results on S\&P500 daily returns ($\sim$six significant digits)}
  \begin{tabular}{c|c|cccc|c}
    \hline
    \multicolumn{7}{c}{$c=0.5$ (NIG)} \\ \hline
    \multirow{11}[11]{*}{\textbf{type I}} & $n$ & $\hat{a}_{n}$ & $\hat{b}_{n}$ & $\hat{\mu}_{n}$ & $\hat{\theta}_{n}$ & PLL \\ \hline
    & 0 & 1.3704 & 432.849 & 0.777245 & $-5.43847$ & 667.041$\star$ \\
    & 0.5 & 1.72512 & 343.139 & 0.783233 & $-5.49118$ & 666.37 \\
    & 1 & 2.26522 & 296.003 & 0.790516 & $-5.5553$ & 665.447 \\
    & 1.5 & 3.01643 & 262.598 & 0.797581 & $-5.61754$ & 664.544 \\
    & 2 & 4.04629 & 236.4 & 0.804581 & $-5.67924$ & 663.674 \\
    & 2.5 & 5.45309 & 214.807 & 0.811644 & $-5.74153$ & 662.832 \\
    & 3 & 7.37333 & 196.483 & 0.818859 & $-5.80518$ & 662.016 \\
    & 3.5 & 9.99496 & 180.635 & 0.826291 & $-5.87079$ & 661.22 \\
    & 4 & 13.5762 & 166.743 & 0.833993 & $-5.93881$ & 660.442 \\
    & 4.5 & 18.472 & 154.447 & 0.842012 & $-6.00967$ & 659.68 \\
    & 5 & 25.17 & 143.48 & 0.850391 & -6.08376 & 658.933 \\ \hline
    \multirow{11}[11]{*}{\textbf{type II}} & $n$ & $\hat{a}_{n}$ & $\hat{b}_{n}$ & $\hat{\mu}_{n}$ & $\hat{\theta}_{n}$ & PLL \\ \hline
    & 0 & 1.3704 & 432.849 & 0.777245 & $-5.43847$ & 667.041$\star$ \\
    & 0.5 & 1.76152 & 357.786 & 0.783616 & $-5.49455$ & 666.261 \\
    & 1 & 2.26522 & 296.003 & 0.790516 & $-5.5553$ & 665.447 \\
    & 1.5 & 2.9143 & 245.126 & 0.797999 & $-5.62123$ & 664.604 \\
    & 2 & 3.75125 & 203.209 & 0.806132 & $-5.69292$ & 663.738 \\
    & 2.5 & 4.83121 & 168.655 & 0.814989 & $-5.77105$ & 662.853 \\
    & 3 & 6.22584 & 140.155 & 0.824658 & $-5.85639$ & 661.951 \\
    & 3.5 & 8.02838 & 116.634 & 0.835242 & $-5.94987$ & 661.036 \\
    & 4 & 10.3604 & 97.2105 & 0.846859 & $-6.05256$ & 660.108 \\
    & 4.5 & 13.3806 & 81.1608 & 0.859651 & $-6.16575$ & 659.17 \\
    & 5 & 17.2969 & 67.8906 & 0.87379 & $-6.29097$ & 658.222 \\ \hline
    \multirow{11}[11]{*}{\textbf{type III}} & $n$ & $\hat{a}_{n}$ & $\hat{b}_{n}$ & $\hat{\mu}_{n}$ & $\hat{\theta}_{n}$ & PLL \\ \hline
    & 0 & 1.3704 & 432.849 & 0.777245 & $-5.43847$ & 667.041$\star$ \\
    & 0.5 & 1.68675 & 371.306 & 0.784277 & $-5.50037$ & 666.26 \\
    & 1 & 2.26522 & 296.003 & 0.790516 & $-5.5553$ & 665.447 \\
    & 1.5 & 3.173 & 210.414 & 0.7947 & $-5.59216$ & 664.868 \\
    & 2 & 4.58808 & 135.007 & 0.797632 & $-5.61799$ & 664.453 \\
    & 2.5 & 6.81669 & 79.3114 & 0.799788 & $-5.63699$ & 664.144 \\
    & 3 & 10.3757 & 43.1666 & 0.801435 & $-5.65151$ & 663.907 \\
    & 3.5 & 16.1436 & 21.9721 & 0.802734 & $-5.66296$ & 663.719 \\
    & 4 & 25.6301 & 10.5381 & 0.803784 & $-5.67221$ & 663.567 \\
    & 4.5 & 41.459 & 4.7913 & 0.804649 & $-5.67984$ & 663.442 \\
    & 5 & 68.2414 & 2.07545 & 0.805375 & $-5.68624$ & 663.337 \\ \hline
  \end{tabular}\\
  NIG: normal inverse Gaussian | PLL: profile log-likelihood
\end{table}

\vspace*{0.7in}
\begin{table}[H]\small
  \centering
  \ContinuedFloat
  \caption{\label{tab:3d} Tempered stable estimation results on S\&P500 daily returns ($\sim$six significant digits)}
  \begin{tabular}{c|c|cccc|c}
    \hline
    \multicolumn{7}{c}{$c=0.75$} \\ \hline
    \multirow{11}[11]{*}{\textbf{type I}} & $n$ & $\hat{a}_{n}$ & $\hat{b}_{n}$ & $\hat{\mu}_{n}$ & $\hat{\theta}_{n}$ & PLL \\ \hline
    & 0 & 0.127463 & 248.604 & 0.852286 & $-6.10039$ & 654.027$\star$ \\
    & 0.1 & 0.134541 & 233.914 & 0.853433 & $-6.11053$ & 653.976 \\
    & 0.2 & 0.142421 & 222.685 & 0.855778 & $-6.13128$ & 653.863 \\
    & 0.3 & 0.151034 & 213.54 & 0.858568 & $-6.15597$ & 653.721 \\
    & 0.4 & 0.160364 & 205.785 & 0.861531 & $-6.18219$ & 653.566 \\
    & 0.5 & 0.170423 & 199.021 & 0.864556 & $-6.20896$ & 653.405 \\
    & 0.6 & 0.181235 & 193.002 & 0.867597 & $-6.23588$ & 653.242 \\
    & 0.7 & 0.192836 & 187.563 & 0.870632 & $-6.26276$ & 653.079 \\
    & 0.8 & 0.205268 & 182.593 & 0.873656 & $-6.28954$ & 652.917 \\
    & 0.9 & 0.218579 & 178.007 & 0.876666 & $-6.3162$ & 652.757 \\
    & 1 & 0.232824 & 173.745 & 0.879663 & $-6.34276$ & 652.599 \\ \hline
    \multirow{11}[11]{*}{\textbf{type II}} & $n$ & $\hat{a}_{n}$ & $\hat{b}_{n}$ & $\hat{\mu}_{n}$ & $\hat{\theta}_{n}$ & PLL \\ \hline
    & 0 & 0.127463 & 248.604 & 0.852286 & $-6.10039$ & 654.027$\star$ \\
    & 0.1 & 0.135368 & 239.776 & 0.854794 & $-6.12257$ & 653.885 \\
    & 0.2 & 0.143766 & 231.278 & 0.857349 & $-6.14518$ & 653.743 \\
    & 0.3 & 0.152687 & 223.097 & 0.859953 & $-6.16822$ & 653.601 \\
    & 0.4 & 0.162164 & 215.22 & 0.862607 & $-6.19171$ & 653.458 \\
    & 0.5 & 0.172233 & 207.637 & 0.865313 & $-6.21566$ & 653.316 \\
    & 0.6 & 0.182929 & 200.335 & 0.868072 & $-6.24008$ & 653.173 \\
    & 0.7 & 0.194293 & 193.306 & 0.870884 & $-6.26499$ & 653.03 \\
    & 0.8 & 0.206367 & 186.537 & 0.873753 & $-6.29039$ & 652.887 \\
    & 0.9 & 0.219194 & 180.02 & 0.876679 & $-6.31631$ & 652.743 \\
    & 1 & 0.232824 & 173.745 & 0.879663 & $-6.34276$ & 652.599 \\ \hline
    \multirow{11}[11]{*}{\textbf{type III}} & $n$ & $\hat{a}_{n}$ & $\hat{b}_{n}$ & $\hat{\mu}_{n}$ & $\hat{\theta}_{n}$ & PLL \\ \hline
    & 0 & 0.127463 & 248.604 & 0.852286 & $-6.10039$ & 654.027$\star$ \\
    & 0.1 & 0.132234 & 240.172 & 0.853793 & $-6.11372$ & 653.96 \\
    & 0.2 & 0.13829 & 233.905 & 0.856772 & $-6.14007$ & 653.818 \\
    & 0.3 & 0.14551 & 228.168 & 0.86014 & $-6.16988$ & 653.65 \\
    & 0.4 & 0.153881 & 222.222 & 0.863506 & $-6.19967$ & 653.475 \\
    & 0.5 & 0.163442 & 215.729 & 0.86672 & $-6.22812$ & 653.305 \\
    & 0.6 & 0.174271 & 208.561 & 0.869727 & $-6.25475$ & 653.143 \\
    & 0.7 & 0.186477 & 200.706 & 0.872516 & $-6.27944$ & 652.991 \\
    & 0.8 & 0.200194 & 192.219 & 0.875092 & $-6.30226$ & 652.85 \\
    & 0.9 & 0.21558 & 183.194 & 0.877469 & $-6.32332$ & 652.72 \\
    & 1 & 0.232824 & 173.745 & 0.879663 & $-6.34276$ & 652.599 \\ \hline
  \end{tabular}\\
  PLL: profile log-likelihood
\end{table}

\vspace*{0.7in}
\begin{table}[H]\small
  \centering
  \ContinuedFloat*
  \caption{\label{tab:4a} Tempered stable estimation results on Bitcoin daily returns ($\sim$six significant digits)}
  \begin{tabular}{c|c|cccc|c}
    \hline
    \multicolumn{7}{c}{$c=0$ (VG)} \\ \hline
    \multirow{11}[11]{*}{\textbf{type I}} & $n$ & $\hat{a}_{n}$ & $\hat{b}_{n}$ & $\hat{\mu}_{n}$ & $\hat{\theta}_{n}$ & PLL \\ \hline
    & 0 & 27.7898 & 56.0074 & 3.00165 & $-3.21585$ & 455.411 \\
    & 1 & 39.9891 & 40.7099 & 3.10861 & $-3.46658$ & 524.494 \\
    & 2 & 67.5402 & 34.7769 & 3.23098 & $-3.75877$ & 601.924 \\
    & 3 & 119.117 & 31.0558 & 3.36652 & $-4.08912$ & 655.587 \\
    & 4 & 215.901 & 28.5552 & 3.52426 & $-4.48257$ & 691.131 \\
    & 5 & 400.41 & 26.9375 & 3.71367 & $-4.96792$ & 715.837 \\
    & 6 & 759.406 & 26.0819 & 3.94698 & $-5.58529$ & 729.963 \\
    & 7 & 1474.97 & 25.9914 & 4.24121 & $-6.39498$ & 737.741 \\
    & 8 & 2941.94 & 26.7808 & 4.62072 & $-7.49154$ & 741.75 \\
    & 9 & 6047.17 & 28.7032 & 5.12027 & $-9.02652$ & 742.155$\star$ \\
    & 10 & 12850.6 & 32.2046 & 5.78719 & $-11.2431$ & 740.616 \\ \hline
    \multirow{11}[11]{*}{\textbf{type II}} & $n$ & $\hat{a}_{n}$ & $\hat{b}_{n}$ & $\hat{\mu}_{n}$ & $\hat{\theta}_{n}$ & PLL \\ \hline
    & 0 & 27.7898 & 56.0074 & 3.00165 & $-3.21585$ & 455.411 \\
    & 1 & 39.9891 & 40.7099 & 3.10861 & $-3.46658$ & 524.494 \\
    & 2 & 58.4843 & 30.162 & 3.24722 & $-3.79825$ & 591.811 \\
    & 3 & 87.4167 & 22.9374 & 3.43303 & $-4.25497$ & 645.474 \\
    & 4 & 134.707 & 18.1054 & 3.69371 & $-4.91968$ & 691.284 \\
    & 5 & 217.137 & 15.123 & 4.08334 & $-5.96699$ & 717.942 \\
    & 6 & 375.668 & 13.859 & 4.72003 & $-7.82461$ & 730.62 \\
    & 7 & 730.375 & 14.9641 & 5.89275 & $-11.7664$ & 730.654$\ast$ \\
    & 8 & 1689.11 & 21.4225 & 8.28461 & $-22.3332$ & 716.365 \\
    & 9 & 4447.82 & 42.7097 & 12.7363 & $-55.7046$ & 675.673 \\
    & 10 & 11431.1 & 114.716 & 19.1751 & $-182.6$ & 558.882 \\ \hline
    \multirow{11}[11]{*}{\textbf{type III}} & $n$ & $\hat{a}_{n}$ & $\hat{b}_{n}$ & $\hat{\mu}_{n}$ & $\hat{\theta}_{n}$ & PLL \\ \hline
    & 0 & 27.7898 & 56.0074 & 3.00165 & $-3.21585$ & 455.411 \\
    & 1 & 39.9891 & 40.7099 & 3.10861 & $-3.46658$ & 524.494 \\
    & 2 & 56.2539 & 19.1976 & 3.16875 & $-3.60939$ & 573.498 \\
    & 3 & 73.0303 & 6.24998 & 3.20178 & $-3.6884$ & 598.04 \\
    & 4 & 89.9752 & 1.543 & 3.22249 & $-3.73814$ & 613.948 \\
    & 5 & 106.996 & 0.306219 & 3.23666 & $-3.77226$ & 623.336 \\
    & 6 & 124.059 & 0.05077 & 3.24696 & $-3.79711$ & 630.002 \\
    & 7 & 141.146 & 0.00722557 & 3.25478 & $-3.816$ & 635.146 \\
    & 8 & 158.249 & 0.000900634 & 3.26092 & $-3.83085$ & 639.394 \\
    & 9 & 175.363 & 0.0000998491 & 3.26587 & $-3.84282$ & 642.235 \\
    & 10 & 192.485 & $9.9672\times10^{-6}$ & 3.26994 & $-3.85268$ & 644.439$\ast$ \\ \hline
  \end{tabular}\\
  VG: variance gamma | PLL: profile log-likelihood
\end{table}

\vspace*{0.7in}
\begin{table}[H]\small
  \centering
  \ContinuedFloat
  \caption{\label{tab:4b} Tempered stable estimation results on Bitcoin daily returns ($\sim$six significant digits)}
  \begin{tabular}{c|c|cccc|c}
    \hline
    \multicolumn{7}{c}{$c=0.25$} \\ \hline
    \multirow{11}[11]{*}{\textbf{type I}} & $n$ & $\hat{a}_{n}$ & $\hat{b}_{n}$ & $\hat{\mu}_{n}$ & $\hat{\theta}_{n}$ & PLL \\ \hline
    & 0 & 7.17612 & 47.0281 & 3.1418 & $-3.54481$ & 641.25 \\
    & 1 & 11.3467 & 35.0184 & 3.28923 & $-3.8994$ & 674.554 \\
    & 2 & 20.2654 & 30.7629 & 3.46274 & $-4.32733$ & 703.597 \\
    & 3 & 37.443 & 28.3513 & 3.66136 & $-4.83148$ & 722.383 \\
    & 4 & 70.7801 & 27.0578 & 3.90086 & $-5.45996$ & 734.063 \\
    & 5 & 136.543 & 26.7025 & 4.19975 & $-6.27619$ & 740.223 \\
    & 6 & 268.857 & 27.3237 & 4.5831 & $-7.3758$ & 743.733 \\
    & 7 & 541.024 & 29.136 & 5.08604 & $-8.91013$ & 744.02$\star$ \\
    & 8 & 1113.88 & 32.5598 & 5.756 & $-11.1203$ & 741.869 \\
    & 9 & 2344.96 & 38.2747 & 6.65067 & $-14.3797$ & 737.46 \\
    & 10 & 5028.11 & 47.2474 & 7.82714 & $-19.2312$ & 730.984 \\ \hline
    \multirow{11}[11]{*}{\textbf{type II}} & $n$ & $\hat{a}_{n}$ & $\hat{b}_{n}$ & $\hat{\mu}_{n}$ & $\hat{\theta}_{n}$ & PLL \\ \hline
    & 0 & 7.17612 & 47.0281 & 3.1418 & $-3.54481$ & 641.25 \\
    & 1 & 11.3467 & 35.0184 & 3.28923 & $-3.8994$ & 674.554 \\
    & 2 & 18.2308 & 26.8011 & 3.48754 & $-4.39002$ & 701.452 \\
    & 3 & 29.9485 & 21.3312 & 3.76679 & $-5.10797$ & 721.23 \\
    & 4 & 50.8037 & 18.0118 & 4.18535 & $-6.24526$ & 733.655 \\
    & 5 & 90.4781 & 16.731 & 4.86847 & $-8.26702$ & 737.997$\ast$ \\
    & 6 & 173.503 & 18.2688 & 6.10689 & $-12.5041$ & 734.271 \\
    & 7 & 364.579 & 25.7088 & 8.51289 & $-23.2474$ & 718.548 \\
    & 8 & 795.808 & 47.4225 & 12.7543 & $-53.8354$ & 682.923 \\
    & 9 & 1616.39 & 110.052 & 18.8531 & $-153.237$ & 606.713 \\
    & 10 & 2565.25 & 544.578 & 26.6876 & $-928.402$ & 320.315 \\ \hline
    \multirow{11}[11]{*}{\textbf{type III}} & $n$ & $\hat{a}_{n}$ & $\hat{b}_{n}$ & $\hat{\mu}_{n}$ & $\hat{\theta}_{n}$ & PLL \\ \hline
    & 0 & 7.17612 & 47.0281 & 3.1418 & $-3.54481$ & 641.25 \\
    & 1 & 11.3467 & 35.0184 & 3.28923 & $-3.8994$ & 674.554 \\
    & 2 & 19.4167 & 16.7381 & 3.37351 & $-4.10569$ & 692.703 \\
    & 3 & 33.5171 & 5.48975 & 3.42025 & $-4.2212$ & 701.522 \\
    & 4 & 58.7421 & 1.36161 & 3.44971 & $-4.29443$ & 706.549 \\
    & 5 & 104.855 & 0.271079 & 3.46994 & $-4.34489$ & 709.757 \\
    & 6 & 190.782 & 0.0450476 & 3.48468 & $-4.38175$ & 711.971 \\
    & 7 & 353.758 & 0.00642237 & 3.49589 & $-4.40984$ & 713.587 \\
    & 8 & 668.049 & 0.000801617 & 3.5047 & $-4.43195$ & 714.816 \\
    & 9 & 1283.77 & 0.0000889697 & 3.51181 & $-4.4498$ & 715.782 \\
    & 10 & 2508.23 & $8.88926\times10^{-6}$ & 3.51766 & $-4.46452$ & 716.56$\ast$ \\ \hline
  \end{tabular}\\
  PLL: profile log-likelihood
\end{table}

\vspace*{0.7in}
\begin{table}[H]\small
  \centering
  \ContinuedFloat
  \caption{\label{tab:4c} Tempered stable estimation results on Bitcoin daily returns ($\sim$six significant digits)}
  \begin{tabular}{c|c|cccc|c}
    \hline
    \multicolumn{7}{c}{$c=0.5$ (NIG)} \\ \hline
    \multirow{11}[11]{*}{\textbf{type I}} & $n$ & $\hat{a}_{n}$ & $\hat{b}_{n}$ & $\hat{\mu}_{n}$ & $\hat{\theta}_{n}$ & PLL \\ \hline
    & 0 & 1.69903 & 40.2719 & 3.4866 & $-4.3844$ & 741.57 \\
    & 0.5 & 2.18658 & 34.0606 & 3.60284 & $-4.6784$ & 744.017 \\
    & 1 & 2.95041 & 31.8517 & 3.75372 & $-5.06738$ & 746.459 \\
    & 1.5 & 4.037 & 30.6647 & 3.91344 & $-5.48857$ & 748.356 \\
    & 2 & 5.56725 & 30.0464 & 4.08646 & $-5.95593$ & 749.25 \\
    & 2.5 & 7.72111 & 29.8558 & 4.27762 & $-6.48574$ & 749.903 \\
    & 3 & 10.7581 & 30.0452 & 4.49173 & $-7.09616$ & 747.875 \\
    & 3.5 & 15.0512 & 30.61 & 4.73399 & $-7.80857$ & 750.611$\star$ \\
    & 4 & 21.1362 & 31.5722 & 5.01006 & $-8.64893$ & 749.197 \\
    & 4.5 & 29.7841 & 32.9755 & 5.32626 & $-9.64906$ & 747.685 \\
    & 5 & 42.1034 & 34.8824 & 5.68939 & $-10.8479$ & 745.767 \\ \hline
    \multirow{11}[11]{*}{\textbf{type II}} & $n$ & $\hat{a}_{n}$ & $\hat{b}_{n}$ & $\hat{\mu}_{n}$ & $\hat{\theta}_{n}$ & PLL \\ \hline
    & 0 & 1.69903 & 40.2719 & 3.4866 & $-4.3844$ & 741.57 \\
    & 0.5 & 2.23523 & 35.6267 & 3.60867 & $-4.69302$ & 744.468 \\
    & 1 & 2.95041 & 31.8517 & 3.75372 & $-5.06738$ & 746.459 \\
    & 1.5 & 3.90955 & 28.8422 & 3.92845 & $-5.5294$ & 748.107 \\
    & 2 & 5.20409 & 26.5277 & 4.14219 & $-6.11131$ & 748.78$\ast$ \\
    & 2.5 & 6.96431 & 24.8739 & 4.40824 & $-6.86179$ & 748.142 \\
    & 3 & 9.37804 & 23.891 & 4.74579 & $-7.85667$ & 746.481 \\
    & 3.5 & 12.7184 & 23.648 & 5.18271 & $-9.21738$ & 747.08 \\
    & 4 & 17.3826 & 24.2981 & 5.75891 & $-11.1428$ & 743.584 \\
    & 4.5 & 23.9386 & 26.1153 & 6.52903 & $-13.9616$ & 739.578 \\
    & 5 & 33.1604 & 29.5378 & 7.56059 & $-18.2114$ & 731.809 \\ \hline
    \multirow{11}[11]{*}{\textbf{type III}} & $n$ & $\hat{a}_{n}$ & $\hat{b}_{n}$ & $\hat{\mu}_{n}$ & $\hat{\theta}_{n}$ & PLL \\ \hline
    & 0 & 1.69903 & 40.2719 & 3.4866 & $-4.3844$ & 741.57 \\
    & 0.5 & 2.14646 & 37.2922 & 3.62438 & $-4.73348$ & 743.909 \\
    & 1 & 2.95041 & 31.8517 & 3.75372 & $-5.06738$ & 746.459 \\
    & 1.5 & 4.19843 & 23.7357 & 3.84507 & $-5.30685$ & 748.233 \\
    & 2 & 6.13874 & 15.7484 & 3.9114 & $-5.48266$ & 748.608 \\
    & 2.5 & 9.19582 & 9.4846 & 3.96141 & $-5.61627$ & 749.061 \\
    & 3 & 14.0854 & 5.26192 & 4.00036 & $-5.72095$ & 749.327 \\
    & 3.5 & 22.0249 & 2.71928 & 4.03151 & $-5.80506$ & 749.486 \\
    & 4 & 35.1087 & 1.32035 & 4.05697 & $-5.87407$ & 749.581 \\
    & 4.5 & 56.9807 & 0.606458 & 4.07815 & $-5.93167$ & 749.634 \\
    & 5 & 94.0523 & 0.264957 & 4.09606 & $-5.98048$ & 749.657$\ast$ \\ \hline
  \end{tabular}\\
  NIG: normal inverse Gaussian | PLL: profile log-likelihood
\end{table}

\vspace*{0.7in}
\begin{table}[H]\small
  \centering
  \ContinuedFloat
  \caption{\label{tab:4d} Tempered stable estimation results on Bitcoin daily returns ($\sim$six significant digits)}
  \begin{tabular}{c|c|cccc|c}
    \hline
    \multicolumn{7}{c}{$c=0.75$} \\ \hline
    \multirow{11}[11]{*}{\textbf{type I}} & $n$ & $\hat{a}_{n}$ & $\hat{b}_{n}$ & $\hat{\mu}_{n}$ & $\hat{\theta}_{n}$ & PLL \\ \hline
    & 0 & 0.304774 & 50.5501 & 5.21772 & $-9.19796$ & 752.733$\star$ \\
    & 0.1 & 0.321949 & 48.3428 & 5.26152 & $-9.33471$ & 752.507 \\
    & 0.2 & 0.341336 & 47.5163 & 5.34882 & $-9.6086$ & 752.06 \\
    & 0.3 & 0.362624 & 47.2887 & 5.4522 & $-9.9359$ & 751.546 \\
    & 0.4 & 0.385726 & 47.3793 & 5.56273 & $-10.2896$ & 750.975 \\
    & 0.5 & 0.410649 & 47.664 & 5.67699 & $-10.6595$ & 750.398 \\
    & 0.6 & 0.437445 & 48.0803 & 5.79359 & $-11.0415$ & 749.811 \\
    & 0.7 & 0.466199 & 48.594 & 5.91194 & $-11.434$ & 749.224 \\
    & 0.8 & 0.497011 & 49.1848 & 6.03186 & $-11.8366$ & 748.614 \\
    & 0.9& 0.530001 & 49.8404 & 6.1533 & $-12.2493$ & 748.005 \\
    & 1 & 0.565299 & 50.5528 & 6.27632 & $-12.6725$ & 747.39 \\ \hline
    \multirow{11}[11]{*}{\textbf{type II}} & $n$ & $\hat{a}_{n}$ & $\hat{b}_{n}$ & $\hat{\mu}_{n}$ & $\hat{\theta}_{n}$ & PLL \\ \hline
    & 0 & 0.304774 & 50.5501 & 5.21772 & $-9.19796$ & 752.733$\star$ \\
    & 0.1 & 0.324226 & 50.3274 & 5.30404 & $-9.46608$ & 752.302 \\
    & 0.2 & 0.344917 & 50.1521 & 5.39414 & $-9.74882$ & 751.851 \\
    & 0.3 & 0.366924 & 50.0249 & 5.48823 & $-10.0472$ & 751.379 \\
    & 0.4 & 0.39033 & 49.9464 & 5.5865 & $-10.3622$ & 750.885 \\
    & 0.5 & 0.41522 & 49.9176 & 5.68916 & $-10.695$ & 750.367 \\
    & 0.6 & 0.441685 & 49.9393 & 5.79643 & $-11.0469$ & 749.825 \\
    & 0.7 & 0.469821 & 50.0124 & 5.90855 & $-11.4192$ & 749.257 \\
    & 0.8 & 0.499731 & 50.1382 & 6.02574 & $-11.8131$ & 748.663 \\
    & 0.9 & 0.531519 & 50.3179 & 6.14824 & $-12.2304$ & 748.041 \\
    & 1 & 0.565299 & 50.5528 & 6.27632 & $-12.6725$ & 747.39 \\ \hline
    \multirow{11}[11]{*}{\textbf{type III}} & $n$ & $\hat{a}_{n}$ & $\hat{b}_{n}$ & $\hat{\mu}_{n}$ & $\hat{\theta}_{n}$ & PLL \\ \hline
    & 0 & 0.304774 & 50.5501 & 5.21772 & $-9.19796$ & 752.733$\star$ \\
    & 0.1 & 0.316506 & 49.8956 & 5.27563 & $-9.37895$ & 752.433 \\
    & 0.2 & 0.331647 & 50.6222 & 5.38789 & $-9.73239$ & 751.857 \\
    & 0.3 & 0.349699 & 51.6632 & 5.51481 & $-10.1368$ & 751.21 \\
    & 0.4 & 0.370556 & 52.6039 & 5.64265 & $-10.5493$ & 750.562 \\
    & 0.5 & 0.394296 & 53.2529 & 5.76595 & $-10.9521$ & 749.939 \\
    & 0.6 & 0.421103 & 53.5212 & 5.88255 & $-11.3377$ & 749.352 \\
    & 0.7 & 0.451243 & 53.3751 & 5.99177 & $-11.7029$ & 748.805 \\
    & 0.8 & 0.485048 & 52.815 & 6.09357 & $-12.0468$ & 748.297 \\
    & 0.9 & 0.522912 & 51.8624 & 6.18827 & $-12.3697$ & 747.826 \\
    & 1 & 0.565299 & 50.5528 & 6.27632 & $-12.6725$ & 747.39 \\ \hline
  \end{tabular}\\
  PLL: profile log-likelihood
\end{table}

\begin{figure}[H]
  \centering
  \ContinuedFloat*
  \includegraphics[scale=0.3]{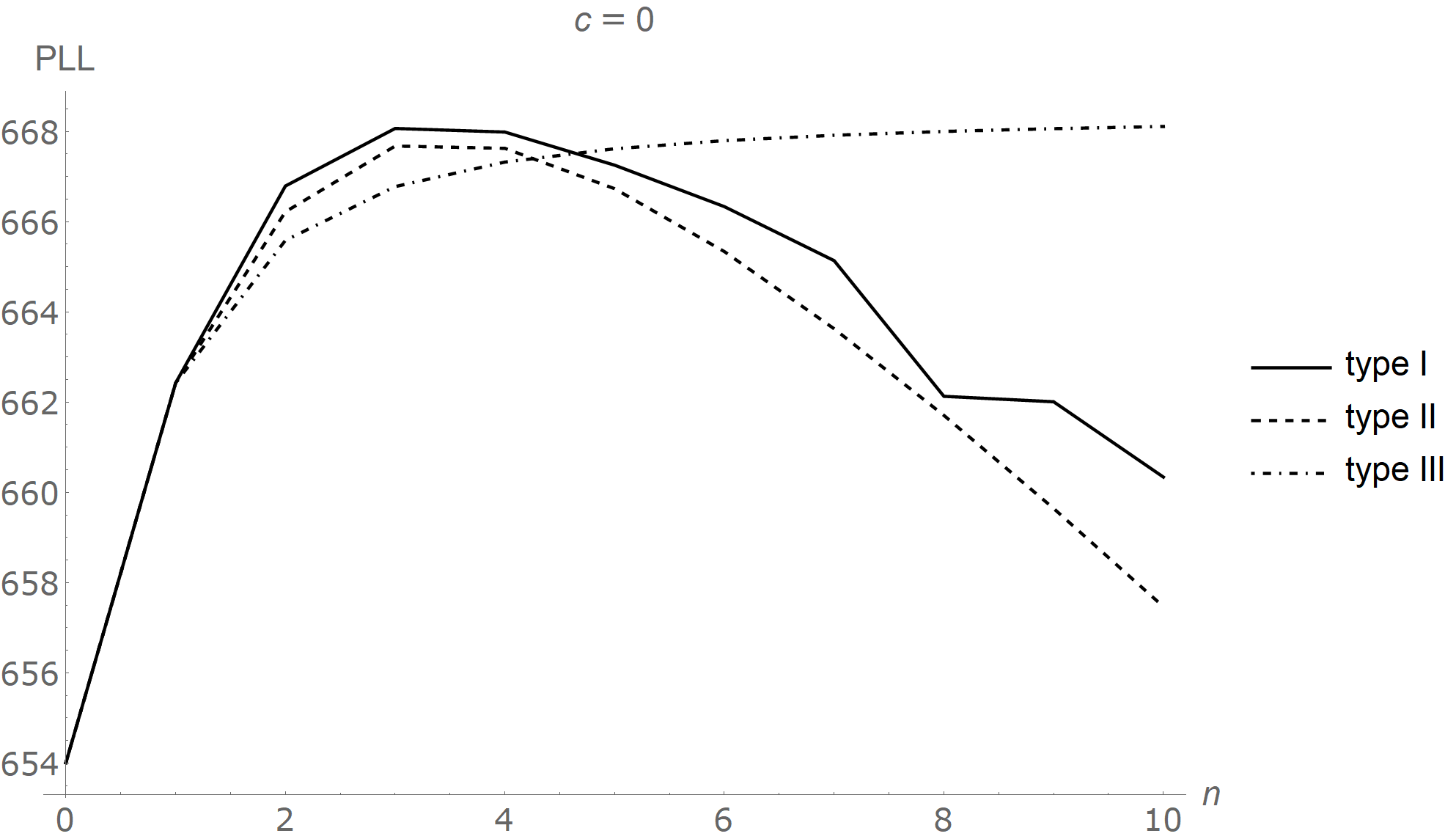}
  \includegraphics[scale=0.3]{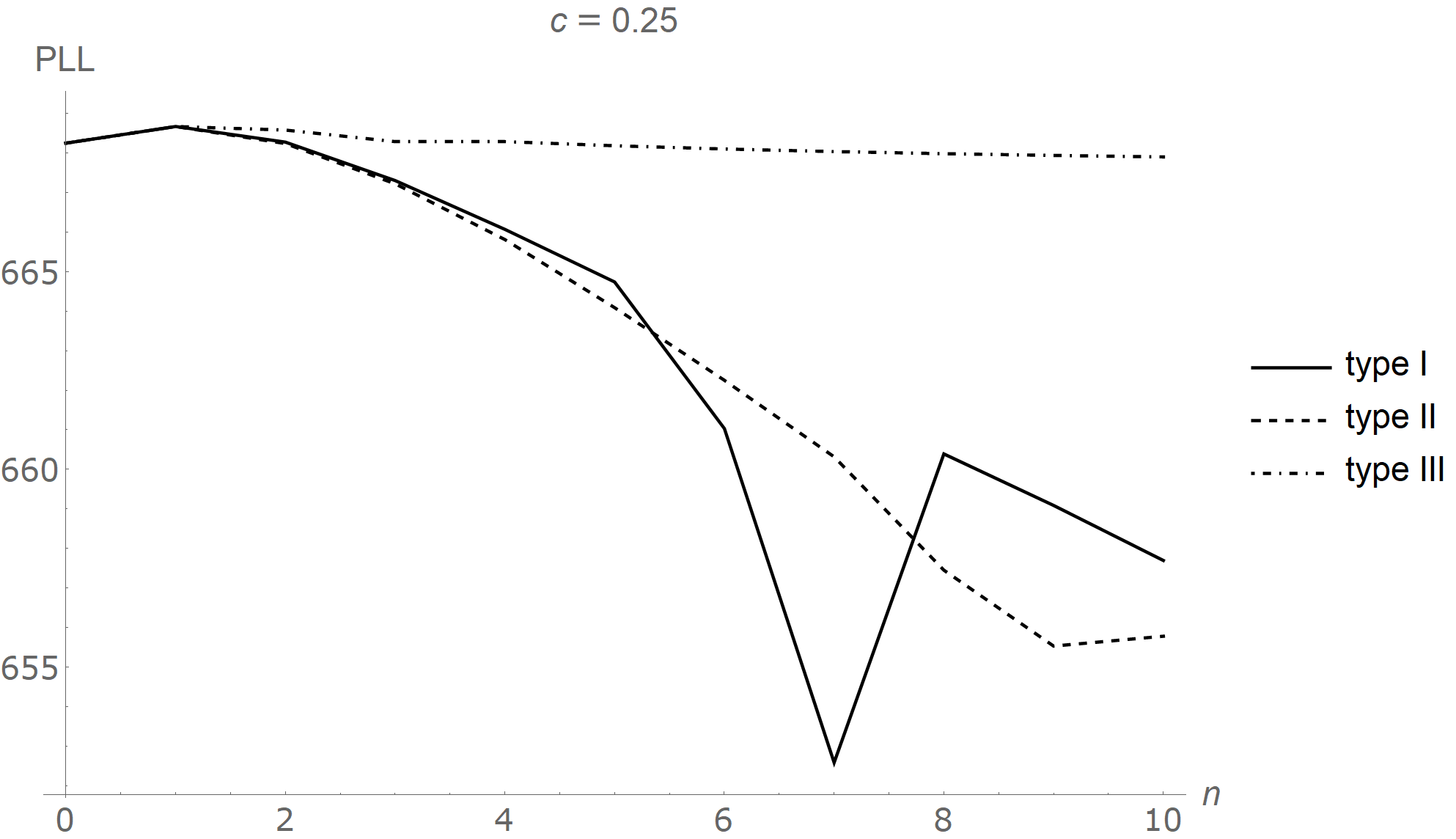}
  \includegraphics[scale=0.3]{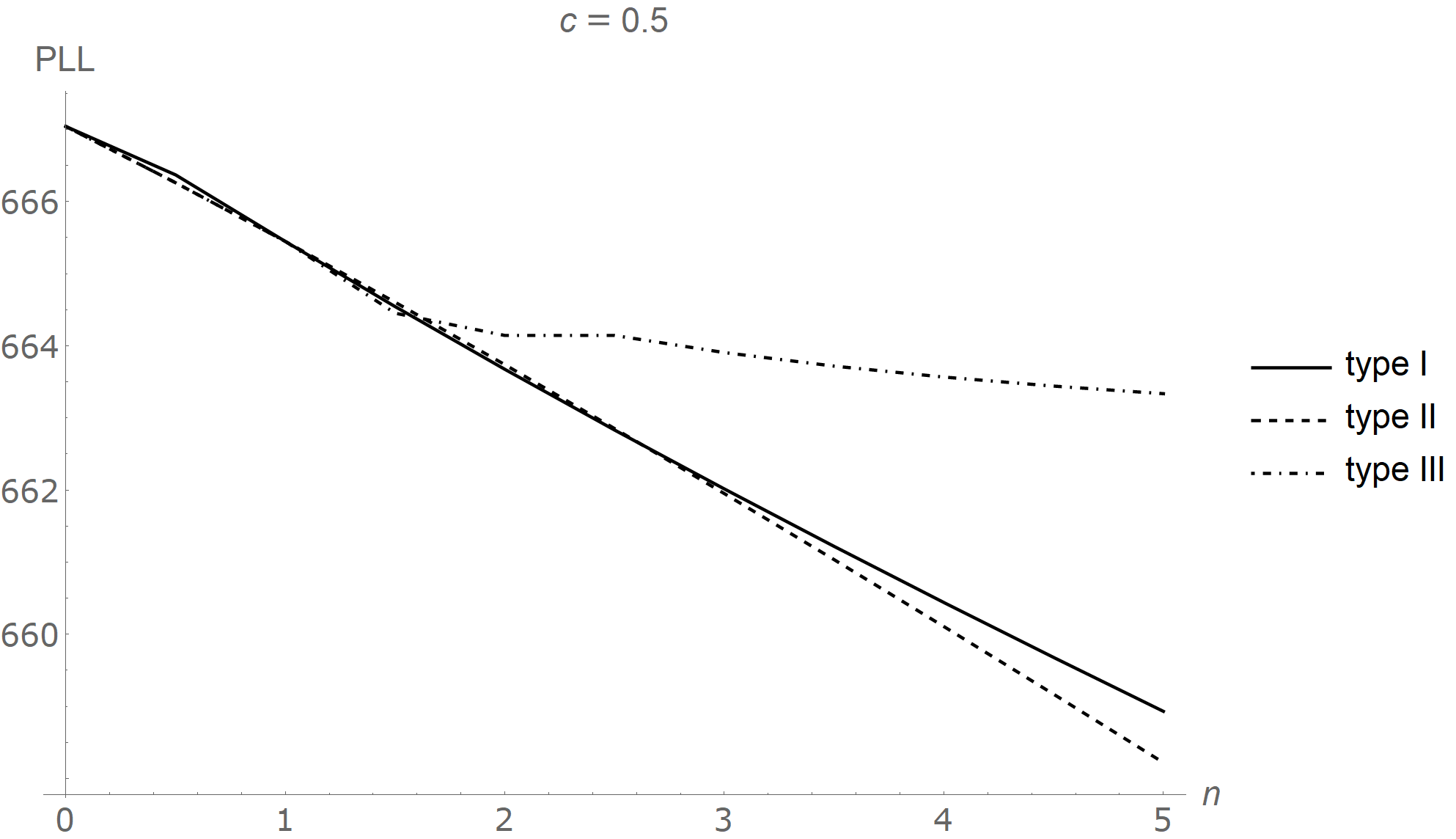}
  \includegraphics[scale=0.3]{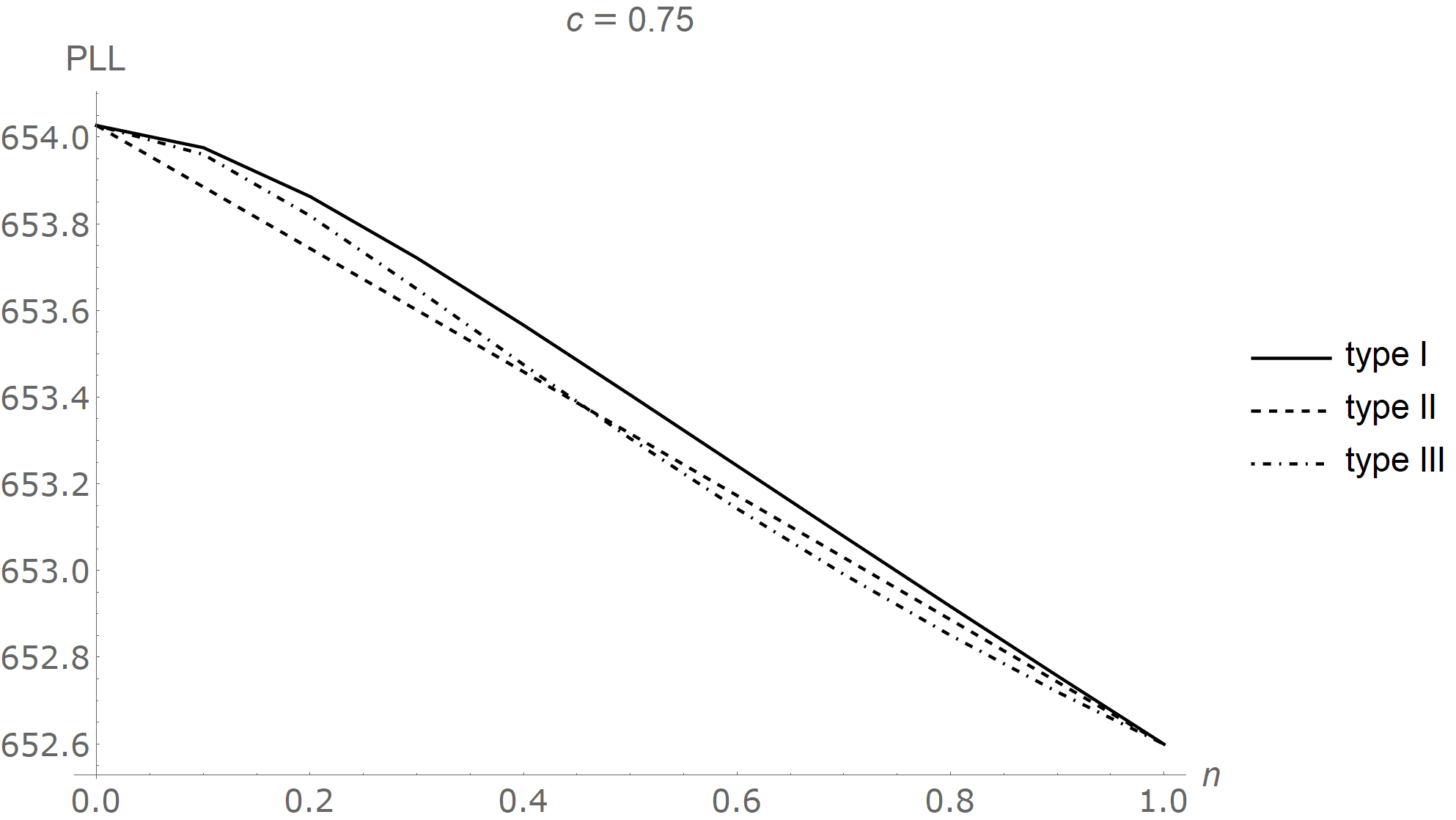}
  \caption{\label{fig:6a} Tempered stable log-likelihood with regulation degrees for S\&P500 daily returns}
\end{figure}

\begin{figure}[H]
  \centering
  \ContinuedFloat
  \includegraphics[scale=0.3]{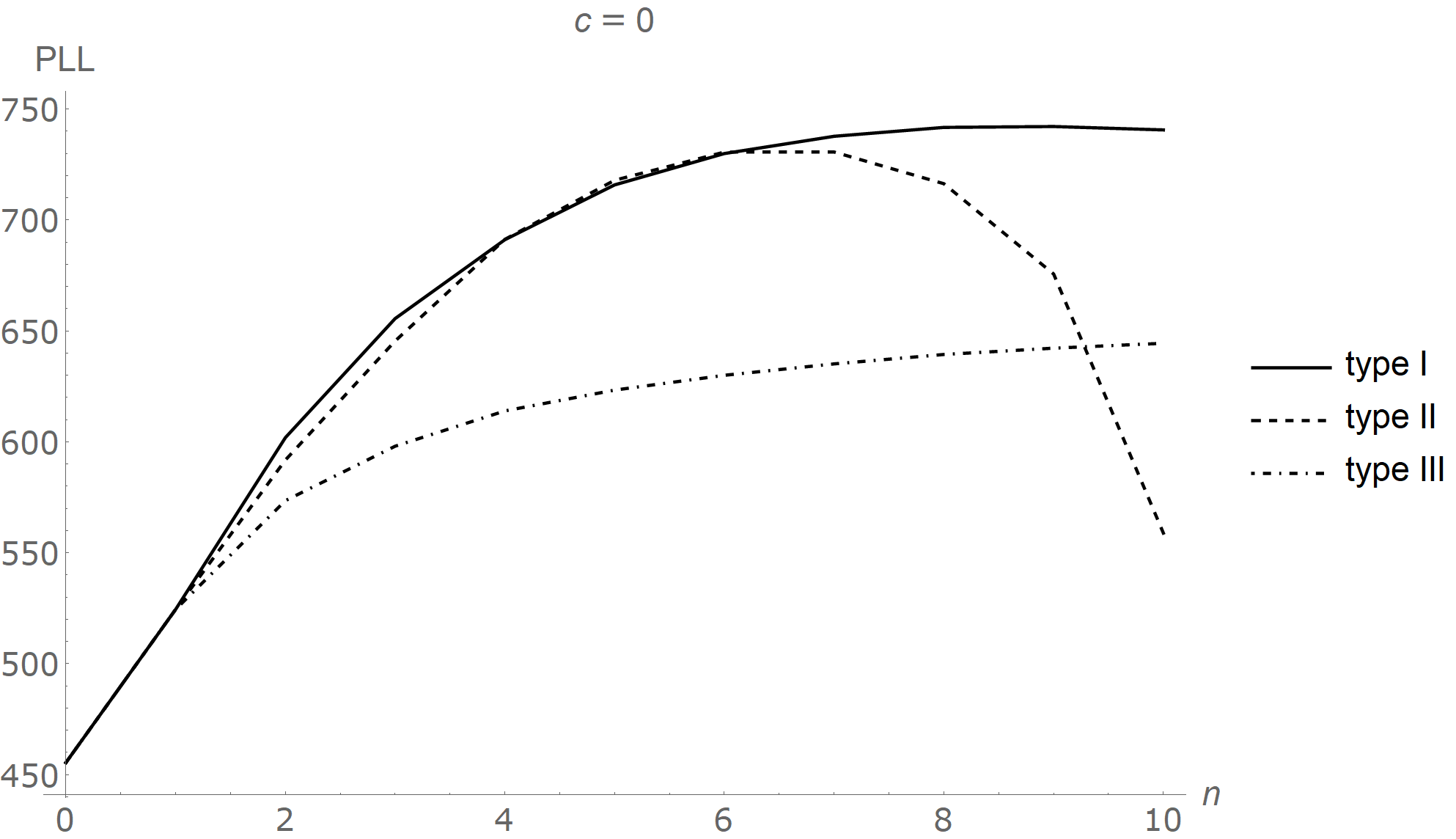}
  \includegraphics[scale=0.3]{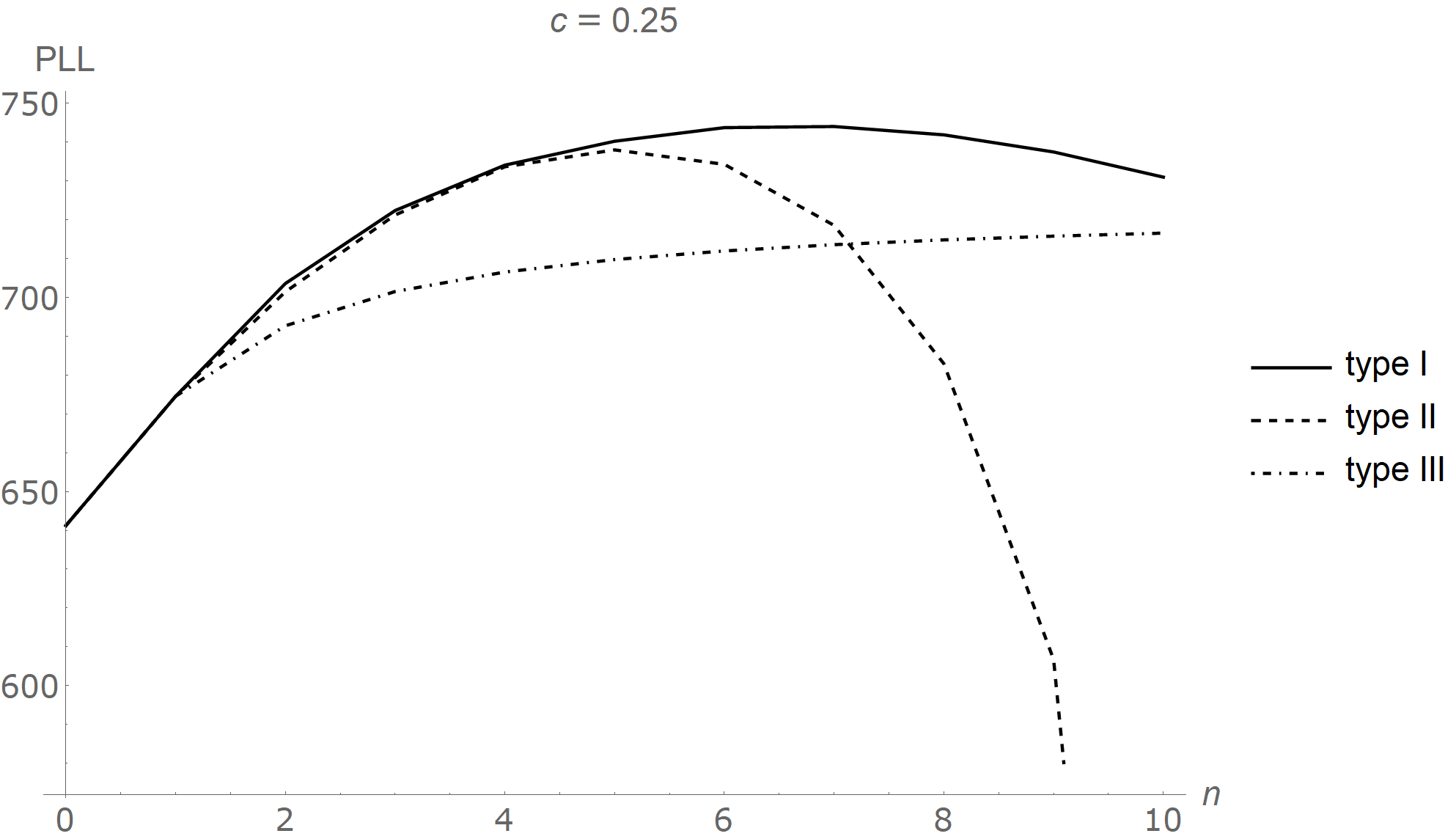}
  \includegraphics[scale=0.3]{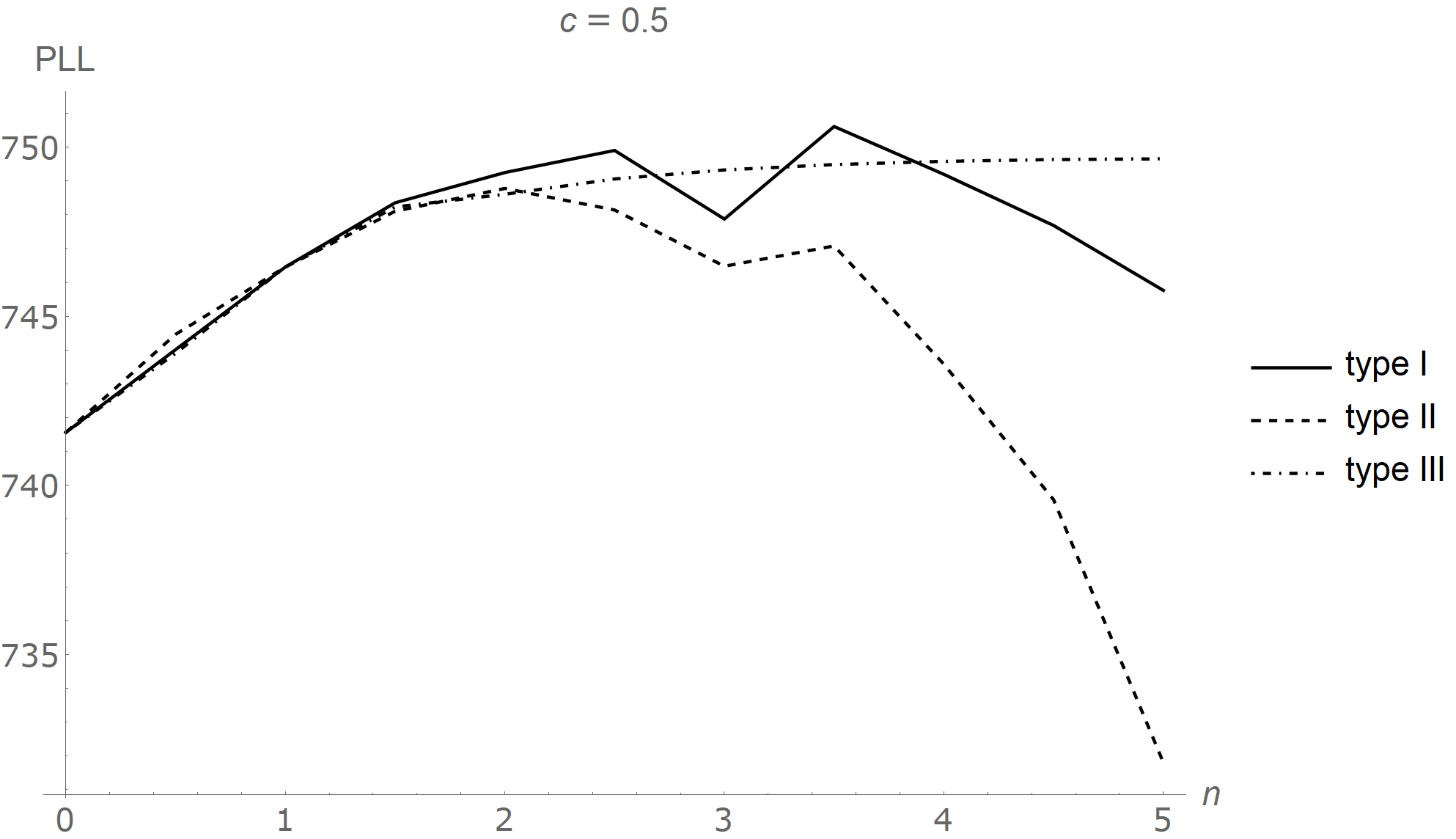}
  \includegraphics[scale=0.3]{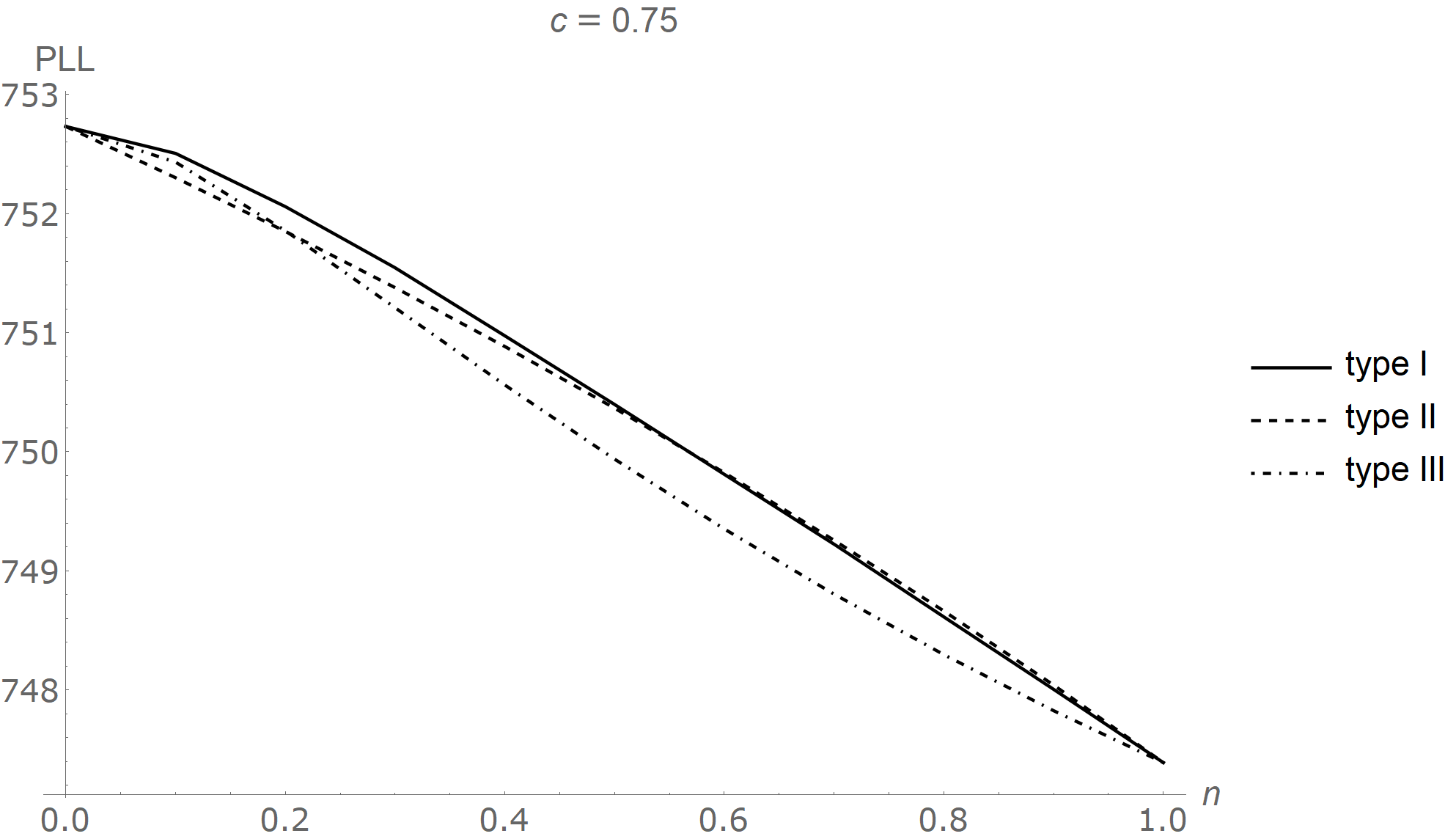}
  \caption{\label{fig:6b} Tempered stable log-likelihood with regulation degrees for Bitcoin daily returns}
\end{figure}

\begin{figure}[H]
  \centering
  \ContinuedFloat*
  \includegraphics[scale=0.39]{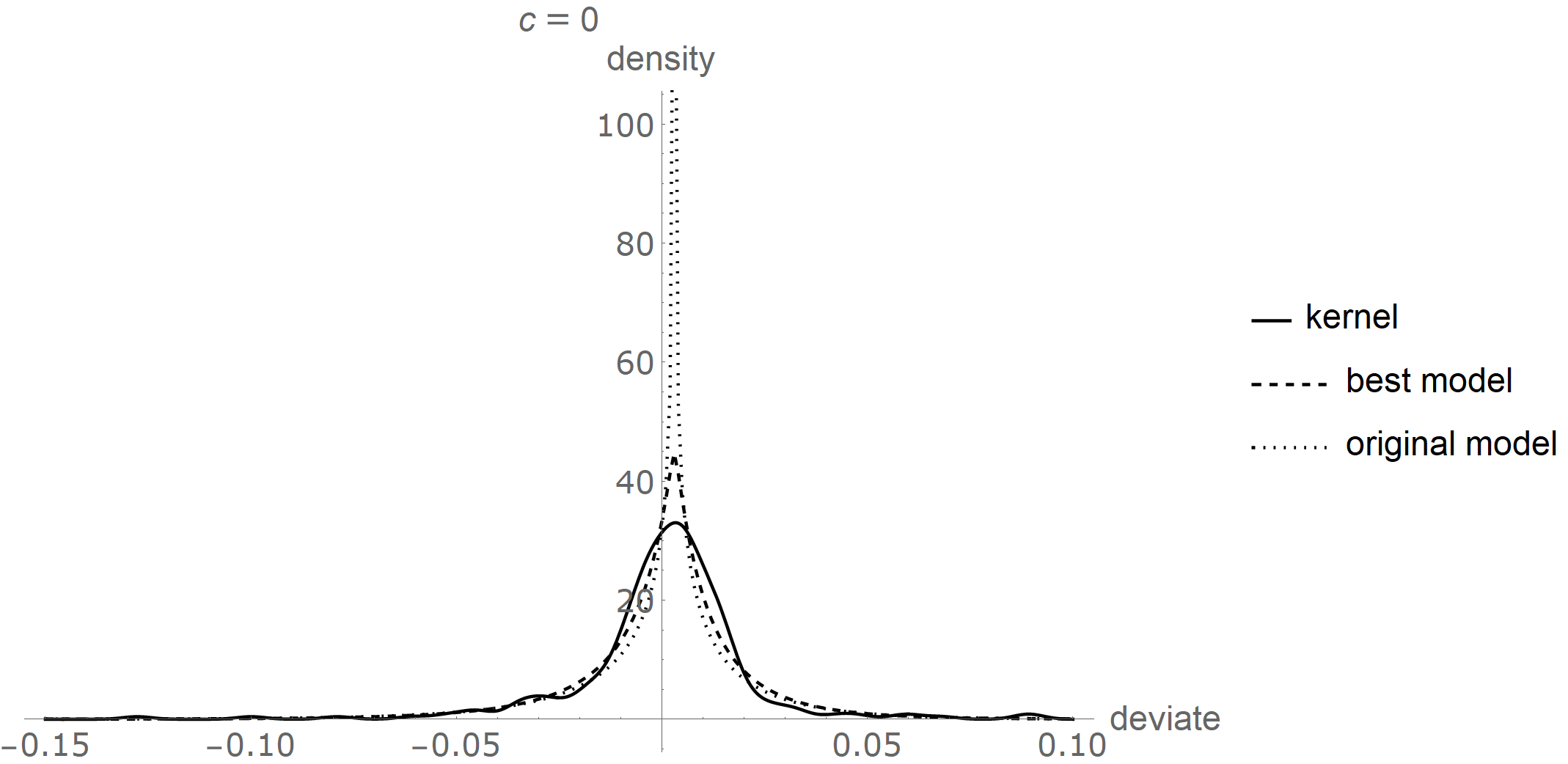}
  \includegraphics[scale=0.39]{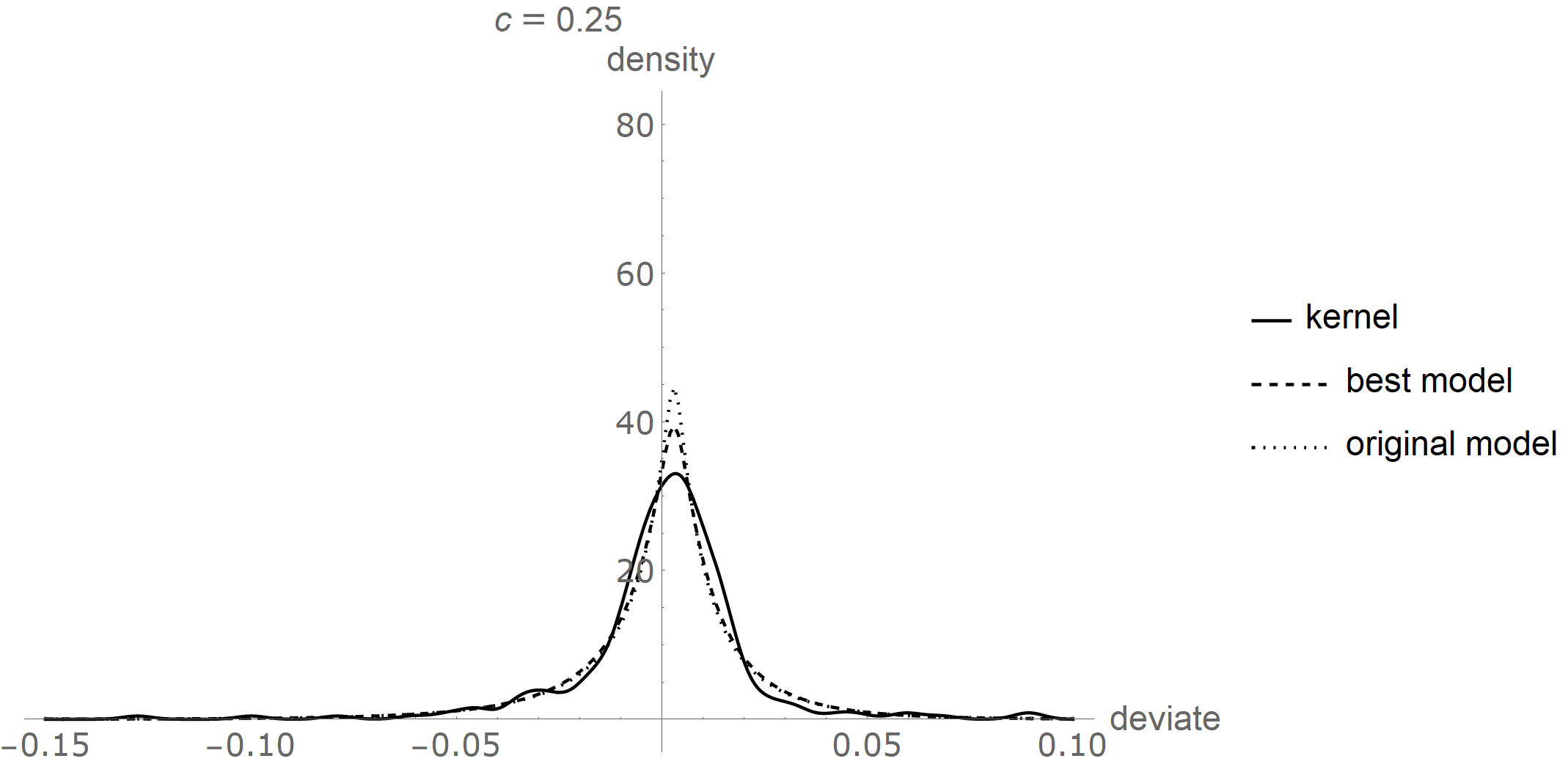}
  \includegraphics[scale=0.39]{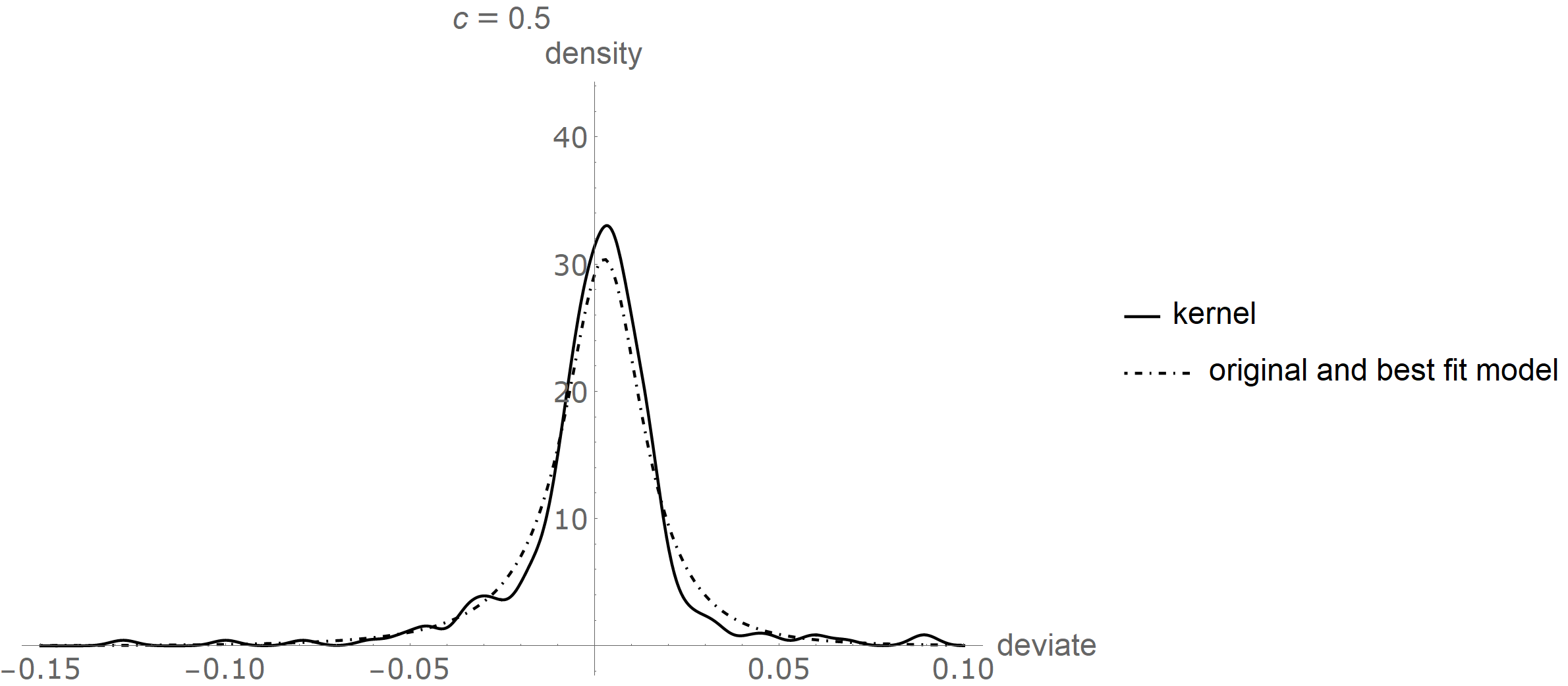}
  \includegraphics[scale=0.39]{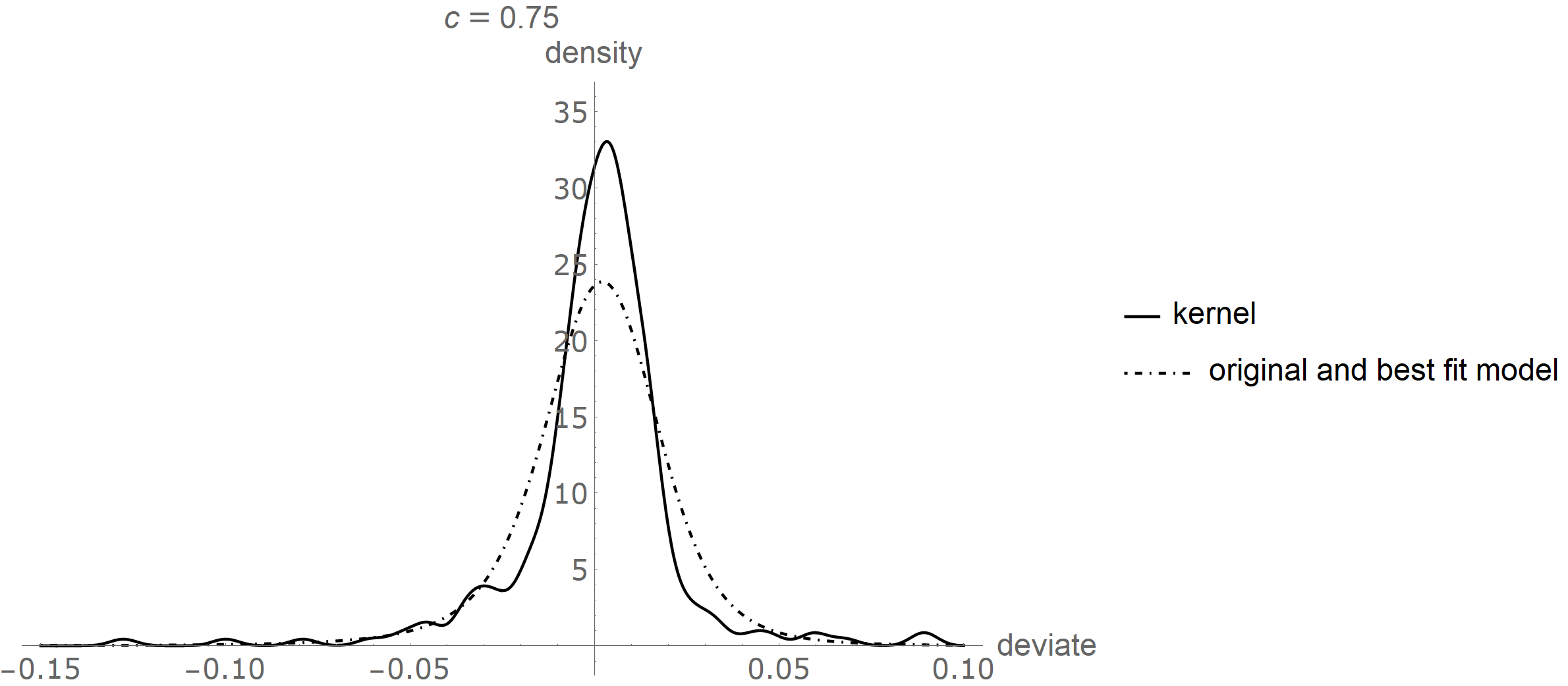}
  \caption{\label{fig:7a} Tempered stable estimated densities with regulation degrees for S\&P500 daily returns}
\end{figure}

\begin{figure}[H]
  \centering
  \ContinuedFloat
  \includegraphics[scale=0.39]{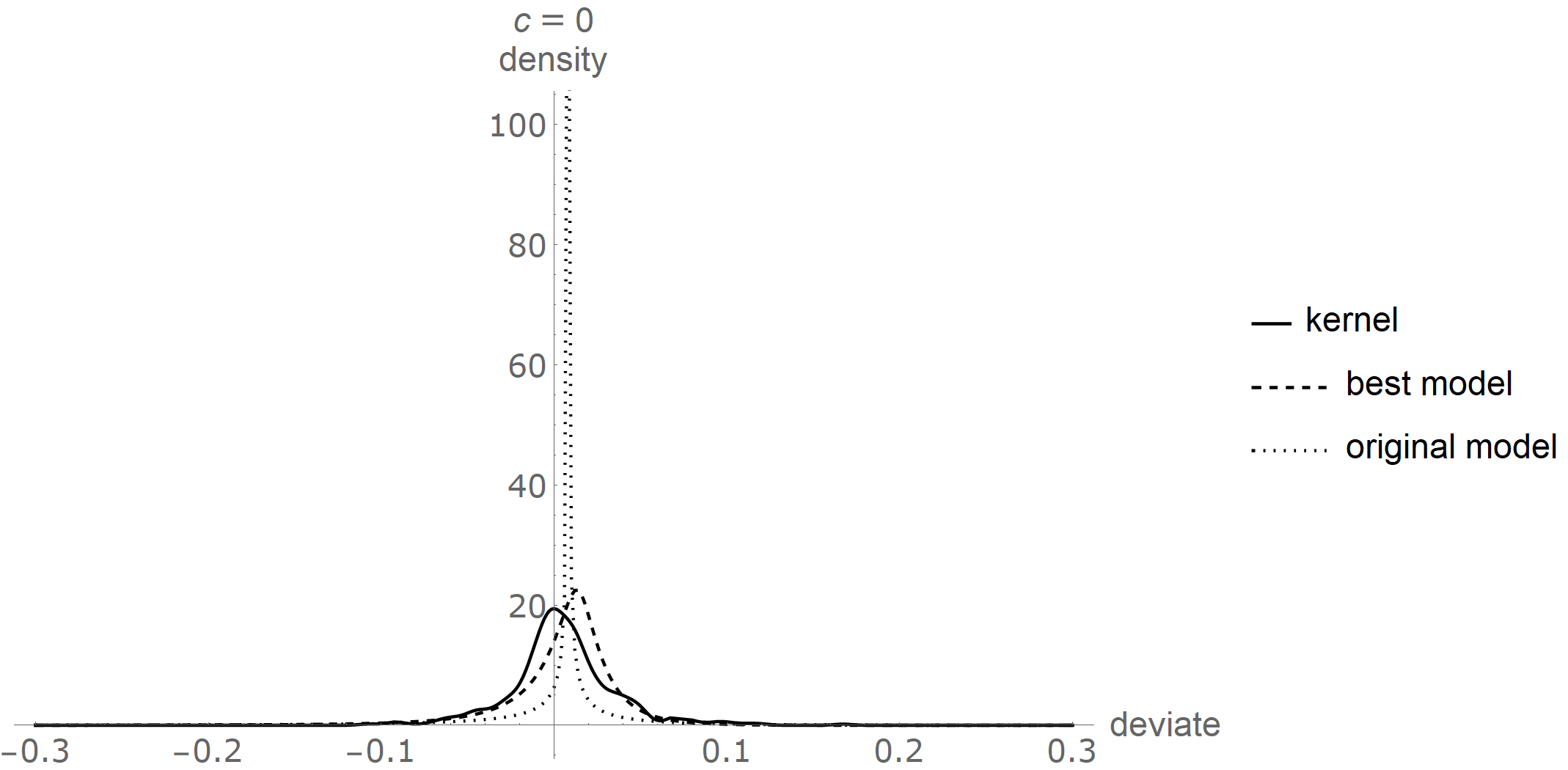}
  \includegraphics[scale=0.39]{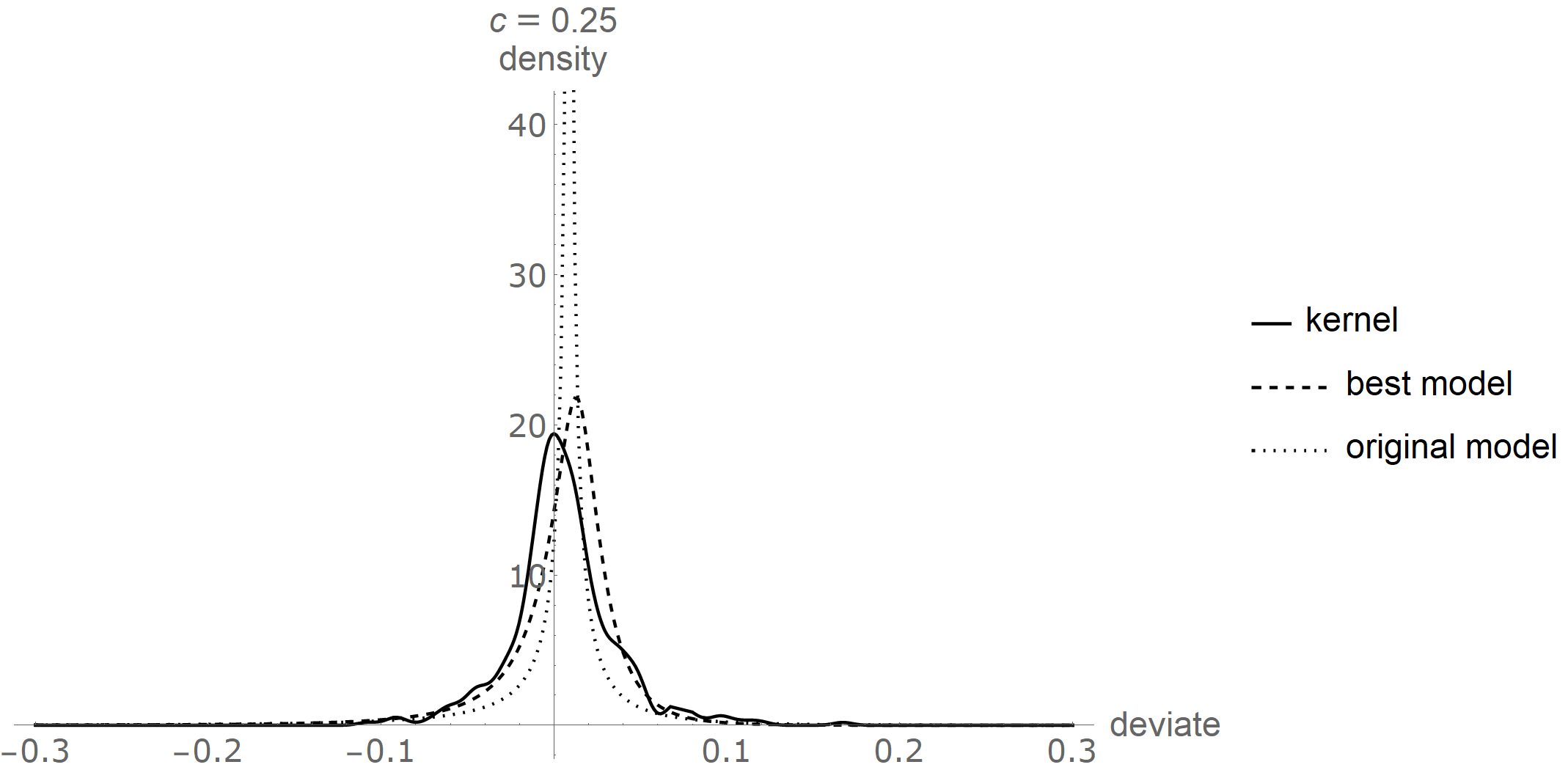}
  \includegraphics[scale=0.39]{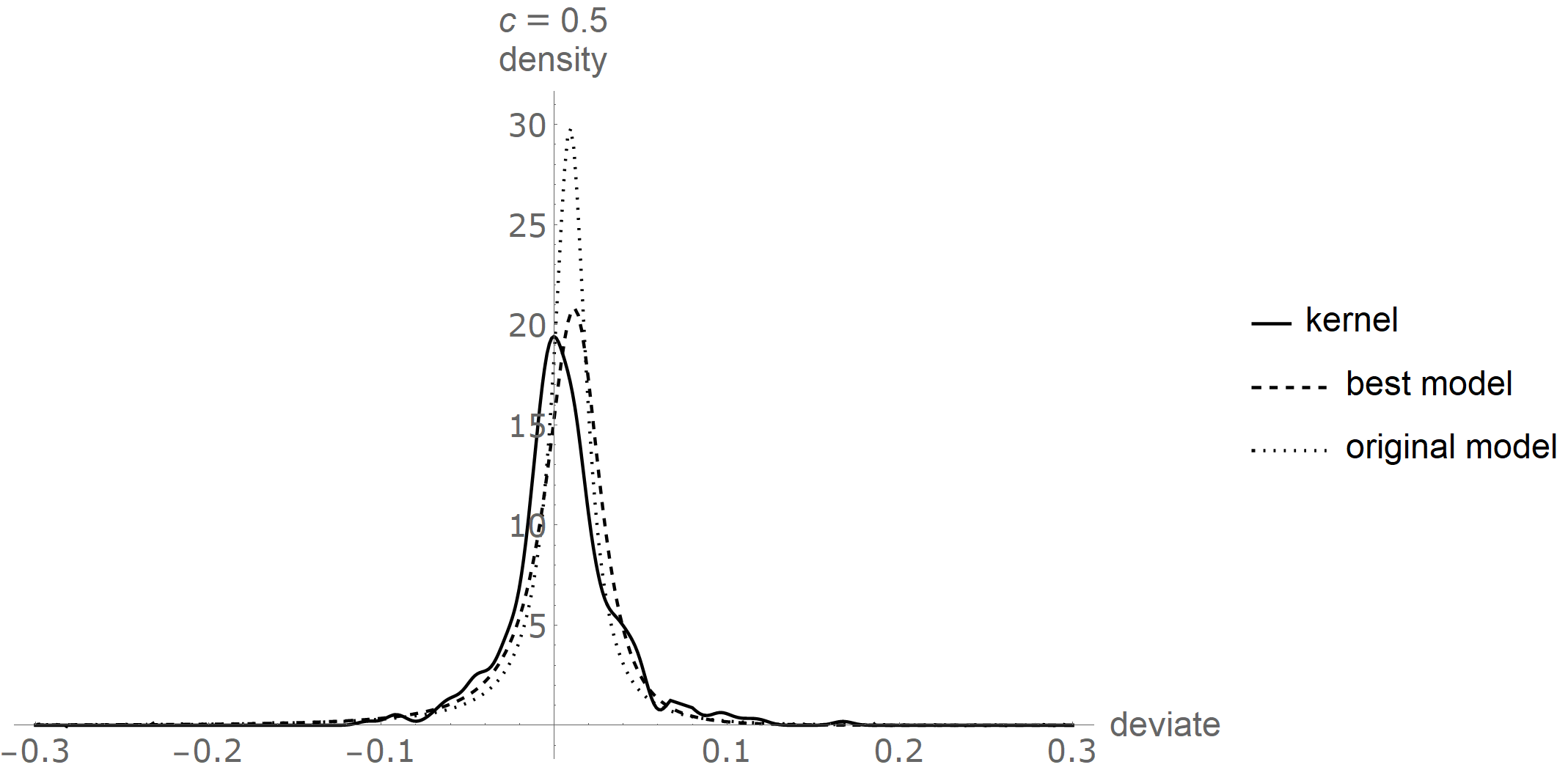}
  \includegraphics[scale=0.39]{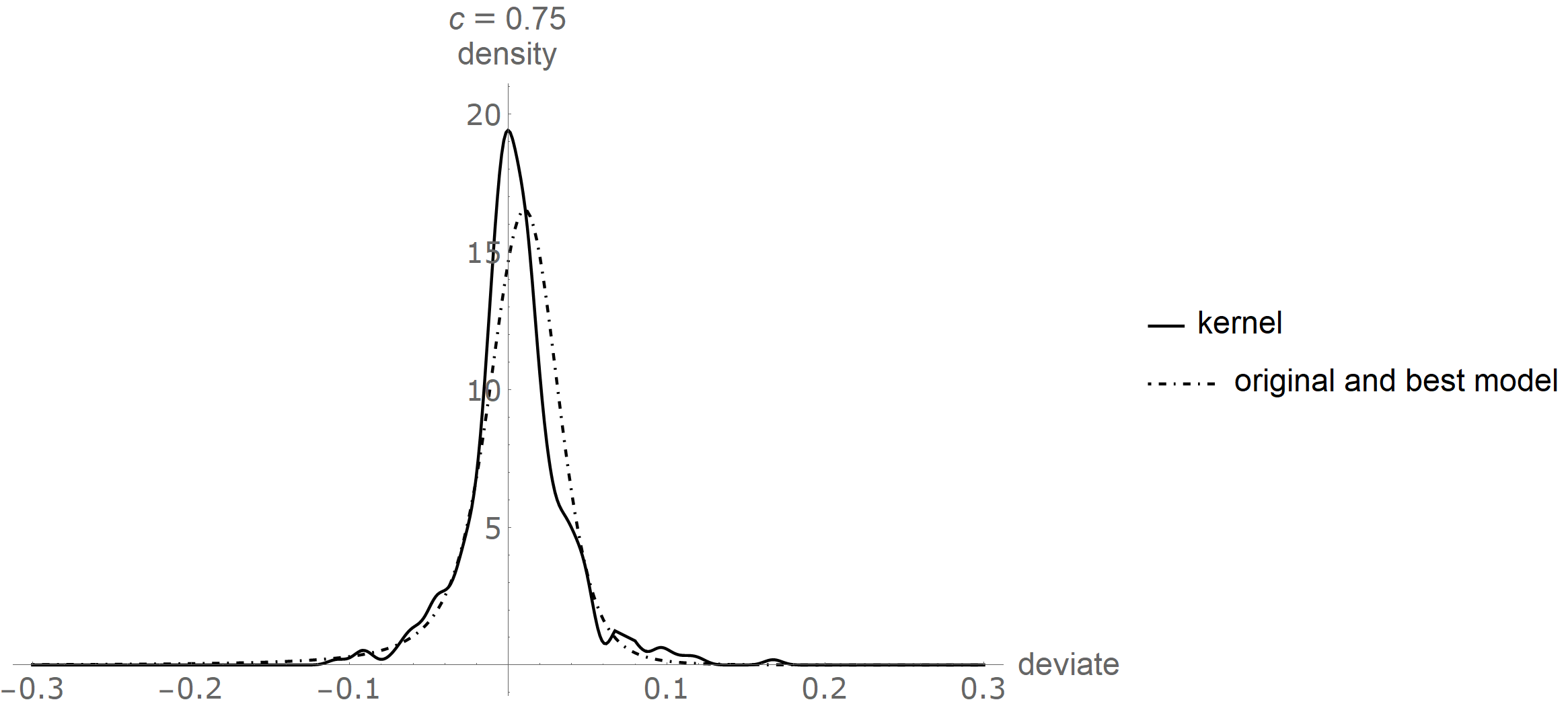}
  \caption{\label{fig:7b} Tempered stable estimated densities with regulation degrees for Bitcoin daily returns}
\end{figure}

For the first exercise involving jump--diffusion models, from Table \ref{tab:2a} and Table \ref{tab:2b}, the steady changes in the parameter values are clear for both data sets, which show robustness of the moment-based estimation. Specifically, while the estimate of the drift parameter, $\hat{\mu}_{n}$, has only marginally increased and that of the Brownian dispersion parameter, $\hat{\sigma}_{n}$, remains fairly stable, the infinite-divisibility parameter estimate $\hat{\lambda}_{n}$ has increased geometrically with the degree $n$ regardless of the regulation type. This is in keeping with our original motivation to not reduce the measure of the number of trades. While decreasing, the rate of change of the rate parameter estimate $\hat{b}_{n}$, on the other hand, does depend on the regulation type, and is the lowest under the type I, moderate under the type II, and the highest (actually in orders of reciprocal factorials) under the type III. This is because the rate parameter plays the role of adjusting the local movement sizes of stochastic clocks in consonance with the decreasing effect from regulation. Speaking of the profile log-likelihood, we see that any type of regulation has resulted in a notable improvement for some $n\geq1$. While the first two types tend to achieve a maximal value relatively quickly for $n\leq10$, the type III seems to be increasing the likelihood much more slowly. A simple explanation is that the third type has the mildest enlargement effect on distributional asymmetry and tail heaviness among the three. Besides, for both data sets the best models (with a ``$\star$'') have all outperformed their counterparts with exponentially distributed jumps, also implying that jump--diffusion models with distorted uniformly distributed jump amplitudes can be used as an ideal replacement of the widely accepted exponential jump--diffusion models in most applications. The density plots in Figure \ref{fig:5} mainly show that despite closeness of the three estimated model densities, the peak of the one of the best fit model tends to lie between those of the models with uniform or exponential jumps, confirming its balancing characteristic mentioned in Subsection \ref{sec:4.1}.

For the second exercise with Gaussian-mixed tempered stable models, it is seen from Table \ref{tab:3a}, Table \ref{tab:3b}, Table \ref{tab:3c}, Table \ref{tab:3d}, Table \ref{tab:4a}, Table \ref{tab:4b}, Table \ref{tab:4c}, and Table \ref{tab:4d} that the parameter estimates also change steadily with the regulation degree, for all regulation types and both data sets. For any chosen value of the family parameter $c$ and any regulation type, apart from marginal increases in the location parameter estimate $\hat{\mu}_{n}$ and the absolute Brownian drift estimate $|\hat{\theta}_{n}|$, the conspicuous value increase of the estimate $\hat{a}_{n}$ of the infinite-divisibility parameter shows that clock regulation is indeed able to cater to a large number of trades while maintaining a significant level of asymmetry and tail heaviness. In the meantime, the interpretation of the value change of the rate parameter estimate $\hat{b}_{n}$ is the same as in the jump--diffusion case in the first exercise. Viewing from the profile log-likelihood, the effect of regulation deteriorates rapidly as $c$ increases, with the variance gamma case ($c=0$) undergoing the greatest improvement. This phenomenon is within expectation, because in the limit as $c\nearrow1$ the tempered stable subordinator will approach the usual drift clock of calendar time, which is by no means affected by regulation (or path averaging); equivalently, its Gaussian mixture will tend to a drifted Brownian motion, by the central limit theorem.\footnote{When minute-level data are employed, the value of $c$ will probably be near 1 for liquid assets (see [Todorov and Tauchen, 2011, Table 5] \cite{TT}), in which case the underlying stochastic clock is close to the calendar time. This in turn shows that clock regulation is particularly useful for daily or lower-frequency data, making up their deficiency relative to the high-frequency counterparts which inevitably come at additional costs.} Such a trend is more salient for the S\&P500 returns than for the Bitcoin returns, depending on the relative extent of asymmetrical and tail risks. To be fair, an intriguing observation is that by increasing the degree $n$ of regulation the same level of log-likelihood (distributional information) has been achieved, whichever $c$-value is being used. Since $c$ has an important dynamical meaning in terms of the local regularity of sample paths, and hence long-term trading intensity (recalling Section \ref{sec:1}), clock regulation has therefore introduced a useful dimension for parameter estimation of a tempered stable model for any fixed value of the family parameter $c$. The fundamental difference, as we shall restate, is that the regulation degree $n$ only changes the sizes of local movements and does not affect the Blumenthal--Getoor index (Corollary \ref{cor:1}). The density plots in Figure \ref{fig:7a} and Figure \ref{fig:7b} further confirm the benefits of increasing the value of $a$ (a large number of trades), as the original model density without clock regulation seems overly ``peaked'' (almost degenerate, or near trading halts) when its moments are matched to the empirical, while proper clock regulation helps get rid of such degeneracy tendency.

To demonstrate the method outlined in Subsection \ref{sec:5.2}, we also do a calibration exercise on Bitcoin options using the Gaussian-mixed tempered stable models. The data set is directly taken from [Xia, 2021, \text{Sect.} 4.2] \cite{X2} and consists of $M=40$ Bitcoin call option prices quoted as of July 11th, 2020. The strike prices $K$ range from \$3,000 to \$32,000 and there are four maturities: $T=19,47,166,257$ days. At the time of quoting, the Bitcoin price stood at $S_{0}=\$9,232.98$. We have $q=r=0$ for digital assets. To keep the demonstration concise, we also only consider four values of the family parameter $c\in\{0,0.25,0.5,0.75\}$ and choose the domain of regulation degrees $\mathfrak{N}=\{0,1,2,3,4,5\}$ (with $|\mathfrak{N}|=6$), which already give rise to 48 independent models. Each model is calibrated according to (\ref{5.2.2}).

\begin{table}[H]\small
  \centering
  \ContinuedFloat*
  \caption{\label{tab:5a} Calibration results on Bitcoin options ($\sim$six significant digits)}
  \begin{tabular}{c|c|ccc|c|c}
    \hline
    \multicolumn{7}{c}{\textbf{type II}} \\ \hline
    \multirow{6}[10]{*}{\tabincell{c}{$c=0$ \\ (VG)}} & $n$ & $\hat{a}_{n}$ & $\hat{b}_{n}$ & $\hat{\theta}_{n}$ & MAPE (\%) & CPU time (sec) \\ \hline
    & 0 & 52.8148 & 97.315 & 1.00541 & 24.7965 & 213.984 \\
    & 1 & 69.312 & 63.5709 & 0.835575 & 24.423 & 221.781 \\
    & 2 & 79.9247 & 34.5063 & $-0.0841175$ & 24.368 & 1208.5 \\
    & 3 & 99.9403 & 21.5024 & $-0.271352$ & 24.2491 & 1011.41 \\
    & 4 & 99.8584 & 10.7058 & $-0.806309$ & 23.7375 & 2647.44 \\
    & 5 & 99.9986 & 5.30486 & $-1.21115$ & 23.3089$\ast$ & 3961.56 \\ \hline
    \multirow{6}[10]{*}{$c=0.25$} & $n$ & $\hat{a}_{n}$ & $\hat{b}_{n}$ & $\hat{\theta}_{n}$ & MAPE (\%) & CPU time (sec) \\ \hline
    & 0 & 5.29036 & 24.9255 & $-1.4834$ & 22.5461 & 347.219 \\
    & 1 & 5.55635 & 9.93689 & $-1.70271$ & 20.484 & 501.266 \\
    & 2 & 6.89235 & 5.1679 & $-1.64386$ & 21.5499 & 3029.7 \\
    & 3 & 2.81083 & 0.476929 & $-1.48728$ & 19.065$\ast$ & 13629.9 \\
    & 4 & - & - & - & - & - \\
    & 5 & - & - & - & - & - \\ \hline
    \multirow{6}[10]{*}{\tabincell{c}{$c=0.5$ \\ (NIG)}} & $n$ & $\hat{a}_{n}$ & $\hat{b}_{n}$ & $\hat{\theta}_{n}$ & MAPE (\%) & CPU time (sec) \\ \hline
    & 0 & 0.485557 & 1.37746 & $-1.59762$ & 19.2856 & 135.625 \\
    & 1 & 0.664932 & 0.572124 & $-1.46531$ & 19.2146 & 378.313 \\
    & 2 & 1.03635 & 0.343634 & $-1.44776$ & 19.1939$\ast$ & 618.609 \\
    & 3 & - & - & - & - & - \\
    & 4 & - & - & - & - & - \\
    & 5 & - & - & - & - & - \\ \hline
    \multirow{6}[10]{*}{$c=0.75$} & $n$ & $\hat{a}_{n}$ & $\hat{b}_{n}$ & $\hat{\theta}_{n}$ & MAPE (\%) & CPU time (sec) \\ \hline
    & 0 & 0.288703 & 9.88632 & $-1.18918$ & 23.3422 & 164.516 \\
    & 1 & 0.331241 & 0.608326 & $-1.32828$ & 20.6779$\ast$ & 304.75 \\
    & 2 & - & - & - & - & - \\
    & 3 & - & - & - & - & - \\
    & 4 & - & - & - & - & - \\
    & 5 & - & - & - & - & - \\ \hline
\end{tabular}\\
  MAPE: mean absolute percentage error | VG: variance gamma | NIG: normal inverse Gaussian
\end{table}

\begin{table}[H]\small
  \centering
  \ContinuedFloat
  \caption{\label{tab:5b} Calibration results on Bitcoin options ($\sim$six significant digits)}
  \begin{tabular}{c|c|ccc|c|c}
    \hline
    \multicolumn{7}{c}{\textbf{type III}} \\ \hline
    \multirow{6}[10]{*}{\tabincell{c}{$c=0$ \\ (VG)}} & $n$ & $\hat{a}_{n}$ & $\hat{b}_{n}$ & $\hat{\theta}_{n}$ & MAPE (\%) & CPU time (sec) \\ \hline
    & 0 & 52.8148 & 97.315 & 1.00541 & 24.7965 & 213.984 \\
    & 1 & 69.312 & 63.5709 & 0.835575 & 24.423 & 221.781 \\
    & 2 & 25.7893 & 7.01917 & $-1.82847$ & 21.9319$\ast$ & 1386.55 \\
    & 3 & 80.8854 & 5.83995 & $-0.551056$ & 23.9882 & 1071.81\\
    & 4 & 90.3674 & 1.28759 & $-0.786714$ & 23.8168 & 968.281 \\
    & 5 & 97.9971 & 0.232557 & $-0.926549$ & 23.6565 & 764.641 \\ \hline
    \multirow{6}[10]{*}{$c=0.25$} & $n$ & $\hat{a}_{n}$ & $\hat{b}_{n}$ & $\hat{\theta}_{n}$ & MAPE (\%) & CPU time (sec) \\ \hline
    & 0 & 5.29036 & 24.9255 & $-1.4834$ & 22.5461 & 347.219 \\
    & 1 & 5.55635 & 9.93689 & $-1.70271$ & 20.484 & 501.266 \\
    & 2 & 4.40503 & 1.49887 & $-1.62404$ & 18.5857$\ast$ & 1442.34 \\
    & 3 & 7.26858 & 0.477655 & $-1.8991$ & 19.6144 & 673.453 \\
    & 4 & 9.61409 & 0.0816453 & $-1.64024$ & 18.9021 & 681.328 \\
    & 5 & 9.94952 & 0.0053706 & $-2.12896$ & 24.8041 & 1469.98 \\ \hline
    \multirow{6}[10]{*}{\tabincell{c}{$c=0.5$ \\ (NIG)}} & $n$ & $\hat{a}_{n}$ & $\hat{b}_{n}$ & $\hat{\theta}_{n}$ & MAPE (\%) & CPU time (sec) \\ \hline
    & 0 & 0.485557 & 1.37746 & $-1.59762$ & 19.2856 & 135.625 \\
    & 1 & 0.664932 & 0.572124 & $-1.46531$ & 19.2146 & 378.313 \\
    & 2 & 1.30094 & 0.242459 & $-1.45114$ & 19.1893$\ast$ & 492.594 \\
    & 3 & 3.02094 & 0.0752527 & $-1.54985$ & 19.2294 & 1282.11 \\
    & 4 & 9.89913 & 0.0457464 & $-1.59855$ & 19.9204 & 815.641 \\
    & 5 & 9.95359 & 0.0000857145 & $-1.25003$ & 23.4282 & 1602.34 \\ \hline
    \multirow{6}[10]{*}{$c=0.75$} & $n$ & $\hat{a}_{n}$ & $\hat{b}_{n}$ & $\hat{\theta}_{n}$ & MAPE (\%) & CPU time (sec) \\ \hline
    & 0 & 0.288703 & 9.88632 & $-1.18918$ & 23.3422 & 164.516 \\
    & 1 & 0.331241 & 0.608326 & $-1.32828$ & 20.6779 & 304.75 \\
    & 2 & 0.978496 & 0.728833 & $-1.24462$ & 22.117 & 762.734 \\
    & 3 & 1.82492 & 0.00378187 & $-1.30428$ & 19.4049$\ast$ & 750.078 \\
    & 4 & 6.33734 & 0.000522405 & $-1.31269$ & 19.43 & 708.563 \\
    & 5 & 27.7777 & 0.000615278 & $-1.37646$ & 19.8326 & 903.156 \\ \hline
  \end{tabular}\\
  MAPE: mean absolute percentage error | VG: variance gamma | NIG: normal inverse Gaussian
\end{table}

As discussed in Subsection \ref{sec:5.2}, the optimal parameter values are subject to minor instability with joint calibration. For this reason, the main goal of this calibration exercise is to demonstrate that tempered stable models with clock regulation can be used with high efficiency and well expected to outperform the originals with unregulated clocks for financial derivatives valuation. The calibration results are summarized in Table \ref{tab:5a} and Table \ref{tab:5b}, along with (locally) minimal pricing error and CPU time (with only a quad-core processor). Notably, under type-II regulated clocks, since the computational complexity of the formula (\ref{4.2.1}) can increase rapidly with the degree of regulation, some models for $c>0$ and $n\geq2$ have not been successfully calibrated within a reasonable time limit (24 hours), and hence their results have been deliberately left out for comparability, while the remaining results are sufficient to serve our conclusions. Likewise, for each chosen value of $c$, the model with the best fit is marked ``$\ast$.''

From Table \ref{tab:5a} and Table \ref{tab:5b} we can see that adopting regulated clocks has significantly improved the calibration results, regardless of the value of $c$. More specifically, for type-II regulation, better performance is generally produced by increasing $n$, though the computation time grows drastically simultaneously. The latter feature is especially notable if $c>0$, when one is forced to apply (\ref{4.2.1}) outside the unit disk, whereas for the variance gamma model with $c=0$, the more elementary formula (\ref{4.2.2}) is used instead. In contrast, under type-III regulation, computation time does not scale with the degree of regulation, but there is oftentimes an optimal value $n<5$ that gives a best fit.

\begin{figure}[H]
  \centering
  \includegraphics[scale=0.39]{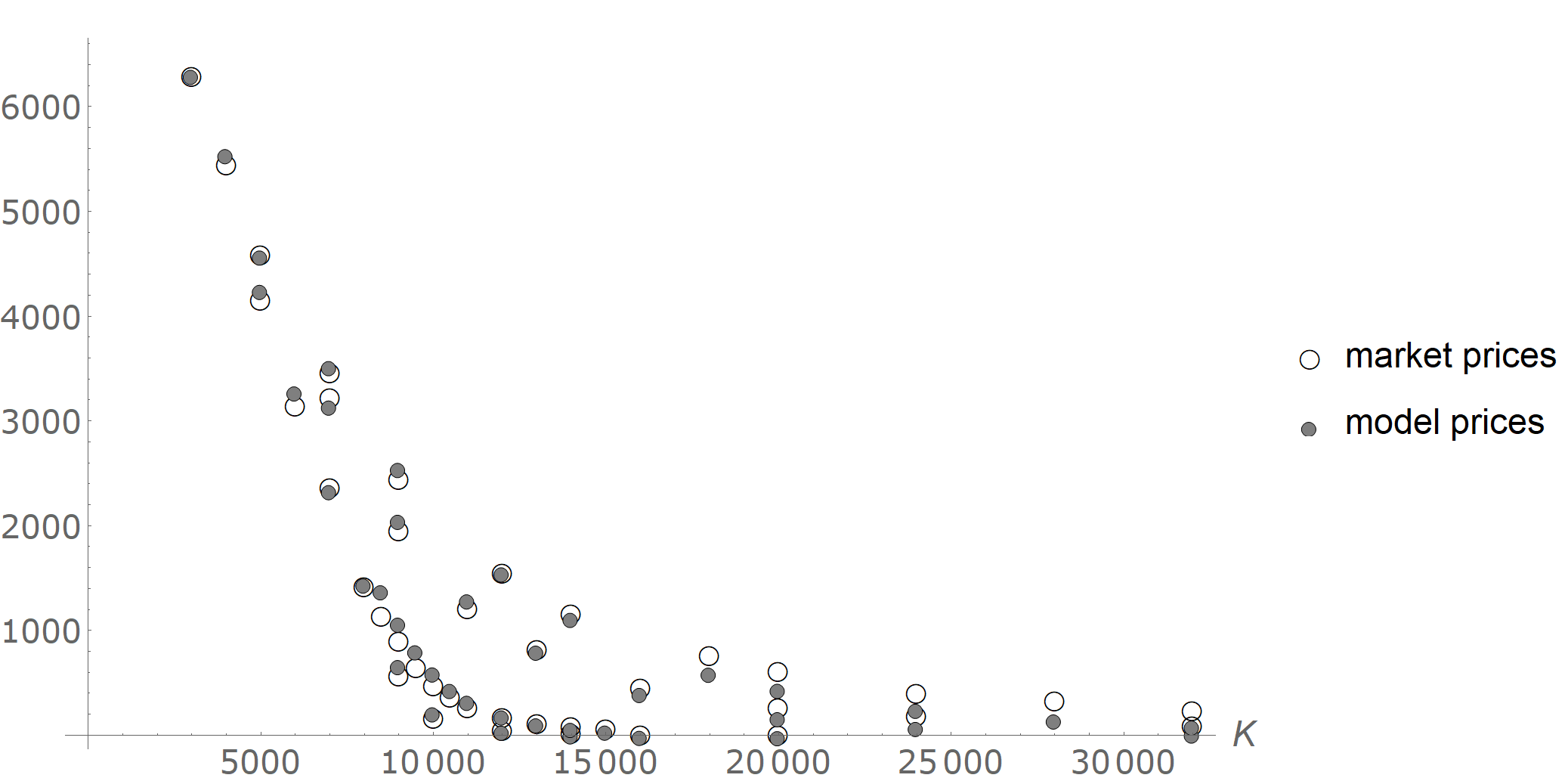}
  \caption{Bitcoin option prices (market vs best fit model)}
  \label{fig:8}
\end{figure}

Speaking of the optimized parameter values, we observe that, for each fixed value of $c$, $\hat{a}_{n}$ usually increases with $n$ while $\hat{b}$ decreases, which is similar to the results of statistical estimation. Besides, $\hat{\theta}_{n}$ seems to be very stable in general, implying, yet again, that the Brownian drift parameter is largely uninfluenced by clock regulation. From Table \ref{tab:5a} and Table \ref{tab:5b} we identify the best fit model as the one with $c=0.25$ and $n=2$ under type-III regulation, whose implied prices are further visually compared with the market prices in Figure \ref{fig:8}.

\begin{figure}[H]
  \centering
  \includegraphics[scale=0.39]{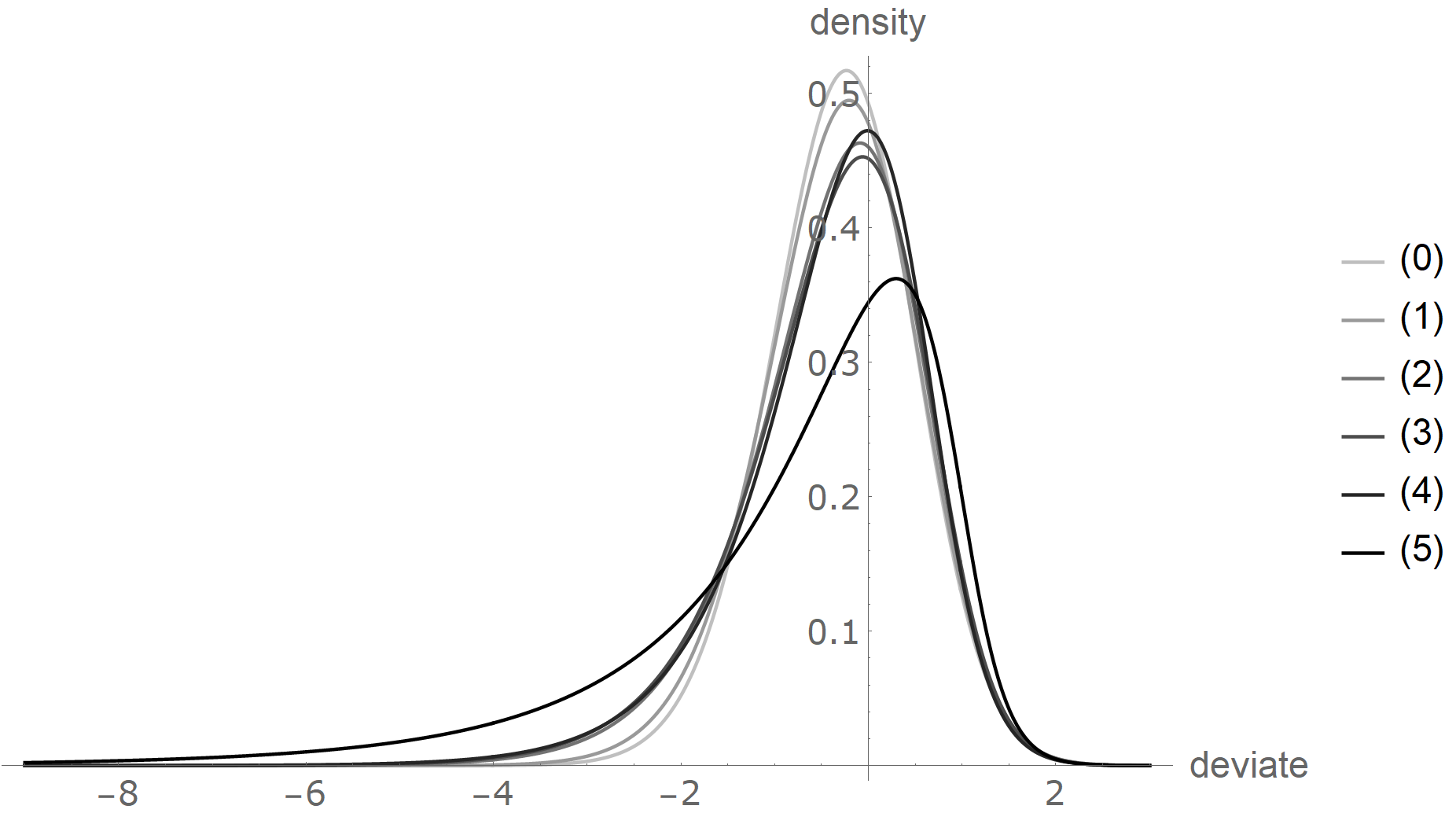}
  \caption{Option-implied risk-neutral densities of Bitcoin log-returns (type III, $c=0.25$)}
  \label{fig:9}
\end{figure}

Furthermore, we compare in Figure \ref{fig:9} the risk-neutral densities (namely the $\Q$-densities) of annualized Bitcoin log-returns (driven by $\Xi^{(n),Z}_{1}$ for $n\in\mathbb{N}\cap[0,5]$) to further illustrate the static effect of clock regulation, though for succinctness the illustration is only done for the subcase ``type III, $c=0.25$'' of which the best fit model is a special case. The formula (\ref{4.2.7}) is applied because $c<1/2$, along with the choice $\mu_{n}=-\log\phi_{\Xi^{(n),Z}_{1}}(-1)$.

\medskip

\section{Concluding remarks}\label{sec:7}

In this paper we have studied the regulation of stochastic clocks built from L\'{e}vy subordinators for the purpose of tackling asymmetrical and tail risks typically observed in financial returns without making undesirable changes on parameters linked to important aspects of the trading activity. These parameters are those measuring infinite divisibility or $\alpha$-stability and so making such an improvement has potential not only in enhancing capacity of the original processes but in balancing various modeling emphases involving the microstructure of data as well. From the viewpoint of static distributions, a direct effect is to arbitrarily enlarge skewness and (excess) kurtosis, while being able to fix the mean and variance and not decrease the jump intensity (or approach degeneracy).

The methodology comes in three concrete recipes. Despite a common motivation from repeated continuous-time averaging, the recipes vary in their impact on the original distributional asymmetry and tail heaviness as well as the computational complexity of resultant distribution formulae, which properties are summarized and compared in Table \ref{tab:6} for the convenience of the reader.

\begin{table}[H]\small
  \centering
  \caption{Summary of properties of regulated stochastic clocks}
  \label{tab:6}
  \begin{tabular}{c|c|c|c}
    \hline
    Category & type I & type II & type III \\ \hline
    induction nature & average of sample paths & average of log-LT & \tabincell{c}{quasi-average of \\ sample paths} \\ \hline
    regulating kernel & incomplete gamma & inverse gamma & Riemann-Liouville \\ \hline
    L\'{e}vy measure distortion & severe & moderate & mild \\ \hline
    Skew/EKurt enlargement & faster than exponential & exponential & \tabincell{c}{slower than linear \\ or bounded} \\ \hline
    \tabincell{c}{explicit formulae \\ (Poisson)} & \tabincell{c}{LT: $n=1$ only \\ LM: any $n>0$} & \tabincell{c}{LT: $n\in\mathbb{N}_{++}$ only \\ LM: any $n>0$} & LT and LM: any $n>0$ \\ \hline
    \tabincell{c}{explicit formulae \\ (tempered stable)} & LT and LM : $n=1$ only & LT and LM: $n\in\mathbb{N}_{++}$ & LT and LM: any $n>0$ \\ \hline
  \end{tabular}\\
  LT: Laplace transform | LM: L\'{e}vy measure
\end{table}

Generally speaking, stochastic clock regulation works well on any L\'{e}vy subordinator as long as the prerequisite (\ref{2.1}) of generic-time log-LT analyticity is satisfied and can be utilized to obtain a good number of newfangled L\'{e}vy models for financial modeling and other uses. It is only futile for processes whose generic-time log-LT is linear or undefined in the neighborhood of zero.\footnote{Note that the log-LT cannot be a finite-order polynomial of degree greater than two; see [Lukacs, 1970, \text{Thm.} 7.3.5] \cite{L}.} Clearly, in the former case the subordinator is degenerate and effectively identical to the clock of calendar time, and in the latter it does not have a finite third-order cumulant. The resultant mixed models then have skewness and excess kurtosis either equal to zero or undefined (including infinity) -- obviously, both zero and infinity cannot be further ``enlarged'' by multiplication.

When applied to a Poisson process, the regulated clocks have led to compound Poisson processes with bounded jumps having a truncated exponential-like distribution and hence corresponding jump--diffusion models inheriting the advantages of both the uniform distribution ([Yan and Hanson, 2006] \cite{YH}) and the (usual) exponential distribution ([Kou, 2002] \cite{K3}). As our empirical study on S\&P500 and Bitcoin returns has shown, when moment-based estimation is employed to match the empirical skewness and kurtosis, the profile likelihood for the new models with regulation degree $n>0$ is indeed improved relative to the uniform and the pure-exponential, which signifies that the distorted uniform distributions are suitable for modeling downside risks in a jump--diffusion framework.

Secondly, operating the regulated clocks on the Gaussian-mixed tempered stable models generates a series of modified processes with unchanged path regularity but enhanced distributional asymmetry and tail heaviness. The empirical results imply that the effect of regulation is most conspicuous when the family parameter $c$ is small, close to the case of variance gamma processes, while it gradually diminishes as $c$ approaches the upper limit 1, when the mixed model tends to Gaussianity. More importantly, by varying the degree $n$ of regulation, one can expect to easily take account of empirical skewness and kurtosis of returns without losing essential information on the actual distribution, regardless of which value of $c$ is in use. Put differently, in the presence of the hyperparameter $n$, the value of $c$ can be confidently fixed for a certain data frequency or taken from other sources of estimation, since $n$, benignly, has no effect on the local regularity of the stochastic process.

The methodology that the present paper has put forward is a jump-off point and so can be advanced in various directions. For instance, as we have mentioned since Section \ref{sec:1}, it can be extended to subordinators of the Sato type (with non-stationary increments) with reasonable effort as well as the establishment of mean-reverting time changes of Ornstein--Uhlenbeck type ([Carr and Wu, 2004] \cite{CW}) for stochastic volatility. Also, an ongoing research is concerned with developments into a multidimensional setting, especially for the Gaussian mixed models, which will be conducive to portfolio analysis and derived optimization problems. We note that multivariate subordination of Brownian motion via tempered stable subordinators has received much attention in the literature; see, e.g., [Luciano and Semeraro, 2010a] \cite{LSa}, [Luciano and Semeraro, 2010b] \cite{LSb}, [P\'{e}rez-Abreu and Stelzer, 2014] \cite{P-AS}, and [Mandoza-Arriaga and Linetsky, 2016] \cite{M-AL}. Of course, the two special cases discussed in Section \ref{sec:4} are also far from exhaustive, and it is possible to consider operating the regulating kernels on other choices of L\'{e}vy subordinators, e.g., compound Poisson-exponential processes or generalized inverse Gaussian processes ([Barndorff-Nielsen and Halgreen, 1977] \cite{B-NH}), and derive the moment-based estimation procedures in the same spirit as in Subsection \ref{sec:5.1}. Last but not least, the empirical aspect may be strengthened by studying other kinds of financial time series, such as individual stocks, exchange rates, volatility indices, etc., in order to explore how the degree of clock regulation is fundamentally reflected in trading volumes or relative illiquidity and its variation with different levels of asymmetrical and tail risks.

\clearpage

\appendix
\gdef\thesection{Appendix \Alph{section}}

\renewcommand{\theequation}{A.\arabic{equation}}

\section{Mathematical proofs}\label{A}

\textbf{Theorem \ref{thm:1}}

\begin{proof}
Let $t>0$ be fixed. By applying integration-by-parts to (\ref{2.1.1}) $n$ times we have the fractional integral representation
\begin{equation}\label{A.1}
  \bar{X}^{(n)}_{t}=\frac{1}{t\Gf(n)}\int^{t}_{0}X_{s}\log^{n-1}\frac{t}{s}\dd s=\int^{t}_{0}\bigg(1-\frac{\Gf(n,\log(t/s))}{\Gf(n)}\bigg)\dd X_{s},
\end{equation}
where the second equality follows from the substitution $\log(t/s)\mapsto s$.\footnote{The first recipe generalizes the fractional integration of nonnegative processes to logarithmic kernels, as for every $n\geq1$ and $t>0$ the integrand $(1/(t\Gf(n))X_{s}\log^{n-1}(t/s))_{s\in(0,t]}$ in (\ref{A.1}) constitutes a nonnegative stochastic process.} The representation (\ref{A.1}) together with the fact that $X$ is a L\'{e}vy process allows to write
\begin{equation*}
  \E e^{-u\bar{X}^{(n)}_{t}}=\prod^{t}_{0}\E\bigg(\exp\bigg(-uX_{1}\bigg(1-\frac{\Gf(n,\log(t/s))}{\Gf(n)}\bigg)\bigg)\bigg)^{\dd s} =\exp\bigg(\int^{t}_{0}\log\phi_{X_{1}}\bigg(\bigg(1-\frac{\Gf(n,\log(t/s))}{\Gf(n)}\bigg)u\bigg)\dd s\bigg),
\end{equation*}
where $\prod^{\cdot}_{\cdot}$ is the geometric integral sign and which leads to the first formula in (\ref{2.1.5}) with the substitution $s/t\mapsto s$.

For the other two formulae it suffices to identify the corresponding fractional integral representations. For (\ref{2.1.3}), observe that for each $n\geq1$, $\tilde{X}^{(n)}$ is the running average of the given L\'{e}vy process $Y^{(n-1)}$, so that the LT of $\tilde{X}^{(n)}_{t}$ at generic time satisfies the recurrence relation
\begin{equation}\label{A.2}
  \phi_{\tilde{X}^{(n)}_{t}}(u)=\exp\bigg(\frac{t}{u}\int^{u}_{0}\log\phi_{\tilde{X}^{(n-1)}_{1}}(v)\dd v\bigg),\quad\Re u\geq0,
\end{equation}
due to the independent increments property of $Y^{(n-1)}$. The recurrence can be resolved easily via integration-by-parts, given the initial condition $\tilde{X}^{(0)}\equiv X$, in the same vein as (\ref{A.1}), yielding the second formula in (\ref{2.1.5}).

Also, since the integrals in (\ref{2.1.4}) are simply iterated, an application of Cauchy's repeated integration formula immediately gives
\begin{equation*}
  \breve{X}^{(n)}_{t}=\frac{1}{t^{n}\Gf(n+1)}\int^{t}_{0}(t-s)^{n}\dd X_{s},
\end{equation*}
which result shows that each $\breve{X}^{(n)}$ can be treated as a (times-scaled) usual fractional L\'{e}vy process under the Riemann--Liouville kernel.
\end{proof}

\medskip

\noindent \textbf{Theorem \ref{thm:2}}

\begin{proof}
Comparing the first formula in (\ref{2.1.5}) with the first formula in (\ref{2.2.1}), we obtain the following Fredholm integral equation:
\begin{equation*}
  \int^{\infty}_{0}(e^{-uz}-1)\bar{\nu}^{(n)}(\dd z)=\int^{1}_{0}\int^{\infty}_{0}\big(e^{-(1-\Gf(n,-\log s)/\Gf(n))uz}-1\big)\nu(\dd z)\dd s,
\end{equation*}
but as $1-\Gf(n,-\log s)/\Gf(n)\geq0$ for all $n\geq1$ and $s\in[0,1]$, the substitution $(1-\Gf(n,-\log s)/\Gf(n))z\mapsto z$ together with the Fubini--Tonelli theorem yields
\begin{equation}\label{A.3}
  \bar{\nu}^{(n)}(\dd z)=\int^{1}_{0}\nu\bigg(\dd\frac{\Gf(n)z}{\Gf(n)-\Gf(n,-\log s)}\bigg)\dd s,\quad z\geq0.
\end{equation}
Since $[0,1]\ni s\mapsto\Gf(n)/(\Gf(n)-\Gf(n,-\log s))\in[1,\infty)$ forms a strictly increasing function for each $n\geq1$, we can apply the further substitution $\Gf(n)/(\Gf(n)-\Gf(n,-\log s))\mapsto y$ to obtain the first formula in (\ref{2.2.2}), after some rearrangement.

Then, for each fixed $n\geq1$, comparing the second formula in (\ref{2.1.5}) with the second formula in (\ref{2.2.1}) and interchanging integrals gives the following integral equation:
\begin{equation*}
  \int^{\infty}_{0}(e^{-uz}-1)\tilde{\nu}^{(n)}(\dd z)=\int^{\infty}_{0}\int^{1}_{0}(e^{-uvz}-1)\dd v\tilde{\nu}^{(n-1)}(\dd z),
\end{equation*}
which is equivalent, on rearranging the right-hand side, to
\begin{equation*}
  \int^{\infty}_{0}(e^{-uz}-1)\tilde{\nu}^{(n)}(\dd z)=\int^{\infty}_{0}(e^{-uz}-1)\int^{\infty}_{1}\frac{\tilde{\nu}^{(n-1)}(\dd(yz))}{y^{2}}\dd y.
\end{equation*}
It leaves us with no choice but
\begin{equation*}
  \tilde{\nu}^{(n)}(\dd z)=\int^{\infty}_{1}\frac{\tilde{\nu}^{(n-1)}(\dd(yz))}{y^{2}}\dd y.
\end{equation*}
This is yet another recurrence equation similar to the ones solved before, subject to the initial condition $\tilde{\nu}^{(0)}\equiv\nu$. Upon integration-by-parts we obtain the second formula in (\ref{2.2.2}).

It is straightforward to obtain the third formula in (\ref{2.2.2}) by mimicking the steps taken for the first.

With absolute continuity with respect to Lebesgue measure, the corresponding formulae for L\'{e}vy densities in (\ref{2.2.3}) follow from direct calculations.
\end{proof}

\medskip

\noindent \textbf{Corollary \ref{cor:1}}

\begin{proof}
According to the definition of the Blumenthal--Getoor index, it suffices to observe that the relations in (\ref{2.2.2}) (or (\ref{2.2.3})) entail integration over domains bounded away from zero. Then the sigma-finiteness of the L\'{e}vy measures along with the Fubini--Tonelli theorem implies that the finiteness of the integrals $\int^{1}_{0}|z|^{p}\bar{\nu}^{(n)}(\dd z)$, $\int^{1}_{0}|z|^{p}\tilde{\nu}^{(n)}(\dd z)$, and $\int^{1}_{0}|z|^{p}\breve{\nu}^{(n)}(\dd z)$, for any $n>0$, are all equivalent to that of $\int^{1}_{0}|z|^{p}\nu(\dd z)$, for any $p>0$.
\end{proof}

\noindent \textbf{Corollary \ref{cor:2}}

\begin{proof}
Fix $n>0$. The construction of the coefficients $C_{m,n}$ and $1/((mn+1)\Gf^{m}(n+1))$, $m\in\mathbb{N}_{++}$, for the type I and type III, respectively, is a direct result of Theorem \ref{thm:1}, or the fractional integral representations (\ref{A.1}) and (\ref{A.2}). For the type II, since induction takes place for each $n$, the coefficients $1/(m+1)^{n}$, $m\in\mathbb{N}_{++}$, are obtained by noting that $C_{m,1}=\int^{1}_{0}(1-s)^{m}\dd s=1/(m+1)$.

The rationality of $C_{m,n}$'s is seen from the fact that, for every $n\in\mathbb{N}_{++}$, $\Gf(n)\equiv(n-1)!$ and the finite series representation $\Gf(n,-\log s)=(n-1)!s\sum^{n-1}_{k=0}(-\log s)^{k}/k!$, and $\int^{1}_{0}s^{j}(-\log s)^{k}\dd s=k!/(j+1)^{k+1}$ for any $j,k\in\mathbb{N}$.
\end{proof}

\medskip

\noindent \textbf{Corollary \ref{cor:3}}

\begin{proof}
The formulae follow immediately by replacing $X$ with its regulated counterparts in (\ref{3.1.2}).
\end{proof}

\medskip

\noindent \textbf{Corollary \ref{cor:4}}

\begin{proof}
According to Corollary \ref{cor:1}, it is enough to consider the constant mixture $\xi^{X}$ based on the L\'{e}vy subordinator $X$. Suppose $\kappa_{1},\kappa_{2}>0$. Following (\ref{3.1.2}) and the L\'{e}vy--Khintchine representation, the L\'{e}vy measure of $\xi^{X}$ is given by
\begin{equation*}
  \nu_{\xi^{X}}(\dd z)=\nu_{X}\bigg(\dd\frac{z}{\kappa_{1}}\bigg)+\nu_{-X}\bigg(\dd\frac{z}{\kappa_{2}}\bigg),\quad z\in\mathbb{R}\setminus\{0\}.
\end{equation*}
Thus, it follows from the inequalities
\begin{equation*}
  \int^{1}_{0}z^{p}\nu_{X}\bigg(\dd\frac{z}{\kappa_{1}}\bigg)\leq\int^{1}_{-1}|z|^{p}\nu_{\xi^{X}}(\dd z)\leq2\max\bigg\{\int^{1}_{0}z^{p}\nu_{X}\bigg(\dd\frac{z}{\kappa_{1}}\bigg),\int^{1}_{0}z^{p}\nu_{X}\bigg(\dd\frac{z}{\kappa_{2}}\bigg)\bigg\}, \quad p>0
\end{equation*}
that $\mathfrak{B}(\xi^{X})=\mathfrak{B}(X)$, as required. If either $\kappa_{1}$ or $\kappa_{2}$ is zero the result is immediate.
\end{proof}

\noindent \textbf{Corollary \ref{cor:5}}

\begin{proof}
The formulae follow immediately by replacing $X$ with its regulated counterparts in (\ref{3.2.2}).
\end{proof}

\medskip

\noindent \textbf{Corollary \ref{cor:6}}

\begin{proof}
Again, with Corollary \ref{cor:1} we only consider the Gaussian mixture $\Xi^{X}$ based on $X$. By construction, the L\'{e}vy measure of $\Xi^{X}$ satisfies
\begin{equation*}
  \nu_{\Xi^{X}}(\dd y)=\int^{\infty}_{0}\nu_{X}(\dd s)\gamma_s(\dd y),\quad y\in\mathbb{R}\setminus\{0\},
\end{equation*}
where $\gamma_{t}$, for $t>0$, represents the Gaussian measure with mean $\theta t$ and variance $t$, that is, the probability measure of the drifted Brownian motion $\theta t+W_{t}$. For every $p>0$, by the Fubini--Tonelli theorem we have
\begin{equation*}
  \int_{-1}^{1} |y|^{p} \int^{\infty}_{0} \nu_{X}(\dd s) \gamma_{s}(\dd y) = \int^{\infty}_{0} \nu_{X}(\dd s) \E\big[|\theta s + \sqrt{s}\zeta|^{p} \mathbb{1}_{A_{s}}\big],
\end{equation*}
$\zeta$ being a standard normal random variable and $A_{s} := \{|\theta s + \sqrt{s}\zeta| \leq 1\}$. From the fact that
\begin{equation*}
  \int^{\infty}_{1}\nu_{X}(\dd s) \E [|\theta s + \sqrt{s}\zeta|^{p} \mathbb{1}_{A_{s}}] \leq \nu_{X}([1,\infty))<\infty,
\end{equation*}
the finiteness of $\int^{\infty}_{0} \nu_{X}(\dd s) \E\big[|\theta s + \sqrt{s} \zeta|^{p} \mathbb{1}_{A_{s}}\big]$ is the same as that of $\int^{1}_{0} \nu_{X}(\dd s) \E \big[|\theta s + \sqrt{s} \zeta|^{p} \mathbb{1}_{A_{s}} \big]$. Towards that end, we first apply the $C^{p}$-inequality to obtain for every $p > 2 \mathfrak{B}(X)$
\begin{align*}
    \int^{1}_{0} \E \big[|\theta s + \sqrt{s} \zeta|^{p} \mathbb{1}_{A_{s}} \big] \nu_{X}(\dd s)&\leq 2^{(p-1)^{+}}\bigg(\theta^{p} \int^{1}_{0} s^{p} \E [\mathbb{1}_{A_{s}}] \nu_{X}(\dd s) + \int^{1}_{0} s^{p/2} \E \big[ |\zeta|^{p} \mathbb{1}_{A_{s}} \big] \nu_{X}(\dd s)\bigg) \\
    &\leq 2^{(p-1)^{+}}\bigg(\theta^{p} \int^{1}_{0} s^{p} \nu_{X}(\dd s) + \E[|\zeta|^{p}] \int^{1}_{0} s^{p/2} \nu_{X}(\dd s)\bigg) \\
    & < \infty,
\end{align*}
so that $\mathfrak{B}(\Xi^{X}) \leq 2 \mathfrak{B}(X)$. Second, for every $p < 2 \mathfrak{B}(X)$, observe that
\begin{align*}
    \int^{1}_{0} \E \big[ |\theta s + \sqrt{s} \zeta|^{p} \mathbb{1}_{A_{s}} \big] \nu_{X}(\dd s) &\geq \int^{1}_{0} \frac{s^{p/2}}{\sqrt{2 \pi}} \int^{\theta \sqrt{s} + 1/\sqrt{s}}_{\theta \sqrt{s}} (x - \theta \sqrt{s})^{p} e^{-z^{2}/2} \dd x \; \nu_{X}(\dd s) \\
    &= \int^{1}_{0} \frac{s^{p/2}}{\sqrt{2 \pi}} \int_{0}^{1/\sqrt{s}} x^{p} \exp\left( -\frac{(x + \theta \sqrt{s})^2}{2}\right) \dd x \; \nu_{X}(\dd s) \\
    &\geq \int^{1}_{0} \frac{s^{p/2}}{\sqrt{2 \pi}}  \int^{1}_{0} x^{p} \exp\left( -\frac{(x + \theta)^2}{2}\right) \dd x \; \nu_{X}(\dd s) \\
    &=  \E \big[ (\zeta - \theta)^{p} \mathbb{1}_{[\theta,\theta + 1]}(\zeta) \big]\int^{1}_{0}s^{p/2}\nu_{X}(\dd s) \\
    &=\infty,
\end{align*}
where the second equality uses the substitution $x-\theta\sqrt{s}\mapsto x$ and which implies that $\mathfrak{B}(\Xi^{X}) \geq 2 \mathfrak{B}(X)$. It is therefore concluded that $\mathfrak{B}(Z)=2\mathfrak{B}(X)$, as required.
\end{proof}

\medskip

\noindent \textbf{Proposition \ref{pro:1}}

\begin{proof}
For the type I, plugging the stated LT of $X_{1}$ into the first formula in (\ref{2.1.5}) and applying substitution as in (\ref{A.3}) the LT of $\bar{X}^{(n)}_{1}$ reads for $n>0$
\begin{align}\label{A.4}
  \phi_{\bar{X}^{(n)}_{1}}(u)&=\exp\bigg(\lambda\bigg(\int^{1}_{0}e^{-(1-\Gf(n,-\log s)/\Gf(n))u}\dd s-1\bigg)\bigg) \nonumber\\
  &=\exp\bigg(\lambda\bigg(\Gf(n)\int^{1}_{0}\frac{e^{-uz}}{\mathrm{Q}^{n-1}(n,1-z)}\dd z-1\bigg)\bigg),
\end{align}
which is clearly an entire function.

The general formula for the type II is immediate from the second formula in (\ref{2.1.5}). If $n\in\mathbb{N}_{++}$, then with $\int^{1}_{0}(-\log v)^{n-1}/\Gf(n)\dd v=1$ we have that
\begin{align*}
  \frac{1}{\Gf(n)}\int^{1}_{0}e^{-uv}(-\log v)^{n-1}\dd v&=\sum^{\infty}_{k=0}\int^{1}_{0}\frac{(-\log v)^{n-1}(-uv)^{k}}{\Gf(n)k!}\dd v\\
  &=\sum^{\infty}_{k=0}\frac{(-u)^{k}}{(k+1)^{n}k!}\\
  &=\sum^{\infty}_{k=0}\frac{(-u)^{k}}{k!}\Bigg(\frac{\prod^{k}_{j=1}j}{\prod^{k}_{j=1}(j+1)}\Bigg)^{n}\\
  &=\;_{n}\F_{n}(1,\dots,1;2\dots,2;-u),
\end{align*}
where the second equality follows from the proof of Corollary \ref{cor:2} and which is also entire.

For the type III, note that
\begin{equation*}
  \phi_{\breve{X}^{(n)}_{1}}(u)=\exp\bigg(\lambda\bigg(\int^{1}_{0}e^{-s^{n}u/\Gf(n+1)}\dd s-1\bigg)\bigg),\quad u\in\mathbb{C}
\end{equation*}
applies to non-integer values of $n$ as well. The integral can be directly evaluated with the substitution $s^{n}\mapsto s$, leading to the third formula in (\ref{4.1.1}).

The formulae (\ref{4.1.2}) for the L\'{e}vy densities are direct consequences of (\ref{2.2.2}) or (\ref{2.2.3}) after taking $\nu(\dd z)=\lambda\delta_{\{1\}}(\dd z)$ with the dirac measure.
\end{proof}

\medskip

\noindent \textbf{Proposition \ref{pro:2}}

\begin{proof}
Given $n\in\mathbb{N}_{++}$, starting with the LT of $Y^{(n)}_{1}$, we have using binomial expansion and the proof of Corollary \ref{cor:2} that for $c>0$,
\begin{align*}
  \frac{1}{\Gf(n)}\int^{1}_{0}(-\log v)^{n-1}(uv+b)^{c}\dd v&=\sum^{\infty}_{k=0}\int^{1}_{0}\binom{c}{k}\frac{b^{c}(-\log v)^{n-1}}{\Gf(n)}\bigg(\frac{uv}{b}\bigg)^{k}\dd v\\
  &=b^{c}\sum^{\infty}_{k=0}\binom{c}{k}\frac{(u/b)^{k}}{(k+1)^{n}}\\
  &=b^{c}\sum^{\infty}_{k=0}\frac{1}{k!}\bigg(\frac{u}{b}\bigg)^{k}\frac{(-1)^{k}(k!)^{n}\prod^{k}_{j=1}(j-c-1)}{((k+1)!)^{n}}\\
  &=b^{c}\;_{n+1}\F_{n}\bigg(1,\dots,1,-c;2,\dots,2;-\frac{u}{b}\bigg),
\end{align*}
where in the first equality the interchange of integration and summation for $u\in(D_{0}(b))^{\complement}\setminus(-\infty,-b]$ is justified by analytic continuation.

The limiting case $c=0$ (corresponding to the gamma process) cannot be immediately gleaned from the last result. Similarly, by performing a separate computation we have
\begin{align*}
  \frac{1}{\Gf(n)}\int^{1}_{0}(-\log v)^{n-1}\log\frac{b}{uv+b}\dd v&=\sum^{\infty}_{k=1}\int^{1}_{0}\frac{(-\log v)^{n-1}}{k\Gf(n)}\bigg(-\frac{uv}{b}\bigg)^{k}\dd v\\
  &=\sum^{\infty}_{k=1}\frac{(-u/b)^{k}}{k(k+1)^{n}}\\
  &=\sum^{\infty}_{k=1}\bigg(-\frac{u}{b}\bigg)^{k}\Bigg(\frac{1}{k}-\sum^{n}_{j=1}(k+1)^{-j}\Bigg)\\
  &=n-\bigg(1+\frac{b}{u}\bigg)\log\frac{u+b}{b}+\frac{b}{u}\sum^{n}_{j=2}\Li_{j}\bigg(-\frac{u}{b}\bigg).
\end{align*}
Thus, after some rearrangement we arrive at the first formula in (\ref{4.2.1}).

Explicit computation of the L\'{e}vy density $\tilde{\ell}^{(n)}$ ($n\in\mathbb{N}_{++}$) is subtler because the integral inside (\ref{2.2.3}) is concentrated on the tail behavior of $\nu$, which obstructs direct expansion. Alternatively, observe that the recurrence relation leading to it, i.e.,
\begin{equation}\label{A.8}
  \tilde{\ell}^{(n)}(z)=\int^{\infty}_{1}\frac{\tilde{\ell}^{(n-1)}(yz)}{y}\dd y,\quad z>0,
\end{equation}
is nothing but an Euler integral equation for the G-function (documented in [Bateman and Erd\'{e}lyi, 1954, \text{Eq.} 20.5.2] \cite{BE}), or
\begin{equation*}
  \int^{\infty}_{0}\frac{1}{y}\G^{l,m}_{p,q}\bigg(
  \begin{array}{cc}
    \alpha_{1},\dots,\alpha_{p} \\
    \beta_{1},\dots,\beta_{q}
  \end{array}
  \bigg|wy\bigg)\dd y=\G^{l+1,m}_{p+1,q+1}\bigg(
  \begin{array}{cc}
    \alpha_{1},\dots,\alpha_{p},1 \\
    0,\beta_{1},\dots,\beta_{q}
  \end{array}
  \bigg|w\bigg)
\end{equation*}
to be precise, where $0\leq m\leq p$ and $0\leq l\leq q$ are all integers and $w\equiv w(z)>0$ is linear. By employing an induction argument the general solution to (\ref{A.8}) reads
\begin{equation*}
  \tilde{\ell}^{(n)}(z)=d\G^{l+n,m}_{p+n,q+n}\bigg(
  \begin{array}{cc}
    \alpha_{1},\dots,\alpha_{p},\overbrace{1,\dots,1}^{n} \\
    \underbrace{0,\dots,0}_{n},\beta_{1},\dots,\beta_{q}
  \end{array}
  \bigg|w\bigg),\quad n\in\mathbb{N},
\end{equation*}
for some yet-to-be-determined numbers $m,l,p,q\in\mathbb{N}$ and $d>0$. Given the initial condition that $\tilde{\ell}^{(0)}(z)=ae^{-bz}/z^{c+1}$, $z>0$, we are forced to choose $m=p=0$ and $l=q=1$, with $\beta_{1}=-c-1$, in which case the G-function specializes for $n=0$ to
\begin{equation*}
  \G^{1,0}_{0,1}\bigg(
  \begin{array}{cc}
     \\
    -c-1
  \end{array}
  \bigg|w\bigg)=\frac{e^{-w}}{w^{c+1}},
\end{equation*}
from where it remains to set $w=bz$ and $d=ab^{c+1}$.

The formulae for the type III are easily obtained by adapting the binomial expansion argument used for the type II.
\end{proof}

\medskip

\noindent \textbf{Proposition \ref{pro:3}}

\begin{proof}
Let $X$ be a Poisson process and consider its constant mixture plus an independent Brownian motion. From Subsection \ref{sec:3.1} and Subsection \ref{sec:4.1} we know that the model cumulants (with regulation degrees) over the time period of $\D$ are given by
\begin{align*}
  &K^{(n)}_{\D}(1)=\D\bigg(\mu_{n}-\rho(1,n)\frac{\lambda_{n}}{b_{n}}\bigg),\quad K^{(n)}_{\D}(2)=\D\bigg(\sigma^{2}_{n}+\rho(2,n)\frac{\lambda_{n}}{b^{2}_{n}}\bigg),\\
  &K^{(n)}_{\D}(3)=-\D\rho(3,n)\frac{\lambda_{n}}{b^{3}_{n}},\quad K^{(n)}_{\D}(4)=\D\rho(4,n)\frac{\lambda_{n}}{b^{4}_{n}},\quad n\in\mathfrak{N},
\end{align*}
where recall that $\mathfrak{N}\subsetneq\mathbb{R}_{+}$ is the truncated and discretized $n$-domain for experimentation, and $\rho(m,n)$'s are specified in (\ref{5.1.1}). We then immediately have the following quotient relation:
\begin{equation*}
  \frac{K^{(n)}_\D(4)}{K^{(n)}_\D(3)} = - \frac{\rho(4, n)}{\rho(3, n)} \frac{1}{b_n},
\end{equation*}
which combined with the four moment conditions in (\ref{5.1.2}) pins down the rate parameter estimate $\hat{b}_{n}$, and subsequently the other estimates in (\ref{5.1.3}).
\end{proof}

\medskip

\noindent \textbf{Proposition \ref{pro:4}}

\begin{proof}
Consider the case of the Gaussian mixture of $X$ being a tempered stable subordinator. Using the results obtained in Subsection \ref{sec:3.2} and Subsection \ref{sec:4.2} we have the following model cumulants:
\begin{align*}
  K_\D^{(n)}(1) & = \Delta( \mu_n + \theta_n \rho(1, n) K_{n;\D}(1) ), \\
  K_\D^{(n)}(2) & = \Delta ( \rho(1, n) K_{n;\D}(1) + \theta^2_n \rho(2, n) K_{n;\D}(2) ) , \\
  K_\D^{(n)}(3) & = \Delta ( 3 \theta_n \rho(2, n) K_{n;\D}(2) + \theta^3_n \rho(3, n) K_{n;\D}(3) ), \\
  K_\D^{(n)}(4) & = \Delta ( 3 \rho(2, n) K_{n;\D}(2) + 6 \theta^2_n \rho(3, n) K_{n;\D}(3) + \theta^4_n \rho(4, n) K_{n;\D}(4) ),
\end{align*}
where
\begin{align*}
  & K_{n;\D}(1) = \Gamma(1 - c) \frac{a_n}{b_n^{1 - c}},\quad K_{n;\D}(2)  = \frac{1 - c}{b_n} K_{n;\D}(1), \\
  & K_{n;\D}(3) = \frac{(2 - c) (1 - c)}{b_n^2} K_{n;\D}(1),\quad K_{n;\D}(4)  = \frac{(3 - c) (2 - c) (1 - c)}{b_n^3} K_{n;\D}(1)
\end{align*}
are the first four cumulants of $X_{\D}$ with regulation degree-subscripted parameters. From here one can deduce the following cumulant quotient relations:
\begin{align*}
  \frac{K_\D^{(n)}(3)}{K_\D^{(n)}(2)}
  & = \theta_n (1 - c) \bigg( 3 \frac{\rho(2, n)}{\rho(1, n)} + \theta_n^2 \frac{\rho(3, n)}{\rho(1, n)} \frac{(2 - c)}{b_n} \bigg) \bigg(b_n +\theta_n^2 \frac{\rho(2, n)}{\rho(1, n)} (1 - c) \bigg)^{-1}, \\
  \frac{K_\D^{(n)}(4)}{K_\D^{(n)}(2)}
  & = (1 - c) \bigg( 3 \frac{\rho(2, n)}{\rho(1, n)} + 6 \theta_n^2 \frac{\rho(3, n)}{\rho(1, n)} \frac{ 2 - c }{b_n} + \theta_n^4 \frac{\rho(4, n)}{\rho(1, n)} \frac{(3 - c) (2 - c) }{b_n^2} \bigg)\bigg(b_n + \theta_n^2 \frac{\rho(2, n)}{\rho(1, n)} (1 - c) \bigg)^{-1},
\end{align*}
which define a rational system for $b_{n}$ and $\theta_{n}$. Note that the latter two conditions in (\ref{5.1.2}) can be equivalently written as $K_\D^{(n)}(3)\big/K_\D^{(n)}(2)=[SK][V]^{1/2}$ and $K_\D^{(n)}(4)\big/K_\D^{(n)}(2)=[EK][V]$; from the first condition the sign of $\theta_{n}$ is seen to be determined by that of $[SK]$, and then using the second the equation (\ref{5.1.5}) governing its absolute value $|\theta_{n}|$ can be easily derived. Notice that, as long as the condition
\begin{equation*}
  \frac{3 - c}{2 - c} \frac{\rho(4, n) \rho(2, n)}{\rho(3, n)^2} < \frac{[EK]}{[SK]^2}
\end{equation*}
is fulfilled, a real solution of (\ref{5.1.5}) always exists and can be easily located numerically; otherwise, the equation may not admit a real solution. Upon rearranging terms we obtain the estimates in (\ref{5.1.4}) as desired.
\end{proof}

\medskip

\section{Alternative skewness and excess kurtosis enlargement for tempered stable subordinators}\label{B}

For each $n\in\mathbb{N}$ (including zero), let $X_{n}$ be a tempered stable subordinator with parametrization $\{a_{n}>0,b_{n}>0,c\in[0,1)\}$, and $X_{0}\equiv X$ ($a_{0}=a$ and $b_{0}=b$). We know from [K\"{u}chler and Tappe, 2013, \text{Eq.} 2.19] \cite{KT} that the mean and variance at fixed $t>0$ are respectively
\begin{equation*}
  \E X_{n,t}=\frac{\Gf(1-c)a_{n}t}{b_{n}^{1-c}},\quad\Var X_{n,t}=\frac{\Gf(2-c)a_{n}t}{b_{n}^{2-c}}.
\end{equation*}
We stick to writing $\#\in\{X,Y,Z\}$ as a placeholder for any type of regulated stochastic clocks, whose $m$th cumulant is $\rho(m,n)\in(0,1]$ times that of $X_{t}$ (see (\ref{5.1.1})). Then, from the quotient relation $\Var X_{n,t}/\E X_{n,t}=(1-c)/b_{n}$ and the first two formulae in (\ref{2.3.3}), (\ref{2.3.5}), and (\ref{2.3.6}) we have
\begin{equation*}
  \frac{\Var\#^{(n)}_t}{\E\#^{(n)}_t}=\frac{(1-c)\rho(2,n)}{b_{n}\rho(1,n)}.
\end{equation*}
Following $\#^{(n)}$ is understood to be $X_{n}$-clocks with parameters $a_{n}$, $b_{n}$, and $c$, as opposed to $X$-clocks. If we keep the mean and variance of $\#^{(n)}_{t}$ to be invariant across $n\geq1$, we need
\begin{align*}
  a_{n}=\frac{\rho^{1-c}(2,n)}{\rho^{2-c}(1,n)}a,\quad b_{n}=\frac{\rho(2,n)}{\rho(1,n)}b.
\end{align*}
It then follows that
\begin{align*}
  \Skew X_{n,t}
  &=\frac{\Gf(3-c)a_{n}/b_{n}^{3-c}}{(\Gf(2-c)a_{n}/b_{n}^{2-c})^{3/2}}=\frac{2-c}{\Gf(2-c)^{1/2}a_{n}^{1/2}b_{n}^{c/2}} =\frac{\rho(1,n)}{\rho^{1/2}(2,n)}\Skew X_{t}, \\
  \EKurt X_{n,t}
  &=\frac{\Gf(4-c)a_{n}/b_{n}^{4-c}}{(\Gf(2-c)a_{n}/b_{n}^{2-c})^{2}}=\frac{(3-c)(2-c)}{\Gf(2-c)a_{n}b_{n}^{c}} =\frac{\rho^{2}(1,n)}{\rho(2,n)}\EKurt X_{t}.
\end{align*}
Therefore, according to the last two formulae in (\ref{2.3.3}), (\ref{2.3.5}), and (\ref{2.3.6}) we obtain for any $n\in\mathbb{N}$
\begin{equation*}
  \Skew\#^{(n)}_{t}=\frac{\rho(1,n)\rho(3,n)}{\rho^{2}(2,n)}\Skew X_{t},\quad
  \EKurt\#^{(n)}_{t}=\frac{\rho^{2}(1,n)\rho(4,n)}{\rho^{3}(2,n)}\EKurt X_{t},\quad t>0,
\end{equation*}
which as well apply to any non-integer values of $n>0$. Specifying the last results for $\rho(m,n)$ according to (\ref{5.1.1}) delivers the results.

\end{document}